\documentclass[a4paper,11pt]{article}

\usepackage[a4paper,left=2.73cm,right=2.7cm,top=3cm,bottom=3.5cm]{geometry}

\usepackage{amsfonts,amsmath,amssymb,amsthm,tabstackengine}
\usepackage[usenames, dvipsnames]{color}
\usepackage{array}
\usepackage{multirow}
\usepackage{makecell}
\usepackage{caption}
\usepackage{rotating}
\usepackage{float}
\usepackage{pdflscape}
\usepackage{placeins}
\usepackage{boldline}

\usepackage{soul}

\allowdisplaybreaks

\usepackage{color, colortbl,xcolor}
\definecolor{mygray}{gray}{0.90}

\usepackage[colorlinks=true,linktocpage=true,linkcolor=red,citecolor=blue]{hyperref}
\usepackage{graphicx}

\usepackage{comment}

\numberwithin{equation}{section}

\newcommand{\be}{\begin{eqnarray}}
\newcommand{\ee}{\end{eqnarray}}

\begin{document}

\begin{titlepage}

\thispagestyle{empty}

\begin{center}

{\Large \textbf{Taxonomy of type II orientifold flux vacua in 3D}}

\vspace{40pt}

\quad{\large \bf \'Alvaro Arboleya}$^{(a,b)}$ \,\,\,\,\,\,\,  \large{,} \,\,\,\,  {\large \bf Gabriele Casagrande}$^{(c)}$

\vspace{8pt}

{\large \bf Adolfo Guarino}$^{(a,b)}$  \,\,\,\,  \large{,} \,\,\,\, {\large \bf Matteo Morittu}$^{(d)}$

\vspace{30pt}

{\normalsize  
${}^{a}$\,Departamento de F\'isica, Universidad de Oviedo,\\
Avda. Federico Garc\'ia Lorca 18, 33007 Oviedo, Spain.}
\\[3mm]

{\normalsize  
${}^{b}$\,Instituto Universitario de Ciencias y Tecnolog\'ias Espaciales de Asturias (ICTEA) \\
Calle de la Independencia 13, 33004 Oviedo, Spain.}
\\[7mm]

{\normalsize
${}^{c}$\,Department of Physics, Ben-Gurion University of the Negev,\\
Be’er-Sheva 84105, Israel.}
\\[7mm]

{\normalsize  
${}^{d}$\,Istituto di Istruzione Superiore ``Artemisia Gentileschi'',\\
Via Carlo Alberto Sarteschi 1, 54033 Carrara, Italy.}\\[15mm]

\texttt{arboleyaalvaro@uniovi.es}  ,  \, 
\texttt{casagran@post.bgu.ac.il} ,  \, \\[1mm]\texttt{adolfo.guarino@uniovi.es} , \,
\texttt{matteomorittu1995@gmail.com}

\vspace{20pt}

\vspace{20pt}

\abstract{

\noindent We complete the study initiated in \cite{Arboleya:2024vnp} and investigate three-dimensional (3D) flux vacua of type II orientifold reductions on twisted tori that include a single type of spacetime-filling O$p$-plane with $\,p=2,\ldots,9$. Restricting to $\textrm{SO}(3)$-invariant setups -- also known as RSTU-models -- and setting axions to zero, we exhaustively chart the landscape of 3D orientifold flux vacua. It consists of $\,56\,$ inequivalent multi-parametric families of AdS$_3$ and Mkw$_3$ flux vacua, all of them without negative masses in the spectrum of scalar fluctuations. Performing T-dualities, all the inequivalent vacua can be found in type IIB with either O9-, O5- or O3-planes. We show that: $i)\,$ type IIB with O$9$ fails to stabilise the volume of the internal space. $ii)\,$ type IIB with O5 realises the only cases of scale-separated AdS$_{3}$ vacua. $iii)\,$ type IIB with O$3$ provides only Mkw$_{3}$ vacua exhibiting a rich variety of supersymmetry-breaking patterns. Assuming that some mechanism (possibly non-perturbative) fixes the unstabilised modulus in type IIB with O$9$, we re-examine the possibility of achieving a weakly-coupled and scale-separated regime.
}

\end{center}

\end{titlepage}

\tableofcontents

\hrulefill
\vspace{10pt}

%\newpage

\baselineskip 4.55mm

\newpage

\section{Motivation and summary of results}

In the context of string compactifications, achieving scale separation between the external spacetime (usually an AdS geometry) and the internal manifold is commonly regarded as essential for obtaining a reliable lower-dimensional effective field theory. Examples of type IIA scale-separated AdS$_{4}$ \cite{Derendinger:2004jn,DeWolfe:2005uu,Camara:2005dc,Cribiori:2021djm} and AdS$_{3}$ \cite{Farakos:2020phe,VanHemelryck:2022ynr,Farakos:2023nms,Farakos:2023wps,Farakos:2025bwf} flux vacua have been constructed using internal manifolds with SU(3) and G$_{2}$ holonomy, respectively. However, the story seemed to be different for type IIB scale-separated AdS flux vacua.

For a long time no type IIB examples were known, and several studies suggested their nonexistence \cite{Emelin:2021gzx,VanHemelryck:2022ynr,Apers:2022vfp}. It was only recently that two examples of scale-separated type IIB AdS$_{3}$ flux vacua were presented in \cite{Arboleya:2024vnp} (see \cite{Arboleya:2025ocb} for a short review). In particular, these two examples were found in a class of type IIB orientifold reductions with a \textit{single type} of spacetime-filling O5-planes. They turned out to be non-supersymmetric vacua -- yet perturbatively (marginally) stable with respect to fluctuations of moduli in both the closed \cite{Arboleya:2024vnp} and open \cite{Arboleya:2025lwu} string sectors --, and they were compatible with an internal manifold with G$_{2}$-structure. Supersymmetric cousins of these two vacua were quickly found to exist, and new supersymmetric examples involving various types of spacetime-filling O5-planes and internal manifolds with G$_{2}$-structure were constructed in \cite{VanHemelryck:2025qok}. The search for scale-separated AdS$_{3}$ flux vacua has then been extended to other T-dual/S-dual setups (we will refer to them as \textit{duality frames}). Examples are the type IIB with O5/O9 (or type I) scale-separated flux vacua of \cite{Miao:2025rgf} which are T-dual to the type IIA with O2/O6 flux vacua constructed in \cite{Farakos:2020phe,Farakos:2025bwf}, or the Heterotic scale-separated flux vacua of \cite{Tringas:2025bwe} which are in turn S-dual to the type I flux vacua.

In this work, rather than tracking individual vacua across different duality frames, we will address the question of how common or rare scale-separated AdS$_3$ vacua are within type II orientifold reductions. In particular, and for the sake of concreteness, we will consider type II orientifold reductions that contain only a \textit{single type} of spacetime-filling O$p$-planes (and possibly also D$p$-branes) with $\,p=2,\ldots,9$. Imposing this restriction on the allowed sources enables a systematic construction of the corresponding flux models \cite{Arboleya:2024vnp}: they describe $\,\mathcal{N}=8\,$ (half-maximal) supergravities in 3D. Moreover, we will restrict to a sub-class of flux models that feature an additional $\textrm{SO}(3)$ symmetry. These models, dubbed RSTU-models in \cite{Arboleya:2024vnp}, describe a consistent $\,\mathcal{N}=2\,$ subsector of the half-maximal supergravities, and are the 3D analogue of the well-studied $\,\mathcal{N}=1\,$ STU-models in 4D (\textit{cf.} \cite{Kachru:2002he,Derendinger:2004jn,DeWolfe:2004ns,Camara:2005dc,Villadoro:2005cu,Derendinger:2005ph,DeWolfe:2005uu,Aldazabal:2006up,Dibitetto:2011gm}).\footnote{The additional complex scalar $\,R\,$ would arise from the dimensional reduction of the four-dimensional metric to three dimensions, which provides a Kaluza--Klein (KK) dilaton-like scalar and a KK vector which, in three dimensions, is dualised into an axion-like scalar.}

\begin{figure}[t]
\centering
\includegraphics[scale=0.45]{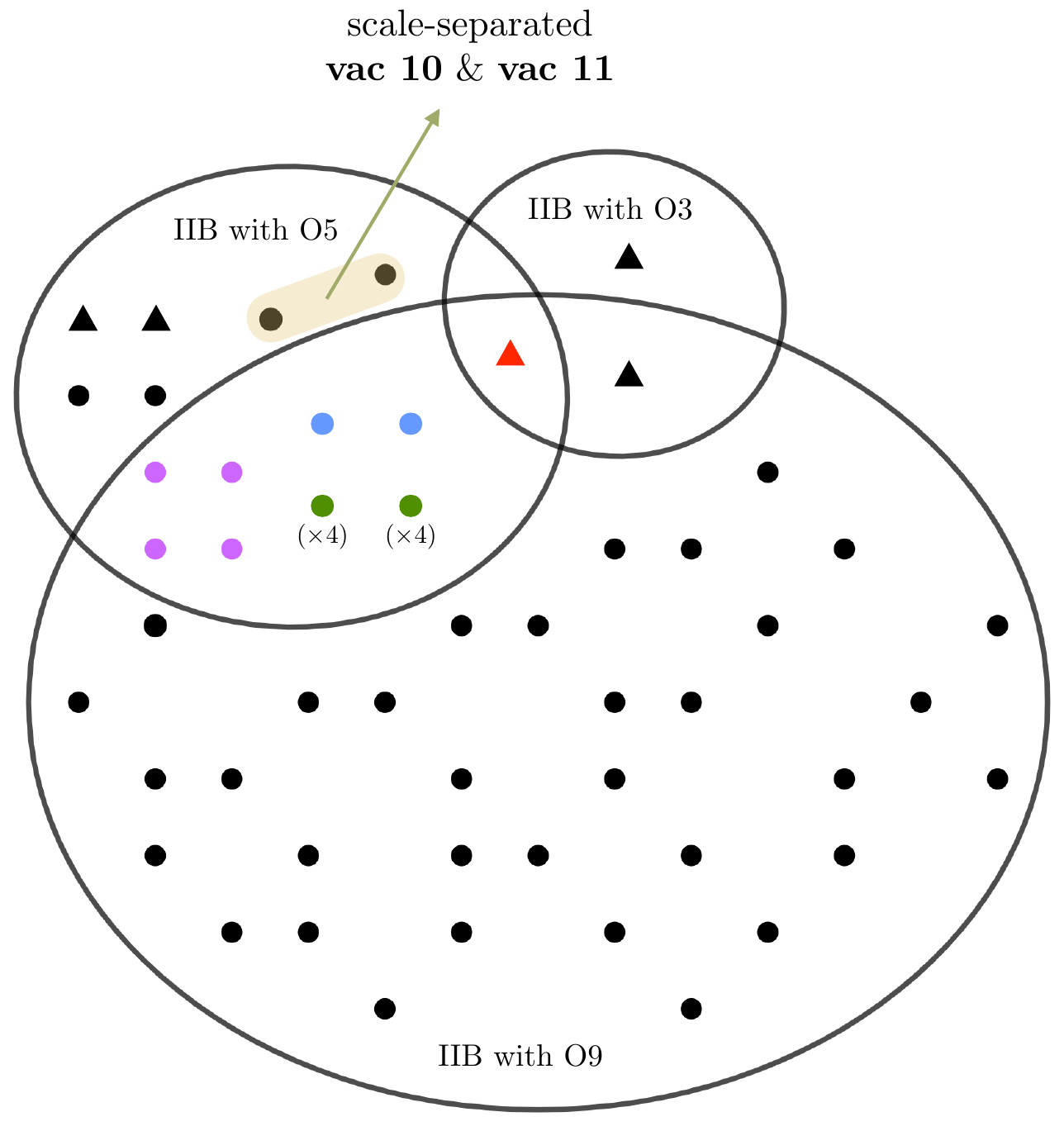}
\caption{Diagram of inequivalent AdS$_{3}$ ($\bullet$) and Mkw$_{3}$ ($\blacktriangle$) flux vacua with vanishing axions in the RSTU-models. All of the vacua arise in type IIB orientifold reductions with O9-, O5- or O3-planes. Some of the vacua (those sitting at intersections) admit different realisations related by T-duality. \textcolor{red}{$\blacktriangle$} (red triangle) is the trivial Mkw$_{3}$ vacuum with vanishing fluxes which, of course, exists in any duality frame. Despite the rich landscape, there are only two instances of weakly-coupled and scale-separated AdS$_3$ flux vacua which arise exclusively in a type IIB orientifold reduction with O$5$-planes.}
\label{picture_vacua}
\end{figure}

In order to be systematic and exhaustive when charting the landscape of vacua of the RSTU-models, we will demand a zero vacuum expectation value (VEV) for the axion-like moduli when extremising the scalar potential.\footnote{Depending on the O$p$-plane duality frame, some of the axions can still be set to zero without loss of generality. However, this is not true in general for the four axions of the RSTU-models \cite{Arboleya:2024vnp}.} This last simplification enables us to be fully exhaustive when scanning the flux vacua. The outcome is displayed in Figure~\ref{picture_vacua} and can be summarised as follows:

\begin{enumerate}

\item There are $\,1 \, \textrm{(trivial, \textcolor{red}{$\blacktriangle$})}+56\,$ inequivalent families of flux vacua, both Minkowski (Mkw$_{3}$) and anti-de Sitter (AdS$_{3}$), that may preserve $\,\mathcal{N}=0,1,2,3,4\,$ and $\,6\,$ supersymmetries within half-maximal supergravity.

\item All the independent flux vacua can be found in type IIB duality frames. In particular, they arise from type IIB orientifold reductions including O$9$-, O$5$- and O$3$-planes.

\begin{itemize}

\item[-] The type IIB with O9 duality frame accommodates the vast majority of the flux vacua. However, none of these vacua succeeds in stabilising all the dilaton-like moduli. In other words, these flux vacua do not come with a well-defined internal volume or string coupling.

\item[-] There are only two examples of AdS$_3$ flux vacua that stabilise all the dilaton-like moduli. Both exhibit scale separation and arise exclusively in the type IIB duality frame with O$5$-planes.

\item[-] The type IIB with O3 duality frame only accommodates Mkw$_{3}$ vacua. These feature a rich variety of supersymmetry-breaking patterns.

\end{itemize}

\item All the scalar masses in all the flux vacua turn out to be non-negative definite, thus showing the absence of perturbative instabilities within half-maximal supergravity. This is of special relevance in the case of non-supersymmetric vacua, for which either negative masses, in the Mkw$_{3}$ case, or masses below the Breitenlohner--Freedman (BF) bound \cite{Breitenlohner:1982bm}, in the AdS$_{3}$ case, are generically expected.

\end{enumerate}

Remarkably and surprisingly (at least to us), the only two examples of AdS$_{3}$ flux vacua that stabilise all the dilaton-like fields turn out to also feature scale separation. These two flux vacua, which belong to the type IIB with O$5$ duality frame, were originally found in \cite{Arboleya:2024vnp}. In there, the vacuum structure of the type IIB with O$5$ duality frame was presented as an appetiser to the landscape of flux vacua of RSTU-models. Now, after having completed the study of flux vacua in all possible RSTU-models arising from the different duality frames, we find that the two scale-separated AdS$_{3}$ vacua of \cite{Arboleya:2024vnp} are actually the only ones. In some sense, we found the needle before seeing the haystack!

The paper is organised as follows. Section~\ref{sec:RSTU_models} contains a review of the $\mathcal{N}=2$ RSTU-models constructed in \cite{Arboleya:2024vnp}, their embedding into three-dimensional $\,\mathcal{N}=8\,$ gauged supergravity and their higher-dimensional origin as type IIA/IIB orientifold reductions on twisted tori with background fluxes and a single type of O$p$-planes. In Section~\ref{sec:charting} we perform an exhaustive charting of the landscape of vacua of the RSTU-models with vanishing axions. We also present the minimal $\mathcal{N}=1$ superpotential formulation of the RSTU-models whenever the fluxes allow for such a description. In Section \ref{sec:discussion} we discuss our results, with a focus on moduli stabilisation and scale separation. We also outline several directions for future work. Three appendices are also included. In Appendix~\ref{App:dimensionsal_reduction} we explicitly perform the type IIB orientifold reductions on seven-dimensional twisted tori giving rise to the RSTU-models with vanishing axions studied in the paper. The definition of some relevant quantities -- like the volume of the internal space or the string coupling -- is presented there. Appendix~\ref{app:Flux_rescalings} collects results regarding the scaling symmetry of the RSTU-models with vanishing axions. Appendix~\ref{App:Mass_spectra} contains the gravitino and scalar mass spectra of all the inequivalent families of type IIB with O$9$, O$5$ and O$3$ flux vacua presented in the main text.

\section{RSTU-models from type II orientifolds}
\label{sec:RSTU_models}

We will study type II orientifold reductions on twisted tori involving background fluxes and only \textit{one type} of spacetime-filling O$p$-plane, with $\,p=2,\ldots,9$, and possibly also D$p$-branes. This compactification scheme has been shown to give rise to $\,\mathcal{N}=8\,$ (half-maximal) gauged supergravities in three-dimensions \cite{Arboleya:2024vnp}.

\subsection{Compactifications on twisted tori with fluxes}

Reductions on twisted tori have been extensively investigated since the seminal paper by Scherk and Schwarz (SS) \cite{Scherk:1979zr}. When combined with background fluxes for the various type II gauge field strengths, they provide an excellent arena where to investigate the issue of moduli stabilisation, one of the central questions in string phenomenology.

In this work we investigate type II reductions on seven-dimensional (7D) twisted tori down to three dimensions. A seven-dimensional twisted torus is defined in terms of Maurer–Cartan one-forms $\,\eta^{m}$, with $\,m=1,\ldots,7$, which obey the structure equation
\begin{equation}
d\eta^{p} + \tfrac{1}{2} \, \omega_{mn}{}^{p} \, \eta^{m} \wedge \eta^{n} = 0 \ ,
\end{equation}
where $\,\omega_{mn}{}^{p}\,$ are commonly referred to as metric fluxes. The metric fluxes $\,\omega_{mn}{}^{p}\,$ are the structure constants of the algebra spanned by the isometry generators of the internal space, \textit{i.e.}, $[X_{m},X_{n}]=\omega_{mn}{}^{p} \, X_{p}$. As such, they must obey the standard group-theoretic Jacobi identity
\begin{equation}
\label{Jacobi_id}
\omega_{[mn}{}^r\omega_{p]r}{}^q=0 \ .
\end{equation}
The Jacobi identity (\ref{Jacobi_id}) comes accompanied by a unimodularity (or traceless) condition
\begin{equation}
\label{trace_cond}
\omega_{mn}{}^{n} = 0 \ ,
\end{equation}
that ensures consistency of the SS reduction at the level of the action \cite{Scherk:1979zr}.

The basis of one-forms $\,\eta^{m}\,$ can then be used to expand the background gauge fluxes. For example, the NS-NS three-form field strength can be expanded as $\,H_{(3)}=\frac{1}{3!} H_{mnp} \, \eta^{m} \wedge \eta^{n} \wedge \eta^{p}\,$ with constant flux components $\,H_{mnp}$. The same expansion can be performed for the various $\,F_{(p)}\,$ field strengths in the RR sector of the massive type IIA ($p=0,2,4,6$) -- where F$_{(0)}$ is the so-called Romans mass parameter -- and the type IIB ($p=1,3,5,7$) supergravities. The various gauge fluxes of type II supergravity are subject to a set of Bianchi identities. Considering a duality frame specified by a given type of O$p$-plane, one has Bianchi identities of the form
\begin{equation}
\label{Bianchi_id_10D}
dH_{(3)}=0  
\hspace{10mm} \textrm{ and } \hspace{10mm} 
dF_{(8-p)}-H_{(3)}\wedge F_{(6-p)} = J_{\textrm{O}p/\textrm{D}p} \ .
\end{equation}
The first condition in (\ref{Bianchi_id_10D}) signals the absence of NS5-branes. The second condition in (\ref{Bianchi_id_10D}) involves a $\,(9-p)$-form $\,J_{\textrm{O}p/\textrm{D}p}\,$ which is non-zero only for the specific type of sources present in the compactification scheme (in the smeared limit). In other words, the second condition in (\ref{Bianchi_id_10D}) signals the absence of other sources different from that single type of O$p$-planes and D$p$-branes.

Performing the twisted torus reduction of type II supergravity in the presence of a single type of O$p$/D$p$ sources and of background gauge fluxes yields a three-dimensional gauged supergravity with $\,\mathcal{N}=8\,$ supersymmetry \cite{Arboleya:2024vnp}. Let us move to briefly discuss the general structure of such a theory.

\subsection{Three-dimensional $\,\mathcal{N}=8\,$ supergravity}

Our starting point is the bosonic Lagrangian of $\mathcal{N}=8$ (half-maximal) gauged supergravity in three dimensions (3D) coupled to eight matter multiplets \cite{Marcus:1983hb, Nicolai:2001ac,deWit:2003ja,Deger:2019tem}. As shown in \cite{Arboleya:2024vnp}, this setup captures the $64$ scalars arising from orientifold reductions of type II supergravity down to 3D including only a single type of spacetime-filling O$p$-planes (and D$p$-branes). In the conventions of \cite{Deger:2019tem}, and focusing on the scalar sector of the theory,\footnote{In three dimensions vectors are dual to scalars and, as such, they do not carry an independent dynamics.} the Lagrangian reads
\begin{equation}
\label{L_N8}
e^{-1} \mathcal{L}_{\mathcal{N}=8} = - \frac{R}{4}  - \frac{1}{32} \, \textrm{Tr} \left( \partial_\mu M \, \partial^{\mu} M^{-1} \right)  -  V_{\mathcal{N}=8}  \ ,
\end{equation}
where $\,e = \sqrt{|\textrm{det} \, g_{\mu\nu}|}$. The first term is the Einstein--Hilbert term for the dreibein field, the second term is the kinetic term for the scalar-dependent matrix $\,M \in \textrm{SO}(8,8)\,$ accounting for the $64$ scalars of the theory, and the last term is a non-trivial scalar potential. The $64$ scalars serve as coordinates on the coset space
\begin{equation}
\label{scalar_geometry_N8}
{\mathcal{M}}_{\mathcal{N}=8} = \frac{{\rm SO}(8,8)}{{\rm SO}(8) \times {\rm SO}(8)}  \ ,
\end{equation}
and the scalar potential in (\ref{L_N8}) is generally a complex function of the scalars, involving a very large number of couplings. Such couplings, which encode the metric and gauge fluxes in the compactification, are all encoded in a single object, called the \textit{embedding tensor}, and denoted by $\,\Theta\,$. This tensor is constant (recall that fluxes were constant) and has a decomposition into irreducible representations (irrep's) of $\,\textrm{SO}(8,8)\,$ of the form
\begin{equation}
\label{ET_N8}
\Theta_{MN|PQ} = \theta_{MNPQ} + 2 \left( \eta_{M[P} \theta_{Q]N} - \eta_{N[P} \theta_{Q]M} \right) + 2 \, \eta_{M[P} \eta_{Q]N} \theta \ ,
\end{equation}
where 
\begin{equation}
\label{ET_irreps}
\theta_{MNPQ}=\theta_{[MNPQ]} \in \mathbf{1820}
\hspace{5mm},\hspace{5mm}
\theta_{MN}=\theta_{(MN)} \in \mathbf{135} \,\,\,\, \textrm{(traceless)}
\hspace{5mm},\hspace{5mm}
\theta \in \mathbf{1} \ .
\end{equation}
The matrix $\,\eta_{MN}\,$ in (\ref{ET_N8}) is the non-degenerate $\textrm{SO}(8,8)$ invariant metric that can be used to raise and lower vector indices $\,M,N=1,\ldots,16\,$ of $\textrm{SO}(8,8)$. Importantly, the consistency of the gauged supergravity requires a set of quadratic constraints (QC's) on the embedding tensor that can be expressed as \cite{Eloy:2024lwn}
\begin{equation}
\label{QC_N8}
\Theta_{KL|[M}{}^{R} \, \Theta_{N]R|PQ} + \Theta_{KL|[P}{}^{R} \, \Theta_{Q]R|MN} = 0 \ .
\end{equation}
From the point of view of the compactification, the QC's in (\ref{QC_N8}) are nothing but the Jacobi and Bianchi identities in (\ref{Jacobi_id}) and (\ref{Bianchi_id_10D}), respectively. More concretely, those involving the sources that are absent in the compactification. Provided an embedding tensor $\,\Theta\,$ that satisfies (\ref{QC_N8}), the scalar potential then takes the schematic form
\begin{equation}
\label{V_N8}
V_{\mathcal{N}=8}(\Theta \, ; M) = g^{2} \,  \Theta  \, \Theta \left( M^{4} + M^{3} + \dots \right) \ ,
\end{equation}
where $\,g\,$ is the gauge coupling of the half-maximal supergravity in 3D. Note that a non-zero value of $\,g\,$ can be reabsorbed in an overall rescaling of the embedding tensor (fluxes), \textit{i.e.}, $\,\Theta \rightarrow \frac{\Theta}{g}$. The scalar potential computed from (\ref{V_N8}) then matches the one arising from the compactification provided the QC's in (\ref{QC_N8}) hold. We refer the reader to \textit{e.g.} \cite{Deger:2019tem} for more details on the Lagrangian (\ref{L_N8}) and its completion to include also the dual vectors.

\subsubsection*{Embedding into maximal supergravity and $\textrm{O}(8,8)$-DFT}

Under certain conditions, a half-maximal gauged supergravity in $D=3$ can be seen as a subsector of a maximal gauged supergravity. As discussed in \cite{Deger:2019tem}, the QC's in (\ref{QC_N8}) must be supplemented with two additional conditions
\begin{equation}
\label{QC_N16_extra}
\begin{array}{rcl}
48 \, \theta \, \theta_{MN} + \theta_{M}{}^{PQR} \, \theta_{NPQR} - \frac{1}{16} \, \eta_{MN} \, \theta^{PQRS} \, \theta_{PQRS} &=& 0 \ , \\[2mm]
\theta_{[M_{1}M_{2}M_{3}M_{4}} \,\, \theta_{M_{5}M_{6}M_{7}M_{8}]}  \Big|_{\textrm{SD}} &=& 0 \ ,
\end{array}
\end{equation}
for a gauged half-maximal supergravity to be secretly maximally supersymmetric. The term SD in the second equation of (\ref{QC_N16_extra}) refers to the self-dual projection of a rank-eight antisymmetric tensor of $\,\textrm{SO}(8,8)$. From the compactification point of view, the extra conditions in (\ref{QC_N16_extra}) ensure the absence of \textit{all} possible sources in the compactification. More concretely,
\begin{equation}
0=J_{\textrm{O}p/\textrm{D}p}
\hspace{10mm} \textrm{for \textit{all} possible $\,p$} \ .
\end{equation}

In addition, it was shown in \cite{Hohm:2017wtr} that a generalised Scherk--Schwarz reduction with twist matrices that obey the \textit{section constraint} of the $\textrm{O($8,8$)}$ enhanced double field theory (DFT) can only reproduce gauged supergravities whose embedding tensors satisfy (\ref{QC_N16_extra}) -- so they are secretly maximally supersymmetric -- as well as an additional condition
\begin{equation}
\label{DFT_constraint}
\theta_{[M_{1}M_{2}M_{3}M_{4}} \,\, \theta_{M_{5}M_{6}M_{7}M_{8}]}  \Big|_{\textrm{ASD}} = 0 \ ,
\end{equation}
where ASD stands for the anti-self-dual projection of the rank-eight antisymmetric tensor. When this happens, there is the possibility of computing the full tower of Kaluza--Klein (KK) masses about a given flux vacuum using the KK spectrometry techniques put forward in \cite{Malek:2019eaz} and applied to three-dimensional gauged supergravity in \cite{Eloy:2020uix}.

\subsection{$\,\mathcal{N}=2\,$ RSTU-models}

The embedding tensor $\Theta$ in (\ref{ET_N8}) lives in the $\mathbf{1820} \,\oplus\, \mathbf{135} \,\oplus\, \mathbf{1}$ (reducible) representation of $\,\textrm{SO}(8,8)\,$ which implies a dependence of the scalar potential (\ref{V_N8}) on almost two thousands different couplings and $\,64\,$ scalars. Extremising it analytically to find all its critical points -- both supersymmetric and non-supersymmetric -- is simply out of computational reach. For the sake of feasibility, it is customary to restrict to a subset of scalars and embedding tensor components so that the resulting models are invariant under the action of some specific subgroup of the duality group $\,\textrm{SO}(8,8)$. This was the strategy followed in \cite{Arboleya:2024vnp}, where an $\textrm{SO}(3) \subset \textrm{SO}(8,8)$ symmetry was imposed from the beginning in all the models.

The particular $\mathrm{SO}(3)$ symmetry used in \cite{Arboleya:2024vnp} is diagonally embedded in the duality group $\,\textrm{SO}(8,8)$ according to
\begin{equation}
\label{SO(8,8)_to_SO(3)_embedding}
\begin{array}{cccccc}
\textrm{SO}(8,8) & \supset & \textrm{SO}(2,2) \times \textrm{SO}(6,6) & \supset & \textrm{SO}(2,2) \times \textrm{SO}(2,2) \times \textrm{SO}(3) \ . 
\end{array}
\end{equation}
The $\textrm{SO}(3)$ symmetry acts on the basis of one-forms $\,\eta^{m}\,$ of the internal space as
\begin{equation}
\label{coord_splitting}
\eta^{m} = (\eta^{a},\eta^{i},\eta^{7})
\hspace{10mm} \textrm{ with } \hspace{10mm}
a=1,3,5 \quad , \quad i=2,4,6 \ ,
\end{equation}
by simultaneously rotating $\,\eta^{a}\,$ and $\,\eta^{i}\,$ as two copies of the vector representation of $\textrm{SO}(3)$, and leaving $\,\eta^{7}\,$ invariant. This decomposition determines the following $\textrm{SO}(3)$-invariant tensors\footnote{In order to clarify the notation, let us observe that the tensors with mixed index ($a$,$i$) structure have to be intended in the following way: being $a=1,3,5$ and $i=2,4,6$, a $\delta_{ia}$ is defined so that it gives $1$ only when, coherently with the imposed $\textrm{SO($3$)}$ symmetry, the positions of the indices in the two lists coincide, \textit{i.e.} $\delta_{12} = \delta_{34} = \delta_{56} = 1$, while all the other components vanish. A similar interpretation applies for the $\epsilon_{abi}$ and $\epsilon_{aij}$ symbols: for instance, $\epsilon_{136} = 1$ or $\epsilon_{146} = 1$.} 
\begin{equation}
\label{SO(3)_inv_tensors}
\delta_{ab} \,\, , \,\, \delta_{ij} \,\, , \,\, \delta_{ai} \,\, , \,\, \delta_{ia} \, \,  , \,\, \epsilon_{abc}  \,\, , \,\,  \epsilon_{abk}  \,\, , \,\, \epsilon_{ajk} \,\, , \,\, \epsilon_{ijk} \ ,
\end{equation}
which are then used to construct the set of $\textrm{SO}(3)$-invariant scalars and fluxes in a given O$p$-plane duality frame. Importantly, the $\,7 \rightarrow 3 + 3+ 1\,$ basis splitting in (\ref{coord_splitting}) sometimes clashes with the O$p$-plane target space involution $\,\sigma_{\textrm{O}p}$. The latter reflects the $\,(9-p)\,$ directions transverse to the O$p$-plane worldvolume and breaks the $\,\textrm{SL}(7)\,$ group of (constant) internal diffeomorphisms down to an $\textrm{SL}(p-2) \times \textrm{SL}(9-p)$ subgroup. This clashing happens specifically for the $\,p=4\,$ and $\,p=7$\, cases for which $\,\sigma_{\textrm{O}p}\,$ reflects $\,5\,$ and $\,2\,$ internal coordinates, respectively, thus being inconsistent with (\ref{coord_splitting}). As a result, RSTU-models cannot be obtained from type IIA reductions with O$4$-planes or type IIB reductions with O$7$-planes. The set of possible RSTU-models and the location of the corresponding spacetime-filling O$p$-planes are summarised in Table~\ref{Table:Op_planes}.

There are $\,158\,$ components inside the embedding tensor $\,\Theta\,$ which are $\textrm{SO}(3)$ singlets \cite{Arboleya:2024vnp}. However, only a few of them have an interpretation in terms of metric and gauge fluxes in a given type II orientifold reduction. The rest are known as ``non-geometric'' fluxes \cite{Shelton:2005cf} and their study goes beyond the purpose of this work. The set of $\mathrm{SO}(3)$-invariant scalars is identified with the commutant of $\textrm{SO}(3)$ inside $\textrm{SO}(8,8)$ in (\ref{SO(8,8)_to_SO(3)_embedding}). Using the fact that $\textrm{SO}(2,2) \sim \textrm{SL}(2) \times \textrm{SL}(2)$, it follows from (\ref{SO(8,8)_to_SO(3)_embedding}) that the scalar geometry of the $\mathrm{SO}(3)$-invariant models is given by four copies of the Poincar\'e disk
\begin{equation}
\label{scalar_geometry_RSTU}
\mathcal{M}_{\mathcal{N}=2} = \left[ \frac{\textrm{SL}(2)}{\textrm{SO}(2)}\right]_{R} \times \left[ \frac{\textrm{SL}(2)}{\textrm{SO}(2)}\right]_{S}  \times \left[ \frac{\textrm{SL}(2)}{\textrm{SO}(2)}\right]_{T}  \times \left[ \frac{\textrm{SL}(2)}{\textrm{SO}(2)}\right]_{U} \subset \frac{{\rm SO}(8,8)}{{\rm SO}(8) \times {\rm SO}(8)}  \ ,
\end{equation}
which we will parameterise in terms of four complex scalars, $R$, $S$, $T$ and $U$, with positive-definite imaginary parts (\textit{i.e.} upper half-plane parameterisation).  These models were dubbed RSTU-models in \cite{Arboleya:2024vnp} and, interestingly, they describe an $\mathcal{N}=2$ sector within the $\mathcal{N}=8$ theory. For the RSTU-models, the Einstein-scalar Lagrangian (\ref{L_N8}) reduces to
\begin{equation}
\label{L_N2}
e^{-1} \mathcal{L}_{\mathcal{N}=2} = - \frac{R}{4}  + \frac{1}{8}  \left[  \frac{\partial R \, \partial \bar{R}}{(\textrm{Im}R)^{2}}  + \frac{\partial S \, \partial \bar{S}}{(\textrm{Im}S)^{2}}  + 3 \frac{\partial T \, \partial \bar{T}}{(\textrm{Im}T)^{2}}   + 3 \frac{\partial U \, \partial \bar{U}}{(\textrm{Im}U)^{2}} \right] -  V_{\mathcal{N}=2}  \ .
\end{equation}
The scalar potential in (\ref{L_N2}) is the one we will extremise in order to chart the landscape of flux vacua in the RSTU-models.

\begin{table}[t]
\begin{center}
\renewcommand{\arraystretch}{1.5}
\begin{tabular}{|c|ccc|cc|cc|cc|c|c|}
\cline{2-11}
\multicolumn{1}{c|}{} & \multicolumn{3}{c|}{External s-t} & \multicolumn{7}{c|}{Internal twisted torus $\,\mathbb{T}_{\omega}^{7}$} \\ 
\cline{2-12}
\multicolumn{1}{c|}{} & $dx^{0}$ & $dx^{1}$ & $dx^{2}$ & $\eta^{1}$  & $\eta^{2}$ & $\eta^{3}$  & $\eta^{4}$ & $\eta^{5}$ & $\eta^{6}$ & $\eta^{7}$ & Internal Diff \\
\hline
\textrm{O2-plane} & $\times$ & $\times$ & $\times$ & &  &  &  &  &  &  & $\textrm{SL}(7)$ \\
\hline
\textrm{O3-plane} & $\times$ & $\times$ & $\times$ & &  &  &  &  &  & $\times$ &  $\textrm{SL}(6)$ \\
\hline
\textrm{O5-plane} & $\times$ & $\times$ & $\times$ & &   $\times$  &  &  $\times$  &  &  $\times$  & & $\textrm{SL}(3) \times \textrm{SL}(4)$ \\
\hline
\textrm{O6-plane} & $\times$ & $\times$ & $\times$ &  $\times$  &  &  $\times$  &  &  $\times$  &  &  $\times$ & $\textrm{SL}(4) \times \textrm{SL}(3)$ \\
\hline
\textrm{O8-plane} & $\times$ & $\times$ &$\times$ & $\times$ & $\times$  & $\times$ & $\times$ & $\times$ & $\times$ & & $\textrm{SL}(6)$ \\
\hline
\textrm{O9-plane} & $\times$ & $\times$ &$\times$ & $\times$ & $\times$  & $\times$ & $\times$ & $\times$ & $\times$ & $\times$ & $\textrm{SL}(7)$ \\
\hline
\end{tabular}
\caption{Single O$p$-plane duality frames compatible with the $\textrm{SO}(3)$ invariance of the RSTU-models. In all the cases the total orientifold action $\,\mathcal{O}_{\mathbb{Z}_{2}}\,$ halves  $\,\mathcal{N}=16\,$ (maximal) into $\,\mathcal{N}=8\,$ (half-maximal) supergravity \cite{Arboleya:2024vnp}. Then, the $\textrm{SO}(3)$ invariance of the RSTU-models further breaks $\,\mathcal{N}=8\,$ down to $\,\mathcal{N}=2$.}
\label{Table:Op_planes}
\end{center}
\end{table}

\subsubsection*{Turning off axions and minimal $\,\mathcal{N}=1\,$ models}

A further simplification that we will make is to set all the axion-like fields to zero when searching for extrema of the scalar potential (\ref{L_N2}). In the upper-half plane parameterisation of the complex scalars, this amounts to set
\begin{equation}
\label{axions_vanish}
\textrm{Re}R=0
\hspace{5mm} , \hspace{5mm} 
\textrm{Re}S=0
\hspace{5mm} , \hspace{5mm} 
\textrm{Re}T=0
\hspace{5mm} , \hspace{5mm} 
\textrm{Re}U=0 \ .
\end{equation}
This naturally raises the question of whether the axions can be consistently turned off at the level of the Lagrangian (\ref{L_N2}). As shown in \cite{Arboleya:2024vnp}, modding out the RSTU-models by an additional $\,\mathbb{Z}_{2}^{*} \subset \textrm{SO}(8,8)\,$ discrete symmetry acting on the basis of one-forms as
\begin{equation}
\label{Z2*_element}
\mathbb{Z}_{2}^{*} \,\,\, : \,\,\, \eta^{i} \rightarrow \eta^{i}
\hspace{4mm} , \hspace{4mm} 
\eta^{a} \rightarrow -\eta^{a}
\hspace{4mm} , \hspace{4mm} 
\eta^{7} \rightarrow -\eta^{7} \ ,
\end{equation}
precisely imposes (\ref{axions_vanish}) already at the level of the Lagrangian. This additional $\,\mathbb{Z}_{2}^{*}\,$ also projects out one of the two gravitini of the RSTU-models yielding a minimal $\,\mathcal{N}=1\,$ supergravity model with Lagrangian
\begin{equation}
\label{L_N1}
e^{-1} \mathcal{L}_{\mathcal{N}=1} = - \frac{R}{4}  + \frac{1}{8}  \left[  \frac{(\partial \textrm{Im}R)^{2}}{(\textrm{Im}R)^{2}}  +  \frac{(\partial \textrm{Im}S)^{2}}{(\textrm{Im}S)^{2}}    + 3 \frac{(\partial \textrm{Im}T)^{2}}{(\textrm{Im}T)^{2}}    + 3  \frac{(\partial \textrm{Im}U)^{2}}{(\textrm{Im}U)^{2}}   \right] -  V_{\mathcal{N}=1}  \ .
\end{equation}
The scalar potential $\,V_{\mathcal{N}=1}\,$ in (\ref{L_N1}) can then be derived from a real superpotential function $\,W_{\mathcal{N}=1}\,$ which, due to supersymmetry, is identified with the mass of the $\,\mathbb{Z}_{2}^{*}$-even gravitino. More concretely, one has the standard formula
\begin{equation}
\label{VfromW}
V_{\mathcal{N}=1} =  2\, \left[\displaystyle\sum_{I=1}^{4}  \frac{(\textrm{Im}\Phi^{I})^{2}}{\, c_{I}} \left(\frac{\partial W_{\mathcal{N}=1}}{\partial\textrm{Im}\Phi^{I}}\right)  \left(\frac{\partial W_{\mathcal{N}=1}}{\partial\textrm{Im}\Phi^{J}}\right) - (W_{\mathcal{N}=1})^{2} \right]\ ,
\end{equation}
where we have collectively denoted $\,\Phi^{I}=(R,S,T,U)$ and $\,c_{I}=(1,1,3,3)\,$. Following the notation of \cite{Arboleya:2024vnp}, and for the sake of comparison with the formulae there, we will rename the dilatons as 
\begin{equation}
\label{dilaton_redef}
\textrm{Im}R=A_{4}
\hspace{5mm} , \hspace{5mm} 
\textrm{Im}S=\mu_{4}
\hspace{5mm} , \hspace{5mm}
\textrm{Im}T=A
\hspace{5mm} \textrm{ and } \hspace{5mm}
\textrm{Im}U=\mu  \ .
\end{equation}

Importantly, the $\,\mathbb{Z}_{2}^{*}\,$ action in (\ref{Z2*_element}) is \textit{not} always compatible with the O$p$-plane internal target space involution $\,\sigma_{\textrm{O}p}\,$ reflecting the directions transverse to the O$p$-plane worldvolume. By inspection of Table~\ref{Table:Op_planes}, one concludes that only the O$6$ and O$2$ cases in type IIA together with the O$9$ and O$5$ cases in type IIB are compatible with the $\,\mathbb{Z}_{2}^{*}\,$ action in (\ref{Z2*_element}) and, therefore, give rise to a consistent $\,\mathcal{N}=1\,$ supergravity model when turning off the axions in the corresponding RSTU-models. In these four cases, by direct computation of the appropriate gravitino mass, we will present the $\,\mathcal{N}=1\,$ superpotential from which the scalar potential follows accordingly to (\ref{VfromW}). Also in these four cases, we can think of the RSTU-models without axions as a class of type II orientifold reductions on $\,\mathbb{Z}_{2}^{3}\,$ orbifolds of seven-dimensional manifolds equipped with a co-closed G$_2$-structure \cite{Arboleya:2024vnp}. It is within this geometric framework that an $\,\mathcal{N}=1\,$ superpotential has been proposed \cite{Emelin:2021gzx} (see also \cite{Miao:2025rgf}).

\section{Charting the landscape of RSTU-models}
\label{sec:charting}

The RSTU-models (\ref{L_N2}) come along with a scalar potential whose features crucially depend on the type of O$p$-planes present in the type II orientifold reduction from which they derive. The reason is that a particular type of O$p$-planes projects out specific types of fluxes in the compactification scheme, thus turning off certain couplings in the scalar potential. For example, if O$2$-planes are present in a type IIA compactification scheme, then metric fluxes $\,\omega\,$ as well as Ramond--Ramond (RR) fluxes $\,F_{(2)}\,$ and $\,F_{(6)}\,$ are projected out and, with them, also the corresponding couplings in the flux-induced scalar potential \cite{Farakos:2020phe}.

The complete set of fluxes that are allowed/forbidden by an O$p$-plane, with $p=2,\ldots,9$, in a type IIA ($p$ even) or type IIB ($p$ odd) orientifold reduction down to 3D was exhaustively identified in \cite{Arboleya:2024vnp}. Also in \cite{Arboleya:2024vnp}, the precise dictionary between the allowed fluxes and the embedding tensor irrep's in (\ref{ET_irreps}) was derived using group-theoretic techniques. Amongst the allowed fluxes, those compatible with the $\textrm{SO}(3)$ symmetry of the RSTU-models were also identified. Equipped with the results of \cite{Arboleya:2024vnp}, here we will first construct the scalar potentials of the RSTU-models for the various O$p$-plane cases, and then extremise them to chart the landscape of three-dimensional flux vacua. The extremisation of the scalar potential must be supplemented with the quadratic constraints in (\ref{QC_N8}) for the three-dimensional RSTU-models to describe a consistent subsector of half-maximal supergravity. As discussed in \cite{Arboleya:2024vnp}, these constraints can be interpreted as the Jacobi identity (\ref{Jacobi_id}) and the Bianchi identity (\ref{Bianchi_id_10D}).

\subsection*{Mission statement}

Following \cite{Arboleya:2024vnp}, we proceed with the construction of the RSTU-models from flux compactifications of type~II string theory with a single type of orientifold planes O$p$, thus breaking the 3D supersymmetry from maximal ($\mathcal{N}=16$) to half-maximal ($\mathcal{N}=8$). While the type~IIB with an O$5$-plane duality frame has been extensively considered in \cite{Arboleya:2024vnp}, here we complete the discussion specifically addressing the remaining cases of Table~\ref{Table:Op_planes}, \textit{i.e.} type~IIA with O$2$, O$6$ or O$8$ and type~IIB with O$3$ or O$9$.
\\

\noindent For each of the O$p$-plane duality frames we will perform the following actions:

\begin{enumerate}

\item We will construct the $V_{\mathcal{N}=2}$ scalar potential in (\ref{L_N2}) for each of the RSTU-models arising from the O$p$-planes in Table~\ref{Table:Op_planes}.  For the cases $\,p=2,5,6,9\,$, we will also present the superpotential $\,W_{\mathcal{N}=1}\,$ of the associated $\,\mathcal{N}=1\,$ minimal models (\ref{L_N1}) obtained upon further modding out by the $\,\mathbb{Z}_{2}^{*}\,$ symmetry in (\ref{Z2*_element}). The scalar potential is invariant under a rescaling of the dilaton-like fields of the form
\begin{equation}
\label{scaling_moduli}
\textrm{Im}\Phi'= \lambda_{\Phi} \, \textrm{Im}\Phi
\hspace{10mm} \textrm{ with } \hspace{10mm}
\Phi=R,S,T,U \ ,
\end{equation}
followed by a suitable rescaling of the flux parameters (see Appendix~\ref{app:Flux_rescalings}). Exploiting this scaling symmetry, we will search for extrema of the scalar potential at $\,\left\langle \, \textrm{Im}\Phi \, \right\rangle=1\,$ (without loss of generality), and further set axions to zero for simplicity, namely, $\,\left\langle \, \textrm{Re}\Phi \, \right\rangle = 0$. We will refer to this specific point in field space
\begin{equation}
\left\langle \,\Phi \, \right\rangle = i 
\hspace{10mm} \textrm{ with } \hspace{10mm}
\Phi=R,S,T,U \ ,
\end{equation}
as the origin of moduli space. The (negative) value of the cosmological constant, $\,\left\langle V \right\rangle\,$, at a given AdS$_{3}$ vacuum sets the value of the corresponding AdS$_{3}$ radius $\,L^2=-2/\left\langle V \right\rangle$. If $\,\left\langle V \right\rangle=0\,$ then we have a Mkw$_{3}$ vacuum.

\item We will extremise the scalar potential at the origin of moduli space using the algebraic software \textsc{singular} \cite{DGPS}. More concretely, we will make use of the GTZ built-in algorithm for primary decomposition of a system of polynomial equations. The outcome of this algorithm is a number of simpler polynomial systems, known as prime factors,  each of which describes a branch of solutions of the original extremisation problem. These prime factors can be solved analytically giving rise to the various families of flux vacua that we find. It is important to remember that the extremisation of the scalar potential must be supplemented with the quadratic constraints in (\ref{QC_N8}) for the three-dimensional RSTU-models to consistently  describe a subsector of half-maximal supergravity.\footnote{Whenever nontrivial, the condition (\ref{trace_cond}) is accounted for by the extremisation procedure of the RSTU scalar potential (\ref{V_N8}). Namely, the whole landscape of RSTU flux vacua corresponds to compactifications of type~II supergravity that obey the Scherk--Schwarz zero-trace requirement on the metric fluxes \cite{Scherk:1979zr}.}

\item The landscape of flux vacua and the associated properties and mass spectra will be presented in tables. The number $\,\mathcal{N}\,$ of supersymmetries preserved by a given vacuum is identified with the number of gravitini with mass $\,m^{2}_{3/2} L^{2}=1\,$ (for an AdS$_{3}$ vacuum) or $\,m^{2}_{3/2}=0\,$ (for a Mkw$_{3}$ vacuum). The $\,\mathcal{N}=p\,$ supersymmetry is realised as $\,\mathcal{N}=(p,0)\,$ or $\,\mathcal{N}=(0,p)\,$ depending on the upper/lower sign choice of the flux parameters in the corresponding vacuum. Lastly, the conformal dimensions $\,\Delta$'s of the would-be dual CFT$_2$ operators are also presented for those AdS$_{3}$ flux vacua in which they turn out to be integer-valued.

\end{enumerate}

\noindent \textbf{Boxes and colour code:} Independent flux vacua are enclosed in boxes, while coloured entries indicate generalisations of vacua with the same colour appearing in other tables.
\\[-2mm]

\noindent \textbf{$*\,$ and $\,\dagger\,$ code:} Flux vacua marked with an asterisk $\,*\,$ satisfy the extra conditions (\ref{QC_N16_extra}) and
admit an embedding into maximal supergravity. Those marked with a dagger $\,\dagger\,$ satisfy the anti-self-dual condition (\ref{DFT_constraint}). If both conditions are satisfied simultaneously then the corresponding flux vacuum can be realised as a generalised Scherk--Schwarz reduction of the $\,\textrm{O}(8,8)\,$ enhanced DFT.

\subsection{Type IIB orientifold flux vacua}

Let us chart the landscape of RSTU-models obtained from type IIB orientifold reduction with O$9$-, O$7$-, O$5$- and O$3$-planes. For the particular O$9$ and O$5$ cases we will also present the minimal $\,\mathcal{N}=1\,$ supergravity models with no axions obtained upon modding out the corresponding RSTU-models by the $\,\mathbb{Z}_{2}^{*}\,$ symmetry in (\ref{Z2*_element}).

\subsubsection{Type~IIB with O$9$-planes}
\label{sec:IIB_O9}

The type IIB with O$9$-planes duality frame, \textit{i.e.} type I theory, was considered in Section~$3.9$ of \cite{Arboleya:2024vnp}. Introducing O$9$-planes is compatible with an $\,\textrm{SL}(7)\,$ covariant description of the fluxes (see Table~\ref{Table:Op_planes}). The O$9$-plane orientifold action $\mathcal{O}_{\mathbb{Z}_{2}} = \Omega_{P} \, \sigma_{\textrm{O}9}$ allows for metric fluxes $\omega$ as well as gauge fluxes $\,F_{(3)}\,$ and $\,F_{(7)}$.\footnote{A dilaton flux $\,H_{(1)}\,$ is also permitted by the orientifold action but this cannot be generated in an ordinary SS reduction.} An explicit computation of the QC's in (\ref{QC_N8}) yields the Jacobi identity (\ref{Jacobi_id}) together with the second Bianchi identity in (\ref{Bianchi_id_10D}), with $\,p=5$, signaling the absence of O$5$/D$5$ sources in the compactification scheme. However, even though half-maximal supersymmetry does not require it, the type IIB with O$9$ flux models:
\begin{itemize}

\item[$i)$] are embeddable into maximal supergravity since the additional quadratic constraints in (\ref{QC_N16_extra}) are satisfied. From the point of view of the compactification, this means that a flux-induced tadpole for the O$9$/D$9$ sources in (\ref{Bianchi_id_10D}) is not permitted (actually, it is not possible), namely, 
\begin{equation}
0=J_{\textrm{O}9/\textrm{D}9} \ ,
\end{equation}

\item[$ii)$] can be obtained as a generalised Scherk--Schwarz reduction of O$(8,8)$-DFT since the condition (\ref{DFT_constraint}) also holds.

\end{itemize}

\noindent These two observations will become relevant in Section~\ref{sec:discussion} where we will address the issues of moduli stabilisation and scale separation in this duality frame.

Using the set of $\textrm{SO}(3)$-invariant tensors in (\ref{SO(3)_inv_tensors}) one can construct the following set of $\textrm{SO}(3)$-invariant metric fluxes
\begin{equation}
\label{metric_fluxes_SO(3)_O9}
\begin{array}{c}
\omega_{ab}{}^{c}=\omega_{1} \, \epsilon_{ab}{}^{c}
 \hspace{3mm} ,  \hspace{3mm} 
\omega_{7a}{}^{b}=\omega_{2} \, \delta_{a}{}^{b}
 \hspace{3mm} ,  \hspace{3mm} 
\omega_{ab}{}^{i}=\omega_{3} \, \epsilon_{ab}{}^{i} \hspace{3mm} , \hspace{3mm} \omega_{ia}{}^{b}=\omega_{4} \, \epsilon_{ia}{}^{b} \ ,\\[2mm]
\omega_{7a}{}^{i}=\omega_{5} \, \delta_{a}{}^{i} \hspace{3mm} , \hspace{3mm} \omega_{ia}{}^{7}=\omega_{6} \, \delta_{ia}  \hspace{3mm} ,  \hspace{3mm}
\omega_{7i}{}^{a}=\omega_{7} \, \delta_{i}{}^{a} \hspace{3mm} , \hspace{3mm}
\omega_{ij}{}^{a}=\omega_{8} \, \epsilon_{ij}{}^{a} \ ,\\[2mm]
\omega_{ai}{}^{j}=\omega_{9} \, \epsilon_{ai}{}^{j} \hspace{3mm} , \hspace{3mm} \omega_{7i}{}^{j}=\omega_{10} \, \delta_{i}{}^{j} \hspace{3mm} , \hspace{3mm}
\omega_{ij}{}^{k}=\omega_{11} \, \epsilon_{ij}{}^{k} \ ,
\end{array}
\end{equation}
as well as gauge fluxes
\begin{equation}
\label{gauge_fluxes_SO(3)_O9}
\begin{array}{c}
F_{abc}= f_{31}  \, \epsilon_{abc}
\hspace{3mm} , \hspace{3mm} 
F_{abi}= f_{32}  \, \epsilon_{abi}
\hspace{3mm} , \hspace{3mm} F_{aij}= f_{33}  \, \epsilon_{aij} 
\hspace{3mm} , \hspace{3mm} 
F_{ijk}= f_{34}  \, \epsilon_{ijk} 
\hspace{3mm} , \hspace{3mm} 
F_{ia7}= f_{35}  \, \delta_{ia} \ , \\[2mm] 
F_{ijkabc7}= f_{7}  \,  \epsilon_{ijk} \epsilon_{abc} \ ,
\end{array}
\end{equation}
adding up to $\,17\,$ arbitrary flux parameters.\footnote{The zero-trace condition in (\ref{trace_cond}) reduces to $\,\omega_{7a}{}^{a} + \omega_{7i}{}^{i} = 0\,$ or, equivalently, $\,\omega_2+\omega_{10} = 0$. Similar linear relations between metric fluxes will appear in other O$p$-plane duality frames. Still we will include both fluxes in tables like Table~\ref{Table:flux_vacua_IIB_O9}, etc.} The $\textrm{SO}(3)$-invariant fluxes (\ref{metric_fluxes_SO(3)_O9}) and (\ref{gauge_fluxes_SO(3)_O9}) induce a non-trivial scalar potential. Its direct extremisation gives the list of critical points in Table~\ref{Table:flux_vacua_IIB_O9} and its continuation.

\subsubsection*{Supersymmetry enhancements} 

The gravitino masses for all the type IIB flux vacua with O$9$-planes in Table~\ref{Table:flux_vacua_IIB_O9} (and its continuation) can be found in Tables~\ref{Table:flux_vacua_IIB_O9_gravitini_spectrum_Mkw} and \ref{Table:flux_vacua_IIB_O9_gravitini_spectrum_AdS} of Appendix~\ref{App:IIB_O9_spectra}. By looking at them, one realises that some of the multi-parametric families of AdS$_{3}$ flux vacua enjoy supersymmetry enhancements when the arbitrary parameters take specific values. 
\\[-2mm]

\noindent More concretely:  

\begin{itemize}

\item[-] \textcolor{ForestGreen}{\textbf{vac~8}} and \textbf{vac~16} exhibit an enhancement to $\,\mathcal{N}=(1,0)\,$/$\,\mathcal{N}=(0,1)\,$ supersymmetry respectively for $\,\kappa = 0\,$ and $\,\xi = 0\,$ depending on the upper/lower sign choice of the fluxes. \\[-2mm]

\item[-]  \textcolor{ForestGreen}{\textbf{vac~9}},  \textcolor{ForestGreen}{\textbf{vac~11}} and \textbf{vac~12} become $\,\mathcal{N}=(4,0)\,$/$\,\mathcal{N}=(0,4)\,$ supersymmetric for $\,\xi = 0\,$ depending on the upper/lower sign choice of the fluxes; the same happens for \textbf{vac~13}, \textbf{vac~17} and \textbf{vac~19} but for $\,\kappa = 0$. \\[-2mm]

\item[-] \textbf{vac~21} presents an enhancement to $\,\mathcal{N}=1\,$ for $\,\kappa = 0\,$: the upper sign choice of the fluxes gives $\,\mathcal{N}=(1,0)\,$ ($\mathcal{N}=(0,1)$) for $\,\xi > 0\,$ ($\xi < 0$), while the lower sign choice leads to $\,\mathcal{N}=(1,0)\,$ ($\mathcal{N}=(0,1)$) for $\xi < 0$ ($\xi > 0$). \\[-2mm]

\item[-] \textbf{vac~26}, \textbf{vac~27}, \textbf{vac~30}, \textbf{vac~34} and \textbf{vac~35} exhibit an enhancement to $\,\mathcal{N}=(4,0)\,$ / {$\,\mathcal{N}=(0,4)\,$} for $\,\xi=-\kappa\,$ depending on the upper/lower sign choice; the same happens for \textbf{vac~32} as $\,\xi=\kappa$. \\[-2mm]

\item[-] \textbf{vac~31} and \textbf{vac~33} become $\,\mathcal{N}=1\,$ supersymmetric as $\,\xi=\kappa\,$ and $\,\xi=-\kappa$, respectively. In particular, the upper choice of signs leads to $\,\mathcal{N}=(1,0)\,$ ($\mathcal{N}=(0,1)$) for $\kappa >0$ ($\kappa<0$), while the lower sign choice gives $\,\mathcal{N}=(1,0)\,$ ($\mathcal{N}=(0,1)$) for $\,\kappa < 0\,$ ($\kappa>0$).\\[-2mm]

\item[-] \textcolor{ForestGreen}{\textbf{vac~46}} and \textcolor{ForestGreen}{\textbf{vac~49}} present an enhancement to $\,\mathcal{N}=4\,$ for $\,\kappa_2 = \pm \sqrt{\kappa_1 \, \kappa_3}\,$. \\[-2mm]

\item[-] \textcolor{ForestGreen}{\textbf{vac~48}} exhibits an enhancement to $\,\mathcal{N}=1\,$ for $\,\kappa_2 = \pm \sqrt{-\frac{\mathcal{D}_1}{2}}\,$. 

\end{itemize}

\subsubsection*{Minimal $\,\mathcal{N}=1\,$ models without axions}

The action of the $\,\mathbb{Z}_2^{*}\,$ element in (\ref{Z2*_element}) forces the axions to vanish but also projects out some of the fluxes in (\ref{metric_fluxes_SO(3)_O9}) and (\ref{gauge_fluxes_SO(3)_O9}). In particular, only non-zero flux parameters 
\begin{equation}
\label{Z2-even_fluxes_O9}
\omega_{3} 
\,\, , \,\, 
\omega_{4} 
\,\, , \,\, 
\omega_{5} 
\,\, , \,\, 
\omega_{6} 
\,\, , \,\, 
\omega_{7} 
\,\, , \,\, 
\omega_{11} 
\,\,\,\,\,\, , \,\,\,\,\,\, 
f_{32} 
\,\, , \,\, 
f_{34} 
\,\, , \,\, 
f_{35} 
\,\, \,\,\,\, \textrm{ and }\,\,\,\, \,\, 
f_{7} \ ,
\end{equation}
are permitted due to their index structure (they are $\,\mathbb{Z}_2^{*}$-even). In this restricted setup, the scalar potential can be derived from the real superpotential
\begin{equation}
\label{WO9}
\begin{array}{rcl}
g^{-1} \, W^{\textrm{O}9}_{\mathcal{N}=1} & = & \pm\frac38\sqrt{\frac{\mu_4}{A\,A_4\,\mu^3}}\,\omega_3\pm\frac34\sqrt{\frac{\mu\,\mu_4}{A\,A_4}}\,\omega_4+\frac{3}{8A_4\,\mu}\,\omega_5-\frac{3\mu_4}{8A}\,\omega_6-\frac{3\mu}{8\,A_4}\,\omega_7 \mp \frac38 \sqrt{\frac{\mu \, \mu_4}{A \, A_4}} \omega_{11}\\
& & \pm\frac38\sqrt{\frac{\mu_4}{A^3\,A_4\,\mu}}\,f_{32}\mp\frac18\sqrt{\frac{\mu^3\,\mu_4}{A^3\,A_4}}\,f_{34}-\frac{3}{8A\,A_4}\,f_{35}+\frac{\mu_4}{8A_4}\,f_7 \ ,
\end{array}
\end{equation}
using the formula (\ref{VfromW}) and the dilatons identification (\ref{dilaton_redef}). It is worth noticing that the flux vacua in Table~\ref{Table:flux_vacua_IIB_O9} (which have vanishing axions) generically require $\,\omega_{1},\,f_{31} \neq 0\,$ and, therefore, they are generically not captured by the $\,\mathcal{N}=1\,$ minimal model in (\ref{WO9}) (which has no axions to begin with). However, some of the flux vacua in Table~\ref{Table:flux_vacua_IIB_O9} become captured by the superpotential  (\ref{WO9}) when some of the free fluxes are properly adjusted. This is the case, for example, for \textcolor{Purple}{\textbf{vac~2}}–\textcolor{Purple}{\textbf{vac~5}} when $\,\xi = 0$, so the $\,\mathbb{Z}_{2}^{*}$-odd fluxes vanish. We will come back to this observation in Section~\ref{sec:discussion}.

Let us close this section with a technical remark. We have obtained the superpotential (\ref{WO9}) by direct computation of the mass of the $\,\mathbb{Z}_{2}^{*} \times \textrm{SO}(3)$-invariant gravitino within half-maximal supergravity.\footnote{\label{footnote_spin(8)}Which one of the two gravitini of the RSTU-models is $\,\mathbb{Z}_{2}^{*}$-even and which one is $\,\mathbb{Z}_{2}^{*}$-odd depends on how the $\,\mathbb{Z}_{2}^{*}\,$ action in (\ref{Z2*_element}) is extended to the gravitini of half-maximal supergravity, \textit{i.e.} $\,\textrm{Spin}(8)\,$ \textit{vs} $\,\textrm{SO}(8)$. Recall that the eight gravitini transform in the spinorial representation of the R-symmetry group $\,\textrm{SO}(8)_{\textrm{R}}$. The choice of the $\,\pm\,$ sign in (\ref{WO9}) selects which of the two gravitini of the RSTU-models is used to construct the superpotential. Only when a sign choice is made in (\ref{WO9}), the $\,\mathbb{Z}_{2}^{*}\,$ action is properly defined on the gravitini. Of course, both choices give rise to the same scalar potential and come along with the same structure of supersymmetric flux vacua. The same sign ambiguity will appear in other O$p$-plane duality frames.} However, since our setup fits into the class of type IIB orientifold reductions on co-closed G$_{2}$-structure manifolds put forward in \cite{Emelin:2021gzx}, the superpotential (\ref{WO9}) must follow also from the general expression provided in \cite{Emelin:2021gzx}.  We have verified that this is indeed the case once the additional $\textrm{SO}(3)$ symmetry of the RSTU-models is imposed.

\begin{landscape}
\begin{table}[]
\scalebox{0.63}{
\renewcommand{\arraystretch}{1.8}
\begin{tabular}{!{\vrule width 1.5pt}l!{\vrule width 1pt}c!{\vrule width 1pt}c!{\vrule width 1pt}ccccccccccc!{\vrule width 1pt}cccc!{\vrule width 1pt}c!{\vrule width 1.5pt}}
\Xcline{4-19}{1.5pt}
\multicolumn{3}{c!{\vrule width 1pt}}{}& \multicolumn{11}{c!{\vrule width 1pt}}{$\omega$} & \multicolumn{4}{c!{\vrule width 1pt}}{$F_{(3)}$} & \multicolumn{1}{c!{\vrule width 1pt}}{$F_{(7)}$} \\ 
\noalign{\hrule height 1.5pt}
     \hspace{6mm} ID & Type &  SUSY & $\omega_{1}$ & $ \omega_{2}$ & $  \omega_{3}$ & $\omega_{4}$ & $ \omega_{5}$ & $\omega_{6}$ & $ \omega_{7}$ & $\omega_{8}$ & $\omega_{9}$ & $\omega_{10}$ & $\omega_{11}$ & $f_{31}$ & $f_{32}$ & $f_{33}$ & $f_{34}$ & $f_7$ \\ 
     \noalign{\hrule height 1pt}
     $ \boxed{\textbf{vac~1}}^{\,\,*\,\dagger}$ & $\textrm{Mkw}_{3}$ & $\mathcal{N}=0$ & $ 0$ & $0 $ & $ 0$ & $ 0$ & $\kappa $ & $0 $ & $ -\kappa$ & $0 $ &  $0 $ & $0 $ & $0 $ & $ 0$ & $0 $  & $ 0$ & $ 0$ & $0 $\\
     \noalign{\hrule height 1pt}
    $\boxed{\textcolor{Purple}{\textbf{vac~2}}}^{\,\,* \,\dagger}$ & \multirow{4}{*}{$\textrm{AdS}_{3}$} & $\mathcal{N}=0$ & $2\xi$ & $\pm\frac{\kappa\xi}{\sqrt{\kappa^2+\xi^2}}$ & $0$ & $\kappa$ & $\mp\frac{\xi^2}{\sqrt{\kappa^2+\xi^2}} $ & $\pm\sqrt{\kappa^2+\xi^2} $ & $\pm\frac{\kappa^2}{\sqrt{\kappa^2+\xi^2}}$ & $0$ &  $\xi$ & $\mp\frac{\kappa\xi}{\sqrt{\kappa^2+\xi^2}}$ & $2\kappa$ & $\frac{2\xi^3}{\kappa^2+\xi^2} $ & $\frac{2\kappa\xi^2}{\kappa^2+\xi^2} $  & $\frac{2\kappa^2\xi}{\kappa^2+\xi^2}$  & $\frac{2\kappa^3}{\kappa^2+\xi^2}$ & $\pm 2\sqrt{\kappa^2+\xi^2}$ 
     \\
     \cline{1-1}\cline{3-19} 
     $\boxed{\textcolor{Purple}{\textbf{vac~3}}}^{\,\,* \,\dagger}$ &  & $\mathcal{N}=0$ & $2\xi$ & $ \pm\frac{\kappa\xi}{\sqrt{\kappa^2+\xi^2}}$ & $0$ & $\kappa$ & $\mp\frac{\xi^2}{\sqrt{\kappa^2+\xi^2}} $ & $\pm\sqrt{\kappa^2+\xi^2} $ & $\pm\frac{\kappa^2}{\sqrt{\kappa^2+\xi^2}}$ & $0$ &  $\xi$ & $\mp\frac{\kappa\xi}{\sqrt{\kappa^2+\xi^2}}$ & $2\kappa$ & $-\frac{2\xi^3}{\kappa^2+\xi^2} $ & $-\frac{2\kappa\xi^2}{\kappa^2+\xi^2} $  & $-\frac{2\kappa^2\xi}{\kappa^2+\xi^2}$  & $-\frac{2\kappa^3}{\kappa^2+\xi^2}$ & $\pm 2\sqrt{\kappa^2+\xi^2}$  \\
    \cline{1-1}\cline{3-19}
    %\noalign{\hrule height 1pt}
     $\boxed{\textcolor{Purple}{\textbf{vac~4}}}^{\,\,* \,\dagger}$ & & $\mathcal{N}=1$ & $2\xi$ & $ \mp\frac{\kappa\xi}{\sqrt{\kappa^2+\xi^2}}$ & $0$ & $\kappa$ & $\pm\frac{\xi^2}{\sqrt{\kappa^2+\xi^2}} $ & $\mp\sqrt{\kappa^2+\xi^2} $ & $\mp\frac{\kappa^2}{\sqrt{\kappa^2+\xi^2}}$ & $0$ &  $\xi$ & $\pm\frac{\kappa\xi}{\sqrt{\kappa^2+\xi^2}}$ & $2\kappa$ & $\frac{2\xi^3}{\kappa^2+\xi^2}$ & $\frac{2\kappa\xi^2}{\kappa^2+\xi^2} $  & $\frac{2\kappa^2\xi}{\kappa^2+\xi^2}$  & $\frac{2\kappa^3}{\kappa^2+\xi^2}$ & $\pm 2\sqrt{\kappa^2+\xi^2}$ 
     \\
     \cline{1-1}\cline{3-19} 
     $\boxed{\textcolor{Purple}{\textbf{vac~5}}}^{\,\,* \,\dagger}$ &  & $\mathcal{N}=4$ & $2\xi$ & $\mp\frac{\kappa\xi}{\sqrt{\kappa^2+\xi^2}}$ & $0$ & $\kappa$ & $\pm\frac{\xi^2}{\sqrt{\kappa^2+\xi^2}} $ & $\mp\sqrt{\kappa^2+\xi^2} $ & $\mp\frac{\kappa^2}{\sqrt{\kappa^2+\xi^2}}$ & $0$ &  $\xi$ & $\pm\frac{\kappa\xi}{\sqrt{\kappa^2+\xi^2}}$ & $2\kappa$ & $-\frac{2\xi^3}{\kappa^2+\xi^2} $ & $-\frac{2\kappa\xi^2}{\kappa^2+\xi^2} $  & $-\frac{2\kappa^2\xi}{\kappa^2+\xi^2}$  & $-\frac{2\kappa^3}{\kappa^2+\xi^2}$ & $\pm 2\sqrt{\kappa^2+\xi^2}$   \\
\noalign{\hrule height 1pt}
      $ \boxed{\textbf{vac~6}}^{\,\,* \,\dagger}$ & \multirow{2}{*}{$\textrm{AdS}_{3}$} & $\mathcal{N}=0$ & $ \kappa$ & $0 $ & $\xi $ & $\xi$ & $ 0$ & $ 0$ & $0 $ & $ 0$ &  $0 $ & $0 $ & $\xi $ & $\kappa $ & $ \xi$  & $0 $ & $ \xi$ & $\pm\sqrt{\kappa^2+4\xi^2} $ 
     \\
     \cline{1-1}\cline{3-19} 
     $ \boxed{\textbf{vac~7}}^{\,\,* \,\dagger}$ &  & $\mathcal{N}=4$ & $ \kappa$ & $0 $ & $\xi $ & $\xi$ & $ 0$ & $ 0$ & $0 $ & $ 0$ &  $0 $ & $0 $ & $\xi $ & $-\kappa $ & $ -\xi$  & $0 $ & $ -\xi$ & $\pm\sqrt{\kappa^2+4\xi^2} $ \\
\noalign{\hrule height 1pt}
     $ \boxed{\textcolor{ForestGreen}{\textbf{vac~8}}}^{\,\,* \,\dagger}$ & \multirow{4}{*}{$\textrm{AdS}_{3}$} & $\mathcal{N}=0,1$ & $\kappa $ & $ 0$ & $ 0$ & $ \xi$ & $0 $ & $0 $ & $ 0$ & $ 0$ &  $0 $ & $ 0$ & $ \xi$ & $ \kappa$ & $ 0$  & $ 0$ & $ \xi$ & $\pm\sqrt{\kappa^2+\xi^2} $ 
     \\
    \cline{1-1}\cline{3-19} 
     $\boxed{\textcolor{ForestGreen}{\textbf{vac~9}}}^{\,\,* \,\dagger}$ &  &  $\mathcal{N}=0,4$ & $\kappa $ & $ 0$ & $ 0$ & $ \xi$ & $0 $ & $0 $ & $ 0$ & $ 0$ &  $0 $ & $ 0$ & $ \xi$ & $ -\kappa$ & $ 0$  & $ 0$ & $ -\xi$ & $\pm\sqrt{\kappa^2+\xi^2} $    \\
     \cline{1-1}\cline{3-19} 
     $ \boxed{\textcolor{ForestGreen}{\textbf{vac~10}}}^{\,\,* \,\dagger}$ &  &  $\mathcal{N}=0$& $\kappa $ & $ 0$ & $ 0$ & $ \xi$ & $0 $ & $0 $ & $ 0$ & $ 0$ &  $0 $ & $ 0$ & $ \xi$ & $ \kappa$ & $ 0$  & $ 0$ & $ -\xi$ & $\pm\sqrt{\kappa^2+\xi^2} $   \\
     \cline{1-1}\cline{3-19} 
     $ \boxed{\textcolor{ForestGreen}{\textbf{vac~11}}}^{\,\,* \,\dagger}$ &  &  $\mathcal{N}=1,4$ & $\kappa $ & $ 0$ & $ 0$ & $ \xi$ & $0 $ & $0 $ & $ 0$ & $ 0$ &  $0 $ & $ 0$ & $ \xi$ & $ -\kappa$ & $ 0$  & $ 0$ & $ \xi$ & $\pm\sqrt{\kappa^2+\xi^2} $\\
      \noalign{\hrule height 1pt}
     $ \boxed{\textbf{vac~12}}^{\,\,* \,\dagger}$ &  \multirow{4}{*}{$\textrm{AdS}_{3}$}  & $\mathcal{N}=0,4$ & $\kappa $ & $0 $ & $0 $ & $ 0$ & $ 0$ & $ 0$ & $ 0$ & $ 0$ &  $ 0$ & $ 0$ & $ \xi$ & $-\kappa $ & $0 $  & $0 $ & $\xi$ & $ \pm\sqrt{\kappa^2+\xi^2}$\\ 
     \cline{1-1}\cline{3-19}
     $\boxed{\textbf{vac~13}}^{\,\,* \,\dagger}$ &  & $\mathcal{N}=0,4$ & $\kappa $ & $0 $ & $0 $ & $ 0$ & $ 0$ & $ 0$ & $ 0$ & $ 0$ &  $ 0$ & $ 0$ & $ \xi$ & $\kappa $ & $0 $  & $0 $ & $-\xi$ & $ \pm\sqrt{\kappa^2+\xi^2}$\\
     \cline{1-1}\cline{3-19}
     $ \boxed{\textbf{vac~14}}^{\,\,* \,\dagger}$ & & $\mathcal{N}=0$ & $\kappa $ & $0 $ & $0 $ & $ 0$ & $ 0$ & $ 0$ & $ 0$ & $ 0$ &  $ 0$ & $ 0$ & $ \xi$ & $\kappa $ & $0 $  & $0 $ & $\xi$ & $ \pm\sqrt{\kappa^2+\xi^2}$\\
     \cline{1-1}\cline{3-19}
     $ \boxed{\textbf{vac~15}}^{\,\,* \,\dagger}$ &  & $\mathcal{N}=4$ & $\kappa $ & $0 $ & $0 $ & $ 0$ & $ 0$ & $ 0$ & $ 0$ & $ 0$ &  $ 0$ & $ 0$ & $ \xi$ & $-\kappa $ & $0 $  & $0 $ & $-\xi$ & $ \pm\sqrt{\kappa^2+\xi^2}$
      \\ 
      \noalign{\hrule height 1pt} 
     $\boxed{\textbf{vac~16}}^{\,\,* \,\dagger}$ & \multirow{4}{*}{$\textrm{AdS}_{3}$} & $\mathcal{N}=0,1$ & $\kappa $ & $0 $ & $0 $ & $ 0$ & $ 0$ & $ 0$ & $ 0$ & $ 0$ &  $ \kappa$ & $ 0$ & $ \xi$ & $\kappa $ & $0 $  & $0 $ & $\xi$ & $ \pm\sqrt{\kappa^2+\xi^2}$  \\ 
     \cline{1-1}\cline{3-19} 
     $ \boxed{\textbf{vac~17}}^{\,\,* \,\dagger}$ &  & $\mathcal{N}=0,4$ & $\kappa $ & $0 $ & $0 $ & $ 0$ & $ 0$ & $ 0$ & $ 0$ & $ 0$ &  $ \kappa$ & $ 0$ & $ \xi$ & $-\kappa $ & $0 $  & $0 $ & $-\xi$ & $ \pm\sqrt{\kappa^2+\xi^2}$
      \\
     \cline{1-1}\cline{3-19}
     $ \boxed{\textbf{vac~18}}^{\,\,* \,\dagger}$ &  & $\mathcal{N}=0$ & $\kappa $ & $0 $ & $0 $ & $ 0$ & $ 0$ & $ 0$ & $ 0$ & $ 0$ &  $ \kappa$ & $ 0$ & $ \xi$ & $-\kappa $ & $0 $  & $0 $ & $\xi$ & $ \pm\sqrt{\kappa^2+\xi^2}$  \\ 
     \cline{1-1}\cline{3-19} 
     $ \boxed{\textbf{vac~19}}^{\,\,* \,\dagger}$ &  & $\mathcal{N}=1,4$ & $\kappa $ & $0 $ & $0 $ & $ 0$ & $ 0$ & $ 0$ & $ 0$ & $ 0$ &  $ \kappa$ & $ 0$ & $ \xi$ & $\kappa $ & $0 $  & $0 $ & $-\xi$ & $ \pm\sqrt{\kappa^2+\xi^2}$
     \\
     \noalign{\hrule height 1pt}
     $\boxed{\textbf{vac~20}}^{\,\,* \,\dagger}$ & \multirow{2}{*}{$\textrm{AdS}_{3}$} & $\mathcal{N}=0$ & $\frac{\kappa^2+\xi^2}{\xi}$ & $0$ & $- \kappa$ & $0 $ & $0 $ & $0 $ & $0 $ &  $0 $ & $ \xi$ & $0$ &$  \kappa$ &  $\frac{-\kappa^4-\xi^4}{\xi\left(\kappa^2+\xi^2\right)}$  & $ \frac{\kappa\left(\kappa^2-\xi^2\right)}{\kappa^2+\xi^2}$ & $\frac{-2\kappa^2\xi}{\kappa^2+\xi^2} $ & $ \frac{\kappa\left(\xi^2-\kappa^2\right)}{\kappa^2+\xi^2}$ & $\pm\frac{\kappa^2+\xi^2}{\xi} $
     \\
     \cline{1-1}\cline{3-19} 
     $\boxed{\textbf{vac~21}}^{\,\,* \,\dagger}$ &  & $\mathcal{N}=0,1$ & $\frac{\kappa^2+\xi^2}{\xi}$ & $0$ & $-\kappa$ & $0 $ & $0 $ & $0 $ & $0 $ &  $0 $ & $ \xi$ & $0$ &$  \kappa$ &   $ \frac{\kappa^4+\xi^4}{\xi\left(\kappa^2+\xi^2\right)}$  & $ \frac{\kappa\left(\xi^2-\kappa^2\right)}{\kappa^2+\xi^2}$ & $\frac{2\kappa^2\xi}{\kappa^2+\xi^2} $ & $ \frac{\kappa\left(\kappa^2-\xi^2\right)}{\kappa^2+\xi^2}$ & $\pm\frac{\kappa^2+\xi^2}{\xi} $  \\
     \noalign{\hrule height 1pt}
      $\boxed{\textbf{vac~22}}^{\,\,* \,\dagger}$ & \multirow{2}{*}{$\textrm{AdS}_{3}$} & $\mathcal{N}=0$ & $\frac{\kappa^2+\xi^2}{\xi} $ & $0 $ & $ -\kappa$ & $ 0$ & $0 $ & $0 $ & $ 0$ & $0 $ &  $ \xi$ & $0 $ & $ \kappa$ & $\frac{\kappa^2-\xi^2}{\xi} $ & $-\kappa $  & $0 $ & $-\kappa$ &  $\pm\frac{\kappa^2+\xi^2}{\xi}  $ 
     \\
     \cline{1-1}\cline{3-19} 
     $\boxed{\textbf{vac~23}}^{\,\,* \,\dagger}$ &  & $\mathcal{N}=1$ & $\frac{\kappa^2+\xi^2}{\xi} $ & $0 $ & $ -\kappa$ & $ 0$ & $0 $ & $0 $ & $ 0$ & $0 $ &  $ \xi$ & $0 $ & $ \kappa$ & $\frac{\xi^2-\kappa^2}{\xi} $ & $\kappa $  & $0 $ & $ \kappa$ & $\pm\frac{\kappa^2+\xi^2}{\xi}  $
     \\
     \noalign{\hrule height 1pt} 
     $ \boxed{\textbf{vac~24}}^{\,\,* \,\dagger}$ & \multirow{2}{*}{$\textrm{AdS}_{3}$} & $\mathcal{N}=0$ & $-\sqrt{\kappa^2+\xi^2} $ & $0 $ & $0 $ & $ 0$ & $ 0$ & $ 0$ & $ 0$ & $-\sqrt{\kappa^2+\xi^2} $ &  $ -\sqrt{\kappa^2+\xi^2}$ & $ 0$ & $ \frac{-2\xi\sqrt{\kappa^2+\xi^2}}{\kappa}$ & $-\xi $ & $\kappa$  & $\xi $ & $\frac{\kappa^2+2\xi^2}{\kappa}$ & $ \pm\frac{2(\kappa^2+\xi^2)}{\kappa}$\\ 
     \cline{1-1}\cline{3-19}
     $ \boxed{\textbf{vac~25}}^{\,\,* \,\dagger}$ &  & $\mathcal{N}=0$ & $\sqrt{\kappa^2+\xi^2} $ & $0 $ & $0 $ & $ 0$ & $ 0$ & $ 0$ & $ 0$ & $ \sqrt{\kappa^2+\xi^2} $ &  $ \sqrt{\kappa^2+\xi^2} $ & $ 0$ & $\frac{2\xi\sqrt{\kappa^2+\xi^2}}{\kappa} $ & $-\xi $ & $\kappa $  & $\xi $ & $ \frac{\kappa^2+2\xi^2}{\kappa}$ & $ \pm\frac{2(\kappa^2+\xi^2)}{\kappa}$  \\
      \noalign{\hrule height 1.5pt}  
\end{tabular}}
\caption{Fluxes in the type~IIB with O$9$-planes duality frame producing a vacuum at the origin of moduli space.}
\label{Table:flux_vacua_IIB_O9}
\end{table}

\begin{table}[]
\scalebox{0.565}{
\renewcommand{\arraystretch}{2}
\begin{tabular}{!{\vrule width 1.5pt}l!{\vrule width 1pt}c!{\vrule width 1pt}c!{\vrule width 1pt}ccccccccccc!{\vrule width 1pt}cccc!{\vrule width 1pt}c!{\vrule width 1.5pt}}
\Xcline{4-19}{1.5pt}
\multicolumn{3}{c!{\vrule width 1pt}}{}& \multicolumn{11}{c!{\vrule width 1pt}}{$\omega$} & \multicolumn{4}{c!{\vrule width 1pt}}{$F_{(3)}$} & \multicolumn{1}{c!{\vrule width 1pt}}{$F_{(7)}$} \\ 
\noalign{\hrule height 1.5pt}
     \hspace{6mm} ID & Type &  SUSY & $\omega_{1}$ & $ \omega_{2}$ & $  \omega_{3}$ & $\omega_{4}$ & $ \omega_{5}$ & $\omega_{6}$ & $ \omega_{7}$ & $\omega_{8}$ & $\omega_{9}$ & $\omega_{10}$ & $\omega_{11}$ & $f_{31}$ & $f_{32}$ & $f_{33}$ & $f_{34}$ & $f_7$ \\ 
     \noalign{\hrule height 1pt}
     \noalign{\hrule height 1pt}
     $ \boxed{\textbf{vac~26}}^{\,\,* \,\dagger}$ & \multirow{4}{*}{$\textrm{AdS}_{3}$} &  $\mathcal{N}=1, 4$& $2\xi+\kappa$ & $ 0$ & $-\xi$ & $\kappa$ & $0$ & $0 $ & $ 0$ & $ -\kappa$ &  $\xi$ & $0$ & $ \xi+2\kappa$ & $\kappa$ & $\xi$  & $\kappa$ & $\xi$ & $\pm 2\sqrt{\kappa^2+\xi^2} $   \\
     \cline{1-1}\cline{3-19} 
     $ \boxed{\textbf{vac~27}}^{\,\,* \,\dagger}$ &  &  $\mathcal{N}=1, 4$ & $2\xi+\kappa$ & $0$ & $\xi$ & $-\kappa$ & $ 0$ & $0$ & $0$ & $-\kappa$ &  $\xi$ & $0$ & $-\xi-2\kappa$ & $\kappa$ & $ -\xi$  & $\kappa$ & $-\xi$ & $\pm 2\sqrt{\kappa^2+\xi^2} $\\
\cline{1-1}\cline{3-19}
     $ \boxed{\textbf{vac~28}}^{\,\,* \,\dagger}$ &  & $\mathcal{N}=0$ & $2\xi-\kappa$ & $0$ & $-\xi$ & $-\kappa$ & $0$ & $0$ & $0$ & $\kappa$ &  $\xi$ & $ 0$ & $\xi-2\kappa$ & $\kappa$ & $ -\xi$  & $\kappa$ & $-\xi$ & $\pm 2\sqrt{\kappa^2+\xi^2} $ 
     \\
\cline{1-1}\cline{3-19} 
     $ \boxed{\textbf{vac~29}}^{\,\,* \,\dagger}$ &  & $\mathcal{N}=0$ & $2\xi-\kappa$ & $0$ & $\xi$ & $\kappa$ & $0$ & $ 0$ & $0$ & $\kappa$ &  $\xi$ & $0 $ & $-\xi+2\kappa$ & $\kappa$ & $\xi$  & $\kappa$ & $ \xi$ & $\pm 2\sqrt{\kappa^2+\xi^2} $ \\
\noalign{\hrule height 1pt}
     $ \boxed{{\textbf{vac~30}}}^{\,\,* \,\dagger}$ & \multirow{4}{*}{$\textrm{AdS}_{3}$} & $\mathcal{N}=0, 4$ & $-\xi-2\kappa$ & $0$ & $-\kappa$ & $\xi$ & $0$ & $0 $ & $0$ & $\xi$ &  $-\kappa$ & $0$ & $2\xi+\kappa$ & $\kappa$ & $-\xi$  & $\kappa$ & $-\xi$ & $\pm 2\sqrt{\kappa^2+\xi^2} $ 
     \\
\cline{1-1}\cline{3-19} 
     $\boxed{\textbf{vac~31}}^{\,\,* \,\dagger}$ &  &  $\mathcal{N}=0, 1$ & $\xi+2\kappa$ & $0$ & $-\kappa$ & $\xi$ & $0$ & $0$ & $0$ & $-\xi$ &  $\kappa$ & $ 0$ & $2\xi+\kappa$ & $\kappa$ & $\xi$  & $\kappa$ & $\xi$ & $\pm 2\sqrt{\kappa^2+\xi^2} $\\
\cline{1-1}\cline{3-19}
     $ \boxed{{\textbf{vac~32}}}^{\,\,* \,\dagger}$ &  & $\mathcal{N}=0, 4$ & $\xi-2\kappa$ & $0$ & $\kappa$ & $\xi$ & $0$ & $0 $ & $0$ & $-\xi$ &  $-\kappa$ & $0$ & $2\xi-\kappa$ & $\kappa$ & $-\xi$  & $\kappa$ & $-\xi$ & $\pm 2\sqrt{\kappa^2+\xi^2} $
     \\
    \cline{1-1}\cline{3-19} 
     $\boxed{\textbf{vac~33}}^{\,\,* \,\dagger}$ &  &  $\mathcal{N}=0, 1$ & $-\xi+2\kappa$ & $0$ & $\kappa$ & $\xi$ & $0$ & $0 $ & $0$ & $\xi$ &  $\kappa$ & $0$ & $2\xi-\kappa$ & $\kappa$ & $\xi$  & $\kappa$ & $\xi$ & $\pm 2\sqrt{\kappa^2+\xi^2} $\\
\noalign{\hrule height 1pt}
     $ \boxed{\textbf{vac~34}}^{\,\,* \,\dagger}$ &  \multirow{2}{*}{$\textrm{AdS}_{3}$}  & $\mathcal{N}=0, 4$ & $\xi$ & $0$ & $\kappa$ & $\kappa$ & $0$ & $0$ & $0$ & $\xi$ &  $\xi$ & $ 0$ & $ \kappa$ & $\kappa $ & $\xi$  & $\kappa $ & $\xi$ & $ \pm 2\sqrt{\kappa^2+\xi^2}$\\ 
     \cline{1-1}\cline{3-19}
     $\boxed{\textbf{vac~35}}^{\,\,* \,\dagger}$ &  & $\mathcal{N}=0, 4$ & $\xi$ & $0 $ & $-\kappa$ & $-\kappa$ & $ 0$ & $ 0$ & $ 0$ & $ \xi$ &  $ \xi$ & $ 0$ & $-\kappa$ & $\kappa $ & $-\xi $  & $\kappa $ & $-\xi$ & $ \pm 2\sqrt{\kappa^2+\xi^2}$\\
\noalign{\hrule height 1pt}
     $ \boxed{\textbf{vac~36}}^{\,\,* \,\dagger}$ & \multirow{2}{*}{$\textrm{AdS}_{3}$}  & $\mathcal{N}=0$ & $\kappa $ & $0 $ & $\xi$ & $ \xi$ & $ 0$ & $ 0$ & $ 0$ & $ \kappa$ &  $ \kappa$ & $ 0$ & $ \xi$ & $\kappa $ & $\xi $  & $\kappa $ & $\xi$ & $ \pm 2\sqrt{\kappa^2+\xi^2}$ \\      
     \cline{1-1}\cline{3-19}
     $ \boxed{\textbf{vac~37}}^{\,\,* \,\dagger}$ &  & $\mathcal{N}=4$ & $-\kappa $ & $0 $ & $\xi $ & $ \xi$ & $ 0$ & $ 0$ & $ 0$ & $-\kappa$ &  $-\kappa$ & $ 0$ & $ \xi$ & $\kappa $ & $-\xi$  & $\kappa $ & $-\xi$ & $ \pm 2\sqrt{\kappa^2+\xi^2}$\\
\noalign{\hrule height 1pt}
     $ \boxed{\textbf{vac~38}}^{\,\,* \,\dagger}$ & \multirow{2}{*}{$\textrm{AdS}_{3}$}  & $\mathcal{N}=0$ & $ \frac{2\kappa^2+\xi^2}{\xi} $ & $0 $ & $\kappa$ & $\kappa$ & $0$ & $0$ & $0$ & $\xi$ &  $\xi$ & $ 0$ & $ -\kappa$ & $ \frac{2\kappa \sqrt{\kappa^2+\xi^2}}{\xi} $ & $ \sqrt{\kappa^2+\xi^2}$  & $ 0 $ & $ \sqrt{\kappa^2+\xi^2}$ & $ \pm \frac{2\left( \kappa^2+\xi^2 \right)}{\xi}$\\
     \cline{1-1}\cline{3-19}
     $ \boxed{\textbf{vac~39}}^{\,\,* \,\dagger}$ &  & $\mathcal{N}=0$ & $ \frac{2\kappa^2+\xi^2}{\xi} $ & $0 $ & $\kappa$ & $\kappa$ & $0$ & $0$ & $0$ & $\xi$ &  $\xi$ & $ 0$ & $ -\kappa$ & $ -\frac{2\kappa \sqrt{\kappa^2+\xi^2}}{\xi} $ & $ -\sqrt{\kappa^2+\xi^2}$  & $ 0 $ & $ -\sqrt{\kappa^2+\xi^2}$ & $ \pm \frac{2\left( \kappa^2+\xi^2 \right)}{\xi}$
     \\         
\noalign{\hrule height 1pt}
     $ \boxed{\textbf{vac~40}}^{\,\,* \,\dagger}$ & \multirow{2}{*}{$\textrm{AdS}_{3}$}  & $\mathcal{N}=0$ & $\frac{\left(\kappa^2-\xi^2 \right)^2-2\xi^4}{2\kappa^2\xi}$ & $0$ & $ \frac{\kappa^2-\xi^2}{2\kappa}$ & $ \kappa$ & $ 0$ & $ 0$ & $ 0$ & $\xi$ &  $\frac{\xi\left(\kappa^2-\xi^2\right)}{2\kappa^2}$ & $0$ & $\frac{\kappa^2+3\xi^2}{2\kappa} $ & $ \frac{\xi^4-\kappa^4}{2\kappa^2\xi}$ & $-\frac{\kappa^2+\xi^2}{2\kappa}$  & $ 0 $ & $-\frac{\kappa^2+\xi^2}{2\kappa}$ & $ \pm \frac{\left(\kappa^2+\xi^2 \right)^2}{2\kappa^2\xi}$\\
     \cline{1-1}\cline{3-19}
     $ \boxed{\textbf{vac~41}}^{\,\,* \,\dagger}$ &  & $\mathcal{N}=0$ & $\frac{\left(\kappa^2-\xi^2 \right)^2-2\xi^4}{2\kappa^2\xi}$ & $0$ & $ \frac{\kappa^2-\xi^2}{2\kappa}$ & $\kappa$ & $0$ & $ 0$ & $ 0$ & $\xi$ &  $\frac{\xi\left(\kappa^2-\xi^2\right)}{2\kappa^2}$ & $0$ & $\frac{\kappa^2+3\xi^2}{2\kappa}$ & $ \frac{\kappa^4-\xi^4}{2\kappa^2\xi}$ & $\frac{\kappa^2+\xi^2}{2\kappa}$  & $ 0 $ & $\frac{\kappa^2+\xi^2}{2\kappa}$ & $ \pm \frac{\left(\kappa^2+\xi^2 \right)^2}{2\kappa^2\xi}$ \\
\noalign{\hrule height 1pt}
     $ \boxed{\textbf{vac~42}}^{\,\,* \,\dagger}$ & \multirow{2}{*}{$\textrm{AdS}_{3}$}  & $\mathcal{N}=0$ & $\frac{\kappa_2 \mathcal{C}_2 +2 \kappa_1 \kappa_3 \sqrt{\mathcal{C}_1}}{\kappa_2^2-\kappa_3^2}$ & 0  & $\kappa_1 $ & $ \kappa_1$ & 0 & 0 & 0 & $\kappa_2$ &  $\kappa_2$ & 0 & $ \frac{\kappa_1 \mathcal{C}_3+2\kappa_2\kappa_3 \sqrt{\mathcal{C}_1}}{\kappa_3^2-\kappa_2^2}$ & $\frac{\kappa_3 \mathcal{C}_2 + 2\kappa_1\kappa_2 \sqrt{\mathcal{C}_1}}{\kappa_3^2-\kappa_2^2} $ & $ - \sqrt{\mathcal{C}_1}$  & $\kappa_3 $ & $\frac{2\kappa_1\kappa_2\kappa_3+\left( \kappa_2^2+\kappa_3^2\right)\sqrt{\mathcal{C}_1}}{\kappa_3^2-\kappa_2^2} $ & $ \pm 2\frac{\sqrt{\left( \kappa_1^2+\kappa_2^2\right) \mathcal{C}_4}}{\kappa_3^2-\kappa_2^2}$\\
     \cline{1-1}\cline{3-19}
     $ \boxed{\textbf{vac~43}}^{\,\,* \,\dagger}$ &  & $\mathcal{N}=0$ & $\frac{\kappa_2 \mathcal{C}_2 -2 \kappa_1 \kappa_3 \sqrt{\mathcal{C}_1}}{\kappa_2^2-\kappa_3^2}$  & 0 & $\kappa_1 $ & $ \kappa_1$ & 0& 0& 0 & $\kappa_2$ &  $\kappa_2$ & 0 & $ \frac{\kappa_1 \mathcal{C}_3-2\kappa_2\kappa_3 \sqrt{\mathcal{C}_1}}{\kappa_3^2-\kappa_2^2}$ & $\frac{\kappa_3 \mathcal{C}_2 - 2\kappa_1\kappa_2 \sqrt{\mathcal{C}_1}}{\kappa_3^2-\kappa_2^2} $ & $  \sqrt{\mathcal{C}_1}$  & $\kappa_3 $ & $\frac{2\kappa_1\kappa_2\kappa_3-\left( \kappa_2^2+\kappa_3^2\right)\sqrt{\mathcal{C}_1}}{\kappa_3^2-\kappa_2^2} $ & $ \pm 2\frac{\sqrt{\left( \kappa_1^2+\kappa_2^2\right) \mathcal{C}_5}}{\kappa_3^2-\kappa_2^2}$
      \\ 
     \noalign{\hrule height 1pt}
     $ \boxed{\textcolor{RoyalBlue}{\textbf{vac~44}}}^{\,\,* \,\dagger}$ &  \multirow{2}{*}{$\textrm{AdS}_{3}$}  & $\mathcal{N}=4$ & $-\frac{\kappa_3^2+\kappa_1\left(\kappa_1-\kappa_2\right)}{\kappa_3}$ & 0 & $\kappa_1$ & $\kappa_1$ & 0& 0& 0 & $-\kappa_3$ &  $-\kappa_3$ & 0 & $ \kappa_2$ & $\frac{\kappa_3^2+\kappa_1\left(\kappa_1-\kappa_2\right)}{\kappa_3} $ & $-\kappa_1$  & $\kappa_3 $ & $-\kappa_2$ & $ \pm \frac{\sqrt{\left[\left(\kappa_1-\kappa_2\right)^2 + 4\kappa_3^2\right]\left(\kappa_1^2 + \kappa_3^2\right)}}{\kappa_3}$\\ 
     \cline{1-1}\cline{3-19}
     $\boxed{\textcolor{RoyalBlue}{\textbf{vac~45}}}^{\,\,* \,\dagger}$ &  & $\mathcal{N}=0$ & $\frac{\kappa_3^2+\kappa_1\left(\kappa_1-\kappa_2\right)}{\kappa_3}$ & 0 & $\kappa_1$ & $\kappa_1$ & 0 & 0 & 0 & $\kappa_3$ &  $\kappa_3$ & 0 &  $ \kappa_2$ & $\frac{\kappa_3^2+\kappa_1\left(\kappa_1-\kappa_2\right)}{\kappa_3} $ & $\kappa_1$  & $\kappa_3 $ & $\kappa_2$ & $ \pm \frac{\sqrt{\left[\left(\kappa_1-\kappa_2\right)^2 + 4\kappa_3^2\right]\left(\kappa_1^2 + \kappa_3^2\right)}}{\kappa_3}$
      \\ 
\noalign{\hrule height 1pt}
     $ \boxed{\textcolor{ForestGreen}{\textbf{vac~46}}}^{\,\,* \,\dagger}$ &  \multirow{2}{*}{$\textrm{AdS}_{3}$}  & $\mathcal{N}=1,4$ & $\frac{\kappa_1^2 \mathcal{D}_1+\kappa_2^2\mathcal{D}_2}{\kappa_1^2\kappa_2}$ & 0  & $\frac{\mathcal{D}_2}{\kappa_1}$ & $\kappa_1$ & 0 & 0 & 0 & $\kappa_2$ &  $\frac{\kappa_2 \mathcal{D}_2}{\kappa_1^2}$ & 0 & $ \kappa_3$ & $\frac{-\kappa_1^2 \mathcal{D}_1+\kappa_2^2\left(\mathcal{D}_2-2\kappa_1^2\right)}{\kappa_1^2\kappa_2} $ & $-\frac{\mathcal{D}_2}{\kappa_1}$  & $-\kappa_2$ & $\frac{\kappa_1\kappa_3-2\kappa_2^2}{\kappa_1}$ & $ \pm \frac{\left(\kappa_1^2+\kappa_2^2\right)\sqrt{ \mathcal{D}_1^2 + \kappa_2^2 \left(\kappa_1^2+\mathcal{D}_1+\mathcal{D}_2\right)}}{\kappa_1^2 \kappa_2}$\\ 
     \cline{1-1}\cline{3-19}
     $\boxed{\textcolor{ForestGreen}{\textbf{vac~47}}}^{\,\,* \,\dagger}$ &  & $\mathcal{N}=0$ & $\frac{\kappa_1^2 \mathcal{D}_1+\kappa_2^2\mathcal{D}_2}{\kappa_1^2\kappa_2}$ & 0  & $\frac{\mathcal{D}_2}{\kappa_1}$ & $\kappa_1$ & 0 & 0 & 0  & $\kappa_2$ &  $\frac{\kappa_2 \mathcal{D}_2}{\kappa_1^2}$ & 0 & $ \kappa_3$ & $\frac{\kappa_1^2 \mathcal{D}_1-\kappa_2^2\left(\mathcal{D}_2-2\kappa_1^2\right)}{\kappa_1^2\kappa_2} $ & $\frac{\mathcal{D}_2}{\kappa_1}$  & $\kappa_2$ & $-\frac{\kappa_1\kappa_3-2\kappa_2^2}{\kappa_1}$ & $\pm \frac{\left(\kappa_1^2+\kappa_2^2\right)\sqrt{ \mathcal{D}_1^2 + \kappa_2^2 \left(\kappa_1^2+\mathcal{D}_1+\mathcal{D}_2\right)}}{\kappa_1^2 \kappa_2}$
      \\ 
\noalign{\hrule height 1pt}
     $ \boxed{\textcolor{ForestGreen}{\textbf{vac~48}}}^{\,\,* \,\dagger}$ &  \multirow{2}{*}{$\textrm{AdS}_{3}$}  & $\mathcal{N}=0,1$ & $\frac{\kappa_1^2 \mathcal{D}_1+\kappa_2^2\mathcal{D}_2}{\kappa_1^2\kappa_2}$ & 0 & $\frac{\mathcal{D}_2}{\kappa_1}$ & $\kappa_1$ & 0 & 0 & 0 & $\kappa_2$ &  $\frac{\kappa_2 \mathcal{D}_2}{\kappa_1^2}$ & 0 &  $ \kappa_3$ & $\frac{\kappa_1^4 \mathcal{D}_1+\kappa_2^4\left(\mathcal{D}_2-\kappa_1^2\right)+2\kappa_1^4\kappa_2^2}{\kappa_1^2\kappa_2\left(\kappa_1^2+\kappa_2^2\right)} $ & $ \frac{4\kappa_1^2-\mathcal{D}_2}{\kappa_1} - \frac{2\kappa_1^2\left(\kappa_1+\kappa_3\right)}{\kappa_1^2+\kappa_2^2}$ & $\frac{\kappa_2\left(2\kappa_2^2-\kappa_1\kappa_3+\mathcal{D}_2\right)}{\kappa_1^2+\kappa_2^2}$   & $\frac{\kappa_1\kappa_3\left( \kappa_1^2-\kappa_2^2 \right)+2\kappa_2^4}{\kappa_1\left( \kappa_1^2+\kappa_2^2\right)}$ & $ \pm \frac{\left(\kappa_1^2+\kappa_2^2\right)\sqrt{ \mathcal{D}_1^2 + \kappa_2^2 \left(\kappa_1^2+\mathcal{D}_1+\mathcal{D}_2\right)}}{\kappa_1^2 \kappa_2}$\\ 
     \cline{1-1}\cline{3-19}
     $\boxed{\textcolor{ForestGreen}{\textbf{vac~49}}}^{\,\,* \,\dagger}$ &  & $\mathcal{N}=0,4$ & $\frac{\kappa_1^2 \mathcal{D}_1+\kappa_2^2\mathcal{D}_2}{\kappa_1^2\kappa_2}$ & 0 & $\frac{\mathcal{D}_2}{\kappa_1}$ & $\kappa_1$ & 0 & 0 & 0 & $\kappa_2$ &  $\frac{\kappa_2 \mathcal{D}_2}{\kappa_1^2}$ & 0 & $ \kappa_3$ & $-\frac{\kappa_1^4 \mathcal{D}_1+\kappa_2^4\left(\mathcal{D}_2-\kappa_1^2\right)+2\kappa_1^4\kappa_2^2}{\kappa_1^2\kappa_2\left(\kappa_1^2+\kappa_2^2\right)} $ & $  \frac{\mathcal{D}_2-4\kappa_1^2}{\kappa_1} + \frac{2\kappa_1^2\left(\kappa_1+\kappa_3\right)}{\kappa_1^2+\kappa_2^2}$  &  $-\frac{\kappa_2\left(2\kappa_2^2-\kappa_1\kappa_3+\mathcal{D}_2\right)}{\kappa_1^2+\kappa_2^2}$ & $-\frac{\kappa_1\kappa_3\left( \kappa_1^2-\kappa_2^2 \right)+2\kappa_2^4}{\kappa_1\left( \kappa_1^2+\kappa_2^2\right)}$ & $ \pm \frac{\left(\kappa_1^2+\kappa_2^2\right)\sqrt{ \mathcal{D}_1^2 + \kappa_2^2 \left(\kappa_1^2+\mathcal{D}_1+\mathcal{D}_2\right)}}{\kappa_1^2 \kappa_2}$
      \\ 
     \noalign{\hrule height 1.5pt}  
\end{tabular}}
\caption*{Continuation of Table~\ref{Table:flux_vacua_IIB_O9}. In order to present \textbf{vac~42,43} we have introduced the quantities $\,\mathcal{C}_{4,5} = \kappa_2^2 \left(\kappa_1^2+\kappa_2^2\right) + \kappa_3^2\left( \kappa_1^2-\kappa_2^2\right) \pm 2\kappa_1\kappa_2\kappa_3 \sqrt{\mathcal{C}_1}$, $\,\mathcal{C}_3 = -\kappa_1^2+2\kappa_3^2 + \mathcal{C}_1\,$ and $\,\mathcal{C}_2 = \kappa_1^2 + \mathcal{C}_1\,$ with $\,\mathcal{C}_1 = \kappa_1^2+\kappa_2^2-\kappa_3^2$. Similarly, in order to present \textcolor{ForestGreen}{\textbf{vac~46--49}} we have introduced the quantities $\,\mathcal{D}_2 = \kappa_2^2 + \kappa_1 \left(\kappa_1-\kappa_3 \right)\,$ and $\,\mathcal{D}_1 = \kappa_1 \left(\kappa_1-\kappa_3 \right)$.}
\label{Table:flux_vacua_IIB_O9_continuation}
\end{table}
\end{landscape}

\subsubsection{Type~IIB with O$7$-planes}

The type IIB with O$7$-planes duality frame was considered in Section~$3.7$ of \cite{Arboleya:2024vnp}. Introducing O$7$-planes is compatible with an $\,\textrm{SL}(2) \times \textrm{SL}(5)\,$ covariant description of the fluxes. The O$7$-plane orientifold action $\mathcal{O}_{\mathbb{Z}_{2}} = \Omega_{P} \, (-1)^{F_{L}} \, \sigma_{\textrm{O}7}$ allows for metric fluxes $\,\omega\,$ as well as gauge fluxes $\,H_{(3)}$, $\,F_{(3)}\,$ and $\,F_{(5)}$.\footnote{A dilaton flux $\,H_{(1)}\,$ and a RR flux $\,F_{(1)}$ are also permitted by the orientifold action but these cannot be generated in an ordinary SS reduction. This will also occur in the type IIB with O$5$ and O$3$ duality frames.} An explicit computation of the QC's in (\ref{QC_N8}) gives rise to the Jacobi identity in (\ref{Jacobi_id}) together with the first and second Bianchi identities in (\ref{Bianchi_id_10D}), with $\,p=3,5$, signaling the absence of NS$5$-branes, O$3$/D$3$ sources and O$5$/D$5$ sources in the compactification scheme. For the O$7$/D$7$-sources in the compactification, a flux-induced tadpole cannot arise from (\ref{Bianchi_id_10D}) in an ordinary Scherk--Schwarz reduction with $\,F_{(1)}=0$.

Unfortunately, the $\,\textrm{SL}(2) \times \textrm{SL}(5)\,$ covariance required by the O$7$-planes does not admit an $\textrm{SO}(3)$ symmetry embedded diagonally as required by the RSTU-models. Equivalently, a $2+5$ splitting of the internal one-form basis is not compatible with the $3+3+1$ splitting in (\ref{coord_splitting}). As a result, there are no RSTU-models in this duality frame.

\subsubsection{Type~IIB with O$5$-planes}

The type IIB with O$5$-planes duality frame was considered in Section~$3.5$ of \cite{Arboleya:2024vnp}. Introducing O$5$-planes is compatible with an $\,\textrm{SL}(3) \times \textrm{SL}(4)\,$ covariant description of the fluxes. The O$5$-plane orientifold action $\,\mathcal{O}_{\mathbb{Z}_{2}} = \Omega_{P} \, \sigma_{\textrm{O}5}\,$ allows for metric fluxes $\,\omega\,$ as well as gauge fluxes $\,H_{(3)}$, $\,F_{(3)}$, $\,F_{(5)}\,$ and $\,F_{(7)}$. An explicit computation of the QC's in (\ref{QC_N8}) gives rise to the Jacobi identity in (\ref{Jacobi_id}) together with the first and second Bianchi identities in (\ref{Bianchi_id_10D}), with $\,p=3,5$, signaling the absence of NS$5$-branes, O$3$/D$3$ sources and O$5$/D$5$ sources (different from the ones in Table~\ref{Table:Op_planes}) in the compactification scheme. For the O$5$/D$5$ sources in Table~\ref{Table:Op_planes}, there is a flux-induced tadpole of the form
\begin{equation}
\label{Tadpole_condition_O5/D5}
dF_{(3)} = J_{\textrm{O}5/\textrm{D}5} \ .
\end{equation}

The $\textrm{SO}(3)$-invariant sector underlying the RSTU-models in this duality frame was put forward in Section~$4.3$ of \cite{Arboleya:2024vnp}. Using the $\textrm{SO}(3)$-invariant tensors in (\ref{SO(3)_inv_tensors}), the set of $\textrm{SO}(3)$-invariant metric fluxes is
\begin{equation}
\label{metric_fluxes_SO(3)_O5}
\begin{array}{c}
\omega_{ab}{}^{k}=\omega_{1} \, \epsilon_{ab}{}^{k}
 \hspace{3mm} ,  \hspace{3mm} 
\omega_{7a}{}^{i}=\omega_{2} \, \delta_{a}{}^{i}
 \hspace{3mm} ,  \hspace{3mm} 
\omega_{i7}{}^{a}=\omega_{3} \, \delta_{i}{}^{a}  , \\[2mm]
\omega_{ai}{}^{7}=\omega_{4} \, \delta_{ai}
 \hspace{3mm} ,  \hspace{3mm} 
\omega_{aj}{}^{c}=-\omega_{5} \, \epsilon_{aj}{}^{c}
 \hspace{3mm} ,  \hspace{3mm} 
\omega_{ij}{}^{k}=\omega_{6} \, \epsilon_{ij}{}^{k}  ,
\end{array}
\end{equation}
whereas the $\textrm{SO}(3)$-invariant gauge fluxes are given by
\begin{equation}
\label{gauge_fluxes_SO(3)_O5}
\begin{array}{c}
H_{abc}= h_{31}  \, \epsilon_{abc}
\hspace{3mm} , \hspace{3mm} 
H_{aij}= h_{32}  \, \epsilon_{aij} 
\hspace{3mm} , \hspace{3mm} 
F_{ijk}= -f_{31}  \, \epsilon_{ijk}
\hspace{3mm} , \hspace{3mm} 
F_{ia7}= f_{32}  \, \delta_{ia} 
\hspace{3mm} , \hspace{3mm}
F_{ibc}= f_{33}  \, \epsilon_{ibc} 
\ , \\[2mm]
F_{abij7}= -f_{5}  \, \delta_{ai} \, \delta_{bj}
\hspace{3mm} , \hspace{3mm} 
F_{abcijk7}= f_{7}  \, \epsilon_{abc} \, \epsilon_{ijk} \ ,
\end{array}
\end{equation}
yielding a total of $\,13\,$ flux parameters. The $\textrm{SO}(3)$-invariant fluxes (\ref{metric_fluxes_SO(3)_O5}) and (\ref{gauge_fluxes_SO(3)_O5}) give rise to a scalar potential also constructed and extremised in \cite{Arboleya:2024vnp}. It contains the set of flux vacua listed in Table~\ref{Table:flux_vacua_IIB_O5}. For historical reasons, we have included there the trivial \textcolor{red}{\textbf{vac~1}}. This is simply the (half-maximal sector of the) KK reduction of type II supergravity on a fluxless seven-torus, which can of course be carried out in any duality frame and produces a Mkw$_{3}$ vacuum (the scalar potential identically vanishes).

\subsubsection*{Supersymmetry enhancements}

The gravitino masses associated with the type IIB with O5 flux vacua in Table~\ref{Table:flux_vacua_IIB_O5} are presented in Table~\ref{Table:flux_vacua_IIB_O5_gravitini_scalar_spectrum}. From there we observe that the Mkw$_{3}$ vacuum labelled as \textbf{vac~2} in Table~\ref{Table:flux_vacua_IIB_O5} has an enhancement to $\,\mathcal{N}=4\,$ supersymmetry whenever $\,\kappa \pm \xi=0$.

\begin{table}[t]
\begin{center}
\scalebox{0.82}{
\renewcommand{\arraystretch}{1.5}
\begin{tabular}{!{\vrule width 1.5pt}l!{\vrule width 1pt}c!{\vrule width 1pt}c!{\vrule width 1pt}cccccc!{\vrule width 1pt}cc!{\vrule width 1pt}ccc!{\vrule width 1pt}c!{\vrule width 1pt}c!{\vrule width 1.5pt}}
\Xcline{4-16}{1.5pt}
\multicolumn{3}{c!{\vrule width 1pt}}{}& \multicolumn{6}{c!{\vrule width 1pt}}{$\omega$} & \multicolumn{2}{c!{\vrule width 1pt}}{$H_{(3)}$} & \multicolumn{3}{c!{\vrule width 1pt}}{$F_{(3)}$} & \multicolumn{1}{c!{\vrule width 1pt}}{$F_{(5)}$} & \multicolumn{1}{c!{\vrule width 1pt}}{$F_{(7)}$}\\ 
\noalign{\hrule height 1.5pt}
     \hspace{5mm} ID & Type &  SUSY & $\omega_{1}$ & $ \omega_{2}$ & $  \omega_{3}$ & $\omega_{4}$ & $ \omega_{5}$ & $ \omega_{6}$ & $ h_{31}$ & $  h_{32}$ & $f_{31}$ & $f_{32}$ & $f_{33}$ & $ f_{5}$ & $  f_{7}$  \\ 
\noalign{\hrule height 1pt}
     $ \boxed{\color{red}{\textbf{vac~1}}}^{\,\,* \,\dagger} $ & \multirow{3}{*}{$\textrm{Mkw}_{3}$} & $\mathcal{N}=8$ &  $ 0 $ & $0$ & $  0 $ & $0$ & $ 0$ & $ 0$ & $ 0$ & $  0$ & $0$ & $ 0$ & $ 0$ & $ 0$ & $  0$  \\
     \cline{1-1}\cline{3-16}
     $ \boxed{\textbf{vac~2}}^{\,\,\dagger} $ &  & $\mathcal{N}=0,4$ & $ \kappa $ & $ \xi$ & $  0 $ & $ 0 $ & $ 0 $ & $ 0 $ & $ 0 $ & $  0 $ & $0$ & $ \kappa$ & $ -\xi$ & $ 0 $ & $  0 $  \\  
     \cline{1-1}\cline{3-16}
     $ \boxed{\textbf{vac~3}}^{\,\,*}$ &  & $\mathcal{N}=0$ & $ 0$ & $ \kappa $ & $  \kappa$ & $0$ & $ 0$ & $ 0$ & $ 0$ & $  0$ & $0$ & $ 0$ & $ 0$ & $ 0$ & $  0$  \\ 
\noalign{\hrule height 1pt}
     $ \textcolor{RoyalBlue}{\textbf{vac~4}}^{\,\,* \,\dagger} $ & \multirow{2}{*}{${\textrm{AdS}_{3}}$} & $\mathcal{N}=4$ & $ 0$ & $ 0$ & $  0$ & $0$ & $ 0$ & $ \kappa$ & $ 0$ & $  0$ & $\pm\kappa$ & $ 0$ & $ 0$ & $ 0$ & $  -\kappa$ 
     \\ 
     \cline{1-1}\cline{3-16}
     $ \textcolor{RoyalBlue}{\textbf{vac~5}}^{\,\,*  \,\dagger} $ &  & $\mathcal{N}=0$ & $ 0$ & $ 0$ & $  0$ & $0$ & $ 0$ & $ \kappa$ & $ 0$ & $  0$ & $\pm\kappa$ & $ 0$ & $ 0$ & $ 0$ & $  \kappa$ 
     \\
\noalign{\hrule height 1pt}
     $ \boxed{\textbf{vac~6}}^{\,\,*} $ & \multirow{2}{*}{$\textrm{AdS}_{3}$} & $\mathcal{N}=3$ & $ \kappa$ & $ 0$ & $  0$ & $0$ & $ -\kappa$ & $ \kappa$ & $ 0$ & $  0$ & $\pm\kappa $ & $ 0$ & $ \mp\kappa$ & $ 0$ & $  -2\kappa$  \\ 
     \cline{1-1}\cline{3-16}
     $ \boxed{\textbf{vac~7}}^{\,\,*} $ &  & $\mathcal{N}=1$ & $ \kappa$ & $ 0$ & $  0$ & $0$ & $ -\kappa$ & $ \kappa$ & $ 0$ & $  0$ & $\mp\kappa $ & $ 0$ & $ \pm\kappa$ & $ 0$ & $  2\kappa$  \\
\noalign{\hrule height 1pt}
     $ \textcolor{ForestGreen}{\textbf{vac~8}}^{\,\,* \,\dagger} $ & \multirow{2}{*}{${\textrm{AdS}_{3}}$} & $\mathcal{N}=1$ & $ 0$ & $ 0$ & $  0$ & $0$ & $ -\kappa$ & $ \kappa$ & $ 0$ & $  0$ & $\pm\kappa$ & $ 0$ & $ 0$ & $ 0$ & $  \kappa$  \\ 
     \cline{1-1}\cline{3-16}
     $ \textcolor{ForestGreen}{\textbf{vac~9}}^{\,\,* \,\dagger} $ &  & $\mathcal{N}=0$ & $ 0$ & $ 0$ & $  0$ & $0$ & $ -\kappa$ & $ \kappa$ & $ 0$ & $  0$ & $\pm\kappa$ & $ 0$ & $ 0$ & $ 0$ & $ - \kappa$  \\ 
\noalign{\hrule height 1pt}
     $ \boxed{\textbf{vac~10}}$ & $\textrm{AdS}_{3}$ & $\mathcal{N}=0$ & $ 0$ & $ 2\kappa$ & $  \kappa$ & $0$ & $ 0$ & $ 0$ & $ 0$ & $  0$ & $\kappa$ & $ \pm\kappa$ & $ -\kappa$ & $ 0$ & $  \pm\kappa$  \\
     \hline 
     $ \boxed{\textbf{vac~11}}$ & $\textrm{AdS}_{3}$ & $\mathcal{N}=0$ & $ 0$ & $ 2\kappa$ & $  \kappa$ & $0$ & $ 0$ & $ 0$ & $ 0$ & $  0$ & $\kappa$ & $ \pm\kappa$ & $ -\kappa$ & $ 0$ & $  \mp\kappa$  \\
\noalign{\hrule height 1pt} 
     $ \textcolor{Purple}{\textbf{vac~12}}^{\,\,* \,\dagger} $ & \multirow{4}{*}{${\textrm{AdS}_{3}}$} & $\mathcal{N}=4$ & $ 0$ & $ 0$ & $  \mp\kappa$ & $\mp\kappa$ & $ \kappa$ & $ -2\kappa$ & $ 0$ & $  0$ & $\mp 2\kappa$ & $ 0$ & $ 0$ & $ 0$ & $  2\kappa$  \\
      \cline{1-1}\cline{3-16}
     $ \textcolor{Purple}{\textbf{vac~13}}^{\,\,* \,\dagger} $ &  & $\mathcal{N}=1$ & $ 0$ & $ 0$ & $  \mp\kappa$ & $\mp\kappa$ & $ \kappa$ & $ -2\kappa$ & $ 0$ & $  0$ & $\mp 2\kappa$ & $ 0$ & $ 0$ & $ 0$ & $  -2\kappa$  \\
     \cline{1-1}\cline{3-16}
     $ \textcolor{Purple}{\textbf{vac~14}}^{\,\,* \,\dagger} $ &  & $\mathcal{N}=0$ & $ 0$ & $ 0$ & $  \pm\kappa$ & $\pm\kappa$ & $ \kappa$ & $ -2\kappa$ & $ 0$ & $  0$ & $\mp 2\kappa$ & $ 0$ & $ 0$ & $ 0$ & $  2\kappa$  \\
     \cline{1-1}\cline{3-16}
     $\textcolor{Purple}{\textbf{vac~15}}^{\,\,* \,\dagger} $ &  & $\mathcal{N}=0$ & $ 0$ & $ 0$ & $  \pm\kappa$ & $\pm\kappa$ & $ \kappa$ & $ -2\kappa$ & $ 0$ & $  0$ & $\mp 2\kappa$ & $ 0$ & $ 0$ & $ 0$ & $  -2\kappa$ \\
\noalign{\hrule height 1.5pt}
\end{tabular}}
\caption{Fluxes in the type~IIB with O$5$-planes duality frame producing a vacuum at the origin of moduli space.}
\label{Table:flux_vacua_IIB_O5}
\end{center}
\end{table}

\subsubsection*{Minimal $\,\mathcal{N}=1\,$ models without axions}

The action of the $\,\mathbb{Z}_2^{*}\,$ element in (\ref{Z2*_element}) now projects out all the $\,H_{(3)}\,$ and $\,F_{(5)}\,$ flux components in (\ref{gauge_fluxes_SO(3)_O5}) while retaining the rest of metric and gauge fluxes in (\ref{metric_fluxes_SO(3)_O5}) and (\ref{gauge_fluxes_SO(3)_O5}). In this restricted setup, and upon setting to zero the axions, the scalar potential can be derived from the real superpotential
\begin{equation}
\label{WO5}
\begin{array}{rcl}
g^{-1} \, W^{\textrm{O}5}_{\mathcal{N}=1} & = & \pm\frac38\sqrt{\frac{\mu}{A^3\,A_4\,\mu_4}}\,\omega_1+\frac{3}{8A^2}\,\omega_2+\frac{3\mu_4}{8A}\,\omega_3+\frac{3\mu}{8A_4}\,\omega_4\mp\frac38\sqrt{\frac{\mu\,\mu_4}{A\,A_4}}\,(2\omega_5+\omega_6) \\[4mm]
     &  & -\,\frac{\mu_4}{8A_4}\,f_{31}\pm\frac38\sqrt{\frac{\mu_4}{A^3\,A_4\,\mu}}\,f_{32}-\frac{3}{8A\,A_4}\,f_{33}\mp\frac18\sqrt{\frac{\mu^3\,\mu_4}{A^3\,A_4}}\,f_7 \ ,
\end{array}
\end{equation}
using again (\ref{VfromW}). Note that this time all the flux vacua in Table~\ref{Table:flux_vacua_IIB_O5} are compatible with $\,h_{31}=h_{32}=f_{5}=0\,$ and, therefore, are captured by the $\,\mathcal{N}=1\,$ minimal model in (\ref{WO5}).

As we did for the type IIB with O9 duality frame, we have obtained the superpotential (\ref{WO5}) from an explicit computation of the mass of the $\,\mathbb{Z}_{2}^{*}\times \textrm{SO}(3)$-invariant gravitino within half-maximal supergravity (see footnote~\ref{footnote_spin(8)} for the sign ambiguity in (\ref{WO5})). Again, this result agrees with the general expression for type IIB orientifold reductions on co-closed G$_2$-structure manifolds given in \cite{Emelin:2021gzx}, once the $\textrm{SO}(3)$ symmetry of the RSTU-models is imposed.

\subsubsection{Type~IIB with O$3$-planes}

The type IIB with O$3$-planes duality frame was considered in Section~$3.3$ of \cite{Arboleya:2024vnp}. Introducing O$3$-planes is compatible with an $\,\textrm{SL}(6)\,$ covariant description of the fluxes. The O$3$-plane orientifold action $\mathcal{O}_{\mathbb{Z}_{2}} = \Omega_{P} \, (-1)^{F_{L}} \,  \sigma_{\textrm{O}3}$ allows for metric fluxes $\,\omega\,$ as well as gauge fluxes $\,H_{(3)}$, $\,F_{(3)}\,$ and $\,F_{(5)}$. An explicit computation of the QC's in (\ref{QC_N8}) gives rise to the Jacobi identity in (\ref{Jacobi_id}) together with the first and second Bianchi identities in (\ref{Bianchi_id_10D}), with $\,p=3,5$, signaling the absence of NS$5$-branes, O$3$/D$3$ sources (different from the ones in Table~\ref{Table:Op_planes}) and O$5$/D$5$ sources in the compactification scheme. For the O$3$/D$3$ sources in Table~\ref{Table:Op_planes}, there is a flux-induced tadpole of the form
\begin{equation}
dF_{(5)} - H_{(3)} \wedge F_{(3)} = J_{\textrm{O}3/\textrm{D}3} \ .
\end{equation}

Using the $\textrm{SO}(3)$-invariant tensors in (\ref{SO(3)_inv_tensors}) one can construct the following set of $\textrm{SO}(3)$-invariant metric fluxes
\begin{equation}
\label{metric_fluxes_SO(3)_O3}
\begin{array}{c}
\omega_{7a}{}^{b}=\omega_{1} \, \delta_{a}{}^{b}
 \hspace{3mm} ,  \hspace{3mm} 
\omega_{7a}{}^{i}=\omega_{2} \, \delta_{a}{}^{i}
 \hspace{3mm} ,  \hspace{3mm} 
\omega_{7i}{}^{a}=\omega_{3} \, \delta_{i}{}^{a}
\hspace{3mm} ,  \hspace{3mm}
\omega_{7i}{}^{j}=\omega_{4} \, \delta_{i}{}^{j}  \hspace{3mm} ,  \hspace{3mm}
\omega_{ai}{}^{7}=\omega_{5} \, \delta_{ai}
\  ,
\end{array}
\end{equation}
as well as gauge fluxes
\begin{equation}
\label{gauge_fluxes_SO(3)_O3}
\begin{array}{c}
H_{abc}= h_{31}  \, \epsilon_{abc}
\hspace{2mm} , \hspace{2mm} 
H_{abi}= h_{32}  \, \epsilon_{abi}
\hspace{2mm} , \hspace{2mm} 
H_{aij}= h_{33}  \, \epsilon_{aij} 
\hspace{2mm} , \hspace{2mm} 
H_{ijk}= h_{34}  \, \epsilon_{ijk} 
\ , \\[2mm] 
F_{abc}= f_{31}  \, \epsilon_{abc}
\hspace{2mm} , \hspace{2mm} 
F_{abi}= f_{32}  \, \epsilon_{abi}
\hspace{2mm} , \hspace{2mm} 
F_{aij}= f_{33}  \, \epsilon_{aij} 
\hspace{2mm} , \hspace{2mm} 
F_{ijk}= f_{34}  \, \epsilon_{ijk} 
\hspace{2mm} , \hspace{2mm} 
F_{abij7}= f_{5}  \, \delta_{ai} \, \delta_{bj} \ ,
\end{array}
\end{equation}
yielding $\,14\,$ arbitrary flux parameters. The $\textrm{SO}(3)$-invariant fluxes (\ref{metric_fluxes_SO(3)_O3}) and (\ref{gauge_fluxes_SO(3)_O3}) produce a non-trivial scalar potential. A direct extremisation of the scalar potential gives the Mkw$_{3}$ vacuum in Table~\ref{Table:flux_vacua_IIB_O3}. In addition, another Mkw$_{3}$ vacuum appears which is T-dual to the type IIB with O9 \textbf{vac~1} in Table~\ref{Table:flux_vacua_IIB_O9} upon use of the correspondence $\,\omega_{2}^{\textrm{O}3} \leftrightarrow \omega_{7}^{\textrm{O}9}\,$ and $\,\omega_{3}^{\textrm{O}3} \leftrightarrow \omega_{5}^{\textrm{O}9}\,$ between fluxes in the two different duality frames.

\begin{table}[]
\begin{center}
\scalebox{0.76}{
\renewcommand{\arraystretch}{1.5}
\begin{tabular}{!{\vrule width 1.5pt}l!{\vrule width 1pt}c!{\vrule width 1pt}c!{\vrule width 1pt}ccccc!{\vrule width 1pt}cccc!{\vrule width 1pt}cccc!{\vrule width 1pt}c!{\vrule width 1pt}c!{\vrule width 1.5pt}}
\Xcline{4-17}{1.5pt}
\multicolumn{3}{c!{\vrule width 1pt}}{}& \multicolumn{5}{c!{\vrule width 1pt}}{$\omega$} & \multicolumn{4}{c!{\vrule width 1pt}}{$H_{(3)}$} & \multicolumn{4}{c!{\vrule width 1pt}}{$F_{(3)}$} & \multicolumn{1}{c!{\vrule width 1pt}}{$F_{(5)}$}\\ 
\noalign{\hrule height 1.5pt}
    \hspace{4mm} ID & Type &  SUSY & $\omega_{1}$ & $ \omega_{2}$ & $  \omega_{3}$ & $\omega_{4}$ & $ \omega_{5}$ & $ h_{31}$ & $  h_{32}$ & $  h_{33}$ & $  h_{34}$ & $f_{31}$ & $f_{32}$ & $f_{33}$ & $  f_{34}$ & $  f_{5}$  \\ 
\noalign{\hrule height 1pt}
     $ \boxed{\textbf{vac~1}}^{\,\,\dagger} $ & \multirow{1}{*}{$\textrm{Mkw}_{3}$} & $\mathcal{N}=0,1,2,3,4,6$ & $0$ & $0$ & $  0 $ & $ 0 $ & $ -\kappa_5 $ & $ -\kappa_4 $ & $  \kappa_3 $ & $ - \kappa_2$ & $ \kappa_1$ & $ \kappa_1$ & $ \kappa_2 $ & $ \kappa_3 $  & $ \kappa_4 $  &  $ \kappa_5 $\\ 
     \cline{1-1}\cline{3-17} 
     \noalign{\hrule height 1.5pt}
\end{tabular}}
\caption{Fluxes in the type~IIB with O$3$-planes duality frame producing a vacuum at the origin of moduli space.}
\label{Table:flux_vacua_IIB_O3}
\end{center}
\end{table}

\subsubsection*{Supersymmetry enhancements}

The gravitino masses associated with the Mkw$_{3}$ vacuum in Table~\ref{Table:flux_vacua_IIB_O3} are presented in Table~\ref{Table:flux_vacua_IIB_O3_gravitini_scalars_spectrum}. They are given by
\begin{equation}
\label{susy_enhancement_O3}
\begin{array}{rll}
\textrm{$(3+3)$ gravitini :} & \hspace{5mm} \sqrt{\left(\kappa_{1}+\kappa_{3}\right)^2 + \left(\kappa_{4}+\kappa_{2}\right)^2} \pm \kappa_5 & , \\[2mm]
\textrm{$(1+1)$ gravitini :}  & \hspace{5mm} \sqrt{\left(\kappa_{1}-3\kappa_{3}\right)^2 + \left(\kappa_{4}-3\kappa_{2}\right)^2} \pm 3\kappa_5 & ,
\end{array}
\end{equation}
where the flux parameter $\,\kappa_{5}$, which determines both the algebra spanned by the isometry generators of the internal space
\begin{equation}
\label{algebra_IIB_O3}
[X_{i},X_{a}] \,=\, \kappa_{5} \, \delta_{ia} \, X_{7}
\end{equation}
and the gauge flux $\,F_{(5)}$, plays a distinguished role. The isometry algebra (\ref{algebra_IIB_O3}) can be viewed as three copies of the three-dimensional Heisenberg algebra $\,\mathfrak{h}_{3}$, all of them sharing a common generator $X_{7}$ (specifying the center). It is a $2$-step nilpotent algebra, and the corresponding seven-dimensional nilmanifold, which is compatible with a co-closed G$_2$-structure, was named ``$17$'' in Table~$3$ of \cite{VanHemelryck:2025qok} (see references therein for the original mathematical literature).

A quick inspection of (\ref{susy_enhancement_O3}) shows that a generic solution is non-supersymmetric. However, supersymmetry enhancements occur when the flux parameters $\,\kappa$'s are appropriately adjusted. In particular, $\,\mathcal{N}=1,2,3,4\,$ and $\,6\,$ supersymmetry can be restored.

\subsection{Type IIA orientifold flux vacua}

In this section we will show that the RSTU-models arising from type IIA orientifold reductions give rise to flux vacua which are \textit{all} T-dual to flux vacua already contained in the type IIB landscape. For the cases with O$6$- and O$2$-planes we present the minimal $\,\mathcal{N}=1\,$ supergravity models with no axions obtained upon modding out the corresponding RSTU-models by the $\,\mathbb{Z}_{2}^{*}\,$ symmetry in (\ref{Z2*_element}).

\subsubsection{Type~IIA with O$8$-planes}

The type IIA with O$8$-planes duality frame was considered in Section~$3.8$ of \cite{Arboleya:2024vnp}. Introducing O$8$-planes is compatible with an $\,\textrm{SL}(6)\,$ covariant description of the fluxes (see Table~\ref{Table:Op_planes}). The O$8$-plane orientifold action ${\mathcal{O}_{\mathbb{Z}_{2}} = \Omega_{P} \, (-1)^{F_{L}} \, \sigma_{\textrm{O}8}}$ allows for metric fluxes $\omega$ as well as gauge fluxes $\,H_{(3)}$, $\,F_{(2)}$, $\,F_{(4)}\,$ and $\,F_{(6)}$.\footnote{We are not considering a dilaton flux $\,H_{(1)}\,$, which is allowed by the orientifold action: this flux is not allowed in an ordinary SS reduction. This will also occur in the type IIA with O$6$ and O$4$ duality frames.} An explicit computation of the QC's in (\ref{QC_N8}) gives rise to the Jacobi identity in (\ref{Jacobi_id}) together with the first and second Bianchi identities in (\ref{Bianchi_id_10D}), with $\,p=2,4,6$, signaling the absence of NS$5$-branes, O$2$/D$2$ sources, O$4$/D$4$ sources and O$6$/D$6$ sources in the compactification scheme. As it happened in the type~IIB with O$9$ duality frame, the embedding tensor configurations associated with the fluxes in the type IIA with O$8$ duality frame specify a consistent maximal gauged supergravity. This implies that no flux-induced tadpole is permitted (actually possible) for the O$8$/D$8$-sources, namely
\begin{equation}
0=J_{\textrm{O}8/\textrm{D}8} \ .
\end{equation}
The ASD condition in (\ref{DFT_constraint}) is also satisfied, so a realisation of these models as a generalised Scherk--Schwarz reduction of O$(8,8)$-DFT is also possible.

The $\textrm{SO}(3)$-invariant tensors in (\ref{SO(3)_inv_tensors}) determine the following set of $\textrm{SO}(3)$-invariant metric fluxes
\begin{equation}
\label{metric_fluxes_SO(3)_O8}
\begin{array}{c}
\omega_{ab}{}^{c}=\omega_{1} \, \epsilon_{ab}{}^{c}
 \hspace{3mm} ,  \hspace{3mm} 
\omega_{ab}{}^{i}=\omega_{2} \, \epsilon_{ab}{}^{i}
 \hspace{3mm} ,  \hspace{3mm} 
\omega_{ia}{}^{b}=\omega_{3} \, \epsilon_{ia}{}^{b} \ ,\\[2mm]
\omega_{ai}{}^{j}=\omega_{4} \, \epsilon_{ai}{}^{j}  \hspace{3mm} ,  \hspace{3mm}
\omega_{ij}{}^{a}=\omega_{5} \, \epsilon_{ij}{}^{a} \hspace{3mm} , \hspace{3mm}
\omega_{ij}{}^{k}=\omega_{6} \, \epsilon_{ij}{}^{k} \ ,
\end{array}
\end{equation}
as well as gauge fluxes
\begin{equation}
\label{gauge_fluxes_SO(3)_O8}
\begin{array}{c}
H_{ia7}= h_{3}  \, \delta_{ia}
\hspace{3mm} , \hspace{3mm} 
F_{ia}= f_{2}  \, \delta_{ia}
\hspace{3mm} , \hspace{3mm} F_{abc7}= f_{41}  \, \epsilon_{abc} 
\hspace{3mm} , \hspace{3mm} 
F_{abi7}= f_{42}  \, \epsilon_{abi} 
\ , \\[2mm] 
F_{aij7}= f_{43}  \, \epsilon_{aij}
\hspace{3mm} , \hspace{3mm} 
F_{ijk7}= f_{44}  \, \epsilon_{ijk}
\hspace{3mm} , \hspace{3mm}
F_{ijkabc}= f_{6}  \, \epsilon_{ijk} \, \epsilon_{abc} 
\ ,
\end{array}
\end{equation}
giving $\,13\,$ arbitrary flux parameters. The $\textrm{SO}(3)$-invariant fluxes (\ref{metric_fluxes_SO(3)_O8}) and (\ref{gauge_fluxes_SO(3)_O8}) produce a non-trivial scalar potential whose extremisation gives a set of flux vacua which are \textit{all} T-dual to flux vacua in the type IIB with O$9$ duality frame. In particular, they are T-dual to the Mkw$_{3}$ \textbf{vac~1} and the AdS$_{3}$ \textbf{vac~6}--\textcolor{ForestGreen}{\textbf{vac~49}} in Table~\ref{Table:flux_vacua_IIB_O9} upon use of the following mapping between fluxes in the two different duality frames
\begin{equation}
    \begin{array}{c}
         \omega_1^{\textrm{O$8$}} \longleftrightarrow \omega_{1}^{\textrm{O$9$}} \hspace{5mm} , \hspace{5mm}  \omega_2^{\textrm{O$8$}} \longleftrightarrow \omega_{3}^{\textrm{O$9$}} \hspace{5mm} , \hspace{5mm} \omega_3^{\textrm{O$8$}} \longleftrightarrow \omega_{4}^{\textrm{O$9$}} \ , \\[2mm] \omega_4^{\textrm{O$8$}} \longleftrightarrow \omega_{9}^{\textrm{O$9$}} \hspace{5mm} , \hspace{5mm} \omega_5^{\textrm{O$8$}} \longleftrightarrow \omega_{8}^{\textrm{O$9$}} \hspace{5mm} , \hspace{5mm} \omega_6^{\textrm{O$8$}} \longleftrightarrow \omega_{11}^{\textrm{O$9$}} \ ,\\[2mm]
         h_3^{\textrm{O$8$}} \longleftrightarrow \omega_{6}^{\textrm{O$9$}} \hspace{5mm} , \hspace{5mm}
         f_2^{\textrm{O$8$}} \longleftrightarrow f_{35}^{\textrm{O$9$}} \ ,\\[2mm]
         f_{41}^{\textrm{O$8$}} \longleftrightarrow f_{31}^{\textrm{O$9$}} \hspace{5mm} , \hspace{5mm} f_{42}^{\textrm{O$8$}} \longleftrightarrow f_{32}^{\textrm{O$9$}} \hspace{5mm} , \hspace{5mm} f_{43}^{\textrm{O$8$}} \longleftrightarrow f_{33}^{\textrm{O$9$}} \hspace{5mm} , \hspace{5mm} f_{44}^{\textrm{O$8$}} \longleftrightarrow f_{34}^{\textrm{O$9$}} \ , \\[2mm]
         f_6^{\textrm{O$8$}} \longleftrightarrow f_{7}^{\textrm{O$9$}} \ .
    \end{array}
\end{equation}
Summarising, no new flux vacua appear in the type IIA with O$8$ duality frame.

\subsubsection{Type~IIA with O$6$-planes}

The type IIA with O$6$-planes duality frame was considered in Section~$3.6$ of \cite{Arboleya:2024vnp}. Introducing O$6$-planes is compatible with an $\,\textrm{SL}(4) \times \textrm{SL}(3)\,$ covariant description of the fluxes (see Table~\ref{Table:Op_planes}). The O$6$-plane orientifold action $\,{\mathcal{O}_{\mathbb{Z}_{2}} = \Omega_{P} \, \sigma_{\textrm{O}6}}\,$ allows for metric fluxes $\,\omega\,$ as well as gauge fluxes $\,H_{(3)}$, $\,F_{(2)}$, $\,F_{(4)}$, $\,F_{(6)}\,$ and a constant Romans mass parameter $\,F_{(0)}=f_{0}$. The computation of the QC's in (\ref{QC_N8}) gives rise to the Jacobi identity in (\ref{Jacobi_id}) together with the first and second Bianchi identities in (\ref{Bianchi_id_10D}), with $\,p=2,4,6$, signaling the absence of NS$5$-branes, O$2$/D$2$ sources, O$4$/D$4$ sources and O$6$/D$6$ sources (different from the ones in Table~\ref{Table:Op_planes}) in the compactification scheme. For the O$6$/D$6$ sources in Table~\ref{Table:Op_planes}, a tadpole cancellation condition of the form
\begin{equation}
dF_{(2)}-H_{(3)} \wedge F_{(0)} = J_{\textrm{O}6/\textrm{D}6} \ ,   
\end{equation}
arises from (\ref{Bianchi_id_10D}).

Using the $\textrm{SO}(3)$-invariant tensors in (\ref{SO(3)_inv_tensors}), the $\textrm{SO}(3)$-invariant metric and gauge fluxes are found to be
\begin{equation}
\label{metric_fluxes_SO(3)_O6}
\begin{array}{c}
\omega_{ab}{}^{c}=\omega_{1} \, \epsilon_{ab}{}^{c}
 \hspace{2mm} ,  \hspace{2mm} 
\omega_{7a}{}^{b}=\omega_{2} \, {\delta_{a}{}^{b}}
 \hspace{2mm} ,  \hspace{2mm}
\omega_{ai}{}^{j}=\omega_{3} \, \epsilon_{ai}{}^{j}
 \hspace{2mm} ,  \hspace{2mm}
\omega_{7i}{}^{j}=\omega_{4} \, {\delta_{i}{}^{j}}  
 \hspace{2mm} ,  \hspace{2mm}
\omega_{ij}{}^{a}=\omega_{5} \, \epsilon_{ij}{}^{a}
\  ,
\end{array}
\end{equation}
and
\begin{equation}
\label{gauge_fluxes_SO(3)_O6}
\begin{array}{c}
H_{iab}= h_{31}  \, \epsilon_{iab}
\hspace{3mm} , \hspace{3mm} 
H_{ia7}= h_{32}  \, \delta_{ia}
\hspace{3mm} , \hspace{3mm}
H_{ijk}= h_{33}  \, \epsilon_{ijk} 
\hspace{3mm} , \hspace{3mm}
f_{0}
\hspace{3mm} , \hspace{3mm}
F_{ia}= f_{2}  \, \delta_{ia} 
\ , \\[2mm] 
F_{abc7}= f_{41}  \, \epsilon_{abc}
\hspace{3mm} , \hspace{3mm}
F_{abij}= f_{42}  \, \delta_{ai} \delta_{bj}
\hspace{3mm} , \hspace{3mm}
F_{ija7}= f_{43}  \, \epsilon_{ija} 
\hspace{3mm} , \hspace{3mm}
F_{ijkabc}= f_{6}  \,   \epsilon_{ijk} \, \epsilon_{abc}
\ ,
\end{array}
\end{equation}
yielding a total of $\,14\,$ flux parameters. The $\textrm{SO}(3)$-invariant fluxes (\ref{metric_fluxes_SO(3)_O6}) and (\ref{gauge_fluxes_SO(3)_O6}) produce a non-trivial scalar potential whose direct extremisation yields four AdS$_3$ flux vacua with $\,f_{0}=0$. These are T-dual to vacua in the type IIB with O5-duality frame, in particular to \textcolor{RoyalBlue}{\textbf{vac~4,5}} and \textcolor{ForestGreen}{\textbf{vac~8,9}} of Table~\ref{Table:flux_vacua_IIB_O5}, upon use of the following mapping between fluxes in the two different duality frames 
\begin{equation}
\label{dictionary:O5_O6}
\begin{array}{c}
\omega_1^{\textrm{O$6$}} \longleftrightarrow \omega_{6}^{\textrm{O$5$}} 
\hspace{5mm} , \hspace{5mm} 
\omega_3^{\textrm{O$6$}} \longleftrightarrow \omega_{5}^{\textrm{O$5$}} 
\hspace{5mm} , \hspace{5mm} 
\omega_5^{\textrm{O$6$}} \longleftrightarrow \omega_{1}^{\textrm{O$5$}} \ , \\[2mm]
h_{31}^{\textrm{O$6$}} \longleftrightarrow h_{32}^{\textrm{O$5$}} 
\hspace{5mm} , \hspace{5mm} 
{h_{32}^{\textrm{O$6$}} \longleftrightarrow \omega_{4}^{\textrm{O$5$}} }
\hspace{5mm} , \hspace{5mm}  
h_{33}^{\textrm{O$6$}} \longleftrightarrow h_{31}^{\textrm{O$5$}} \ , \\[2mm]
f_{2}^{\textrm{O$6$}} \longleftrightarrow f_{32}^{\textrm{O$5$}}
\hspace{3mm} , \hspace{3mm}
f_{41}^{\textrm{O$6$}} \longleftrightarrow f_{31}^{\textrm{O$5$}} 
\hspace{3mm} , \hspace{3mm}
f_{42}^{\textrm{O$6$}} \longleftrightarrow f_{5}^{\textrm{O$5$}} 
\hspace{3mm} , \hspace{3mm}
f_{43}^{\textrm{O$6$}} \longleftrightarrow f_{33}^{\textrm{O$5$}} 
\hspace{3mm} , \hspace{3mm}
f_{6}^{\textrm{O$6$}} \longleftrightarrow f_{7}^{\textrm{O$5$}} \ .
    \end{array}
\end{equation}
In addition to the four AdS$_{3}$ flux vacua, there is also a non-supersymmetric Mkw$_{3}$ vacuum. This vacuum requires a non-vanishing Romans mass parameter, $\,f_{0}\neq0$, whose T-dual in the type IIB with O5 duality frame is a one-form flux $\,F_{(1)}^{\textrm{O}5}$ along the seventh internal direction, which is not allowed in a standard Scherk--Schwarz reduction. This forces us to search for the T-dual of the Mkw$_{3}$ vacuum in some other type IIB duality frame. The winner is the type IIB with O$3$ duality frame. The Mkw$_{3}$ vacuum we are after appears as the T-dual of \textbf{vac~1} in Table~\ref{Table:flux_vacua_IIB_O3} for the particular choice $\,\kappa_{3}=\kappa_{4}=\kappa_{5}=0$. The corresponding fluxes are mapped between the two different duality frames as
\begin{equation}
\label{dictionary:O3_O6}
\omega_{5}^{\textrm{O}6} \leftrightarrow h_{33}^{\textrm{O}3}
\hspace{8mm} , \hspace{8mm}
h_{33}^{\textrm{O}6} \leftrightarrow h_{34}^{\textrm{O}3}
\hspace{8mm} , \hspace{8mm}
f_{0}^{\textrm{O}6} \leftrightarrow f_{31}^{\textrm{O}3}
\hspace{8mm} , \hspace{8mm}
f_{2}^{\textrm{O}6} \leftrightarrow f_{32}^{\textrm{O}3} \ .
\end{equation}
Therefore, all the flux vacua in the type IIA with O$6$ duality frame possess a standard Scherk--Schwarz type IIB dual either with O$5$- or O$3$-planes.

\subsubsection*{Minimal $\,\mathcal{N}=1\,$ models without axions}

The $\,\mathbb{Z}_{2}^{*}\,$ symmetry in (\ref{Z2*_element}) projects out all the metric fluxes in (\ref{metric_fluxes_SO(3)_O6}) and retains the flux parameters
\begin{equation}
h_{31} \,\, , \,\,  
h_{32} \,\, , \,\,  
h_{33} 
\,\,\,\,\,\,\, , \,\,\,\,\,\,\,  
f_{41} \,\, , \,\,  
f_{42} \,\, , \,\,  
f_{43} \,\,\,\,\,\,\, \textrm{ and } \,\,\,\,\,\,\,  
f_{0} \ ,
\end{equation}
in (\ref{gauge_fluxes_SO(3)_O6}). When restricted to the above set of fluxes, the scalar potential (with vanishing axions) can be derived from the real superpotential
\begin{equation}
\label{WO6}
\begin{array}{rcl}
g^{-1} \, W^{\textrm{O}6}_{\mathcal{N}=1} & = & \pm\frac38 \sqrt{\frac{\mu_4}{A \, A_4 \, \mu^3}} h_{31} -\frac{3}{8 A_4 \, \mu} h_{32} \pm \frac18 \sqrt{\frac{1}{A^3 \, A_4 \, \mu^3 \, \mu_4}} h_{33} \\[4mm] &  & - \, \frac{\mu_4}{8 A_4} f_{41} \mp \frac38 \sqrt{\frac{\mu_4}{A_4 \, A^3 \, \mu}} f_{42} - \frac{3}{8 A \, A_4} f_{43} \pm \frac{1}{8} \sqrt{\frac{\mu^3 \, \mu_4}{A^3 \, A_4}} f_0 \ ,
\end{array}
\end{equation}
using the formula (\ref{VfromW}). By looking at the T-duality rules in (\ref{dictionary:O5_O6}), one concludes that the minimal $\,\mathcal{N}=1\,$ model specified by (\ref{WO6}) does not capture the AdS$_{3}$ vacua labelled \textcolor{RoyalBlue}{\textbf{vac~4,5}} and \textcolor{ForestGreen}{\textbf{vac~8,9}} in Table~\ref{Table:flux_vacua_IIB_O5}. However, it captures the Mkw$_{3}$ vacuum labelled \textbf{vac~1} in Table~\ref{Table:flux_vacua_IIB_O3} only when both conditions $\,\kappa_{3}=\kappa_{4}=\kappa_{5}=0\,$ (as explained before) and $\,\kappa_{2}=0\,$ are satisfied.

The superpotential (\ref{WO6}) corresponds to the mass of the $\,\mathbb{Z}_{2}^{*}\times \textrm{SO}(3)$-invariant gravitino within half-maximal supergravity (see footnote~\ref{footnote_spin(8)} for the $\,\pm\,$ sign in (\ref{WO6})). This is precisely the way we have obtained it. Nonetheless, we have verified that it matches the general expression for type IIA orientifold reductions on seven-dimensional manifolds with G$_{2}$-holonomy (including the Romans mass parameter) put forward in \cite{Farakos:2020phe} once the additional $\,\textrm{SO}(3)\,$ symmetry of the RSTU-models is imposed.

\subsubsection{Type~IIA with O$4$-planes}

The type IIA with O$4$-planes duality frame was considered in Section~$3.4$ of \cite{Arboleya:2024vnp}. Introducing O$4$-planes is compatible with an $\,\textrm{SL}(5) \times \textrm{SL}(2)\,$ covariant description of the fluxes. The O$4$-plane orientifold action $\,\mathcal{O}_{\mathbb{Z}_{2}} = \Omega_{P} \, (-1)^{F_{L}} \, \sigma_{\textrm{O}4}\,$ allows for metric fluxes $\,\omega\,$ as well as gauge fluxes $\,H_{(3)}$, $\,F_{(2)}$, $\,F_{(4)}\,$ and $\,F_{(6)}$. The QC's in (\ref{QC_N8}) gives rise to the Jacobi identity in (\ref{Jacobi_id}) together with the first and second Bianchi identities in (\ref{Bianchi_id_10D}), with $\,p=2,4,6$, signaling the absence of NS$5$-branes, O$2$/D$2$ sources, O$4$/D$4$ sources (different from the ones halving maximal supergravity) and O$6$/D$6$ sources in the compactification scheme. For the O$4$/D$4$ sources in Table~\ref{Table:Op_planes}, (\ref{Bianchi_id_10D}) gives a tadpole cancellation condition of the form
\begin{equation}
dF_{(4)} - H_{(3)} \wedge F_{(2)} = J_{\textrm{O}4/\textrm{D}4} \ .
\end{equation}

Unfortunately, the $\,\textrm{SL}(5) \times \textrm{SL}(2)\,$ covariance required by the O$4$-planes does not admit an $\textrm{SO}(3)$ symmetry embedded diagonally as required by the RSTU-models. Equivalently, a $5+2$ splitting of the internal one-form basis is not compatible with the $3+3+1$ splitting in (\ref{coord_splitting}). As a result, there are no RSTU-models in this duality frame.

\subsubsection{Type~IIA with O$2$-planes}

The type IIA with O$2$-planes duality frame was considered in Section~$3.2$ of \cite{Arboleya:2024vnp}. Introducing O$2$-planes is compatible with an $\,\textrm{SL}(7)\,$ covariant description of the fluxes. The O$2$-plane orientifold action $\,\mathcal{O}_{\mathbb{Z}_{2}} = \Omega_{P}  \, \sigma_{\textrm{O}2}\,$ only allows for gauge fluxes $\,H_{(3)}$, $\,F_{(4)}\,$ and a constant Romans mass parameter $\,F_{(0)}=f_{0}$. An explicit computation of the QC's in (\ref{QC_N8}) gives rise to the first and second Bianchi identities in (\ref{Bianchi_id_10D}), with $\,p=4$, signaling the absence of NS$5$-branes and O$4$/D$4$ sources in the compactification scheme. For the O$2$/D$2$ sources in Table~\ref{Table:Op_planes}, a tadpole cancellation condition arises of the form
\begin{equation}
- H_{(3)} \wedge F_{(4)} = J_{\textrm{O}2/\textrm{D}2} \ .
\end{equation}

Using the $\textrm{SO}(3)$-invariant tensors in (\ref{SO(3)_inv_tensors}) one can construct the following set of $\textrm{SO}(3)$-invariant gauge fluxes 
\begin{equation}
\label{gauge_fluxes_SO(3)_O2}
\begin{array}{c}
H_{abc}= h_{31}  \, \epsilon_{abc}
\hspace{2mm} , \hspace{2mm} 
H_{abi}= h_{32}  \, \epsilon_{abi}
\hspace{2mm} , \hspace{2mm} 
H_{aij}= h_{33}  \, \epsilon_{aij} 
\hspace{2mm} , \hspace{2mm} 
H_{ijk}= h_{34}  \, \epsilon_{ijk} 
\hspace{2mm} , \hspace{2mm} 
H_{ai7}= h_{35}  \, \delta_{ai} 
\ , \\[2mm]
F_{abc7}= f_{41}  \, \epsilon_{abc}
\hspace{2mm} , \hspace{2mm} 
F_{abi7}= f_{42}  \, \epsilon_{abi}
\hspace{2mm} , \hspace{2mm} 
F_{aij7}= f_{43}  \, \epsilon_{aij} 
\hspace{2mm} , \hspace{2mm} 
F_{ijk7}= f_{44}  \, \epsilon_{ijk} 
\hspace{2mm} , \hspace{2mm} 
F_{abij}= f_{45} \, \delta_{ai} \, \delta_{bj} \ ,
\end{array}
\end{equation}
together with the Romans mass $F_{(0)} = f_0$. This yields a total of $\,11\,$ arbitrary flux parameters. The $\textrm{SO}(3)$-invariant fluxes (\ref{gauge_fluxes_SO(3)_O2}) produce a non-trivial scalar potential which possesses a multi-parametric Mkw$_{3}$ vacuum with $\,f_{0}=0\,$ which is T-dual to the general type IIB with O$3$ \textbf{vac~1} of Table~\ref{Table:flux_vacua_IIB_O3} upon using the flux mapping
\begin{equation}
\label{T-duality_O2/O3}
\begin{array}{c}
h_{31}^{\textrm{O$2$}} \longleftrightarrow h_{31}^{\textrm{O$3$}} 
\hspace{3mm} , \hspace{3mm} 
h_{32}^{\textrm{O$2$}} \longleftrightarrow h_{32}^{\textrm{O$3$}} 
\hspace{3mm} , \hspace{3mm} h_{33}^{\textrm{O$2$}} \longleftrightarrow h_{33}^{\textrm{O$3$}} 
\hspace{3mm} , \hspace{3mm} 
h_{34}^{\textrm{O$2$}} \longleftrightarrow h_{34}^{\textrm{O$3$}} 
\hspace{3mm} , \hspace{3mm}  
h_{35}^{\textrm{O$2$}} \longleftrightarrow \omega_{5}^{\textrm{O$3$}} \ , \\[2mm]
f_{41}^{\textrm{O$2$}} \longleftrightarrow f_{31}^{\textrm{O$3$}} 
\hspace{3mm} , \hspace{3mm} 
f_{42}^{\textrm{O$2$}} \longleftrightarrow f_{32}^{\textrm{O$3$}} 
\hspace{3mm} , \hspace{3mm} f_{43}^{\textrm{O$2$}} \longleftrightarrow f_{33}^{\textrm{O$3$}} 
\hspace{3mm} , \hspace{3mm} 
f_{44}^{\textrm{O$2$}} \longleftrightarrow f_{34}^{\textrm{O$3$}}
\hspace{3mm} , \hspace{3mm} 
f_{45}^{\textrm{O$2$}} \longleftrightarrow f_{5}^{\textrm{O$3$}} \ .
\end{array}
\end{equation}
Note that $\,f_{0}=0\,$ is consistent with $\,f_{0}\,$ being T-dual to the seventh component of a RR one-form flux $\,F^{\textrm{O3}}_{(1)}$, which must always vanish in an ordinary Scherk--Schwarz reduction. This multi-parametric Mkw$_{3}$ flux vacuum in type IIA with O$2$-planes was investigated in \cite{Arboleya:2024vnp}. The rich variety of possible supersymmetry-breaking patterns was also observed there.

\subsubsection*{Minimal $\,\mathcal{N}=1\,$ models without axions}

Amongst the fluxes in (\ref{gauge_fluxes_SO(3)_O2}), only 
\begin{equation}
h_{32} \,\, , \,\, h_{34} \,\, , \,\,  h_{35} 
\,\,\,\,\,\, \textrm{ and } \,\,\,\,\,\, 
f_{41} \,\, , \,\, f_{43} \,\, , \,\,  f_{45} \ ,
\end{equation}
are compatible with the $\,\mathbb{Z}_{2}^{*}\,$ symmetry in (\ref{Z2*_element}). In addition to them, there is also the Romans mass parameter $\,f_{0}$. When only these fluxes are activated and the axions are set to zero, the scalar potential can be obtained from the real superpotential
\begin{equation}
\label{WO2}
\begin{array}{rcl}
g^{-1} \, W^{\textrm{O}2}_{\mathcal{N}=1} & = & \pm\frac38 \sqrt{\frac{\mu}{A^3 \, A_4 \, \mu_4}} h_{32} \mp \frac18 \sqrt{\frac{1}{A^3 \, \mu^3 \, A_4 \, \mu_4}} h_{34} + \frac{3}{8 A^2} h_{35} \\[4mm]  & & \mp \, \frac18 \sqrt{\frac{\mu^3 \, \mu_4}{A^3 \, A_4}} f_{41}  \pm \frac38 \sqrt{\frac{\mu_4}{A^3 \, A_4 \, \mu}} f_{43} +\frac{3}{8 A \, A_4} f_{45 } - \frac{\mu_4}{8 A_4} f_0 \ ,
\end{array}
\end{equation}
accordingly to the formula (\ref{VfromW}). By inspection of the T-duality rules in (\ref{T-duality_O2/O3}), one sees that the minimal $\,\mathcal{N}=1\,$ model specified by (\ref{WO2}) is able to capture the T-dual version of the Mkw$_{3}$ \textbf{vac~1} of Table~\ref{Table:flux_vacua_IIB_O3} only for the specific choice of parameters $\,\kappa_{2}=\kappa_{4}=0$.
Finally, as it happened for its type IIA with O$6$ counterpart, the real superpotential in (\ref{WO2}) arises as the mass of the $\,\mathbb{Z}_{2}^{*}\times \textrm{SO}(3)$-invariant gravitino (see footnote~\ref{footnote_spin(8)} for the $\,\pm\,$ sign in (\ref{WO2})) within half-maximal supergravity. It also matches the general expression put forward in \cite{Farakos:2020phe} for type II orientifold reductions on seven-dimensional manifolds with G$_{2}$-holonomy, once the $\,\textrm{SO}(3)\,$ symmetry of the RSTU-models is imposed.

\section{Outlook: on moduli stabilisation and scale separation}
\label{sec:discussion}

Type II flux compactifications down to three dimensions in the presence of various types of (intersecting) O$p$-planes, like type IIA with O$2$/O$6$ \cite{Farakos:2020phe} or type IIB with O$5$/O$9$ \cite{Emelin:2021gzx}, have been the focus of recent studies. Having various types of O$p$-planes (possibly also D$p$-branes) in the compactification helps to generate scalar potentials with a richer structure of critical points, ultimately even allowing for the stabilisation of all the dilaton-like moduli fields arising from the closed-string sector of the compactification. Although the stabilisation of the closed-string dilatons is essential for the volume of the internal space and the string coupling constant to be well-defined, a complete characterisation of the landscape of flux vacua is typically not possible when various types of O$p$-planes are present. The reason is that, upon dimensional reduction, the resulting supergravity models have minimal or no supersymmetry, and this leads to scalar potentials with highly intricate structures, rendering their extremisation very difficult and often forcing conclusions to be based on sporadic searches.

In this work, making a combined use of supergravity and algebraic geometry techniques, we have charted the landscape of flux vacua (with vanishing axions) in the RSTU-models that arise from type IIA and type IIB orientifold reductions on twisted tori with a \textit{single type} of spacetime-filling O$p$-planes. The landscape consists of $\,1 \, \textrm{(trivial, \textcolor{red}{$\blacktriangle$})}+56\,$ inequivalent families of both supersymmetric and non-supersymmetric AdS$_{3}$ and Mkw$_{3}$ flux vacua, all of which can be realised, upon implementation of T-dualities, as type IIB orientifold reductions with either O$9$-, O$5$-, or O$3$-planes. In particular, we found the following independent vacua:
\begin{equation*}
\label{independent_solutions}
\begin{array}{lcl}
\textrm{Type~IIB with O$9$} & : & 
\textrm{all vacua of Table~\ref{Table:flux_vacua_IIB_O9}}  \ , \\[2mm]
\textrm{Type~IIB with O$5$} & : &  \textrm{\textrm{\textcolor{red}{\textbf{vac~1}}, \textbf{vac~2,3}, \textbf{vac~6,7} and \textbf{vac~10,11}} of Table~\ref{Table:flux_vacua_IIB_O5}}\ , \\[2mm]
\textrm{Type~IIB with O$3$} & : & \textrm{\textbf{vac~1} of Table~\ref{Table:flux_vacua_IIB_O3}} \ .
\end{array}
\end{equation*}
A Venn diagram is presented in Figure~\ref{picture_vacua}. The fact that the closed-string dilatons can be stabilised and yield a scale-separated and weakly-coupled AdS$_{3}$ vacuum in a type II orientifold reduction with a \textit{single type} of O$5$-plane came as a surprise in \cite{Arboleya:2024vnp}. Here, by fully charting the landscape of flux vacua in all the possible RSTU-models with a single type of O$p$-plane, with $\,p=2,\ldots,9$, we are showing that this is indeed a very rare phenomenon. In what follows we discuss the main properties and salient features of the vacua we have found. We will discuss each of the relevant type IIB duality frames separately.

\subsection{Type IIB with O$9$-planes}

As stated in Section~\ref{sec:IIB_O9}, the type IIB with O$9$ flux models are embeddable into maximal supergravity. This is so because a flux-induced tadpole is not possible for the O$9$/D$9$ sources, namely
\begin{equation}
\label{J_O9/D9=0}
0 = J_{\textrm{O}9/\textrm{D}9} \ ,
\end{equation}
in the Bianchi identity (\ref{Bianchi_id_10D}) for $\,p=9$. In other words, the net charge of O$9$/D$9$ sources must vanish so that there is no contribution to the scalar potential coming from the O$9$/D$9$ DBI action, \textit{i.e.}, $\,V_{\textrm{O$9$/D$9$}}=0\,$ in (\ref{V_DBI}) due to the first line of (\ref{BI_Fp}). When $\,V_{\textrm{O$p$/D$p$}} \ge 0\,$ in a compactification, it was argued on very general grounds in \cite{Tringas:2025uyg} that no weakly-coupled and scale-separated regime is expected to occur in any AdS vacuum. Consequently, we do not expect any of the AdS$_{3}$ flux vacua in Table~\ref{Table:flux_vacua_IIB_O9} to realise such a regime.

\subsubsection*{AdS$_{3}$ vacua: moduli stabilisation and scale separation}

Let us frame the rest of the discussion within the minimal $\,\mathcal{N}=1\,$ RSTU-models with only dilatons described by the real superpotential in (\ref{WO9}). When moving to the standard picture in which moduli fields are expressed in terms of fluxes by using the rescalings in Appendix~\ref{app:Flux_rescalings}, the first thing we observe is that the four dilatons in the RSTU-models \textit{cannot} be simultaneously stabilised in any of the flux vacua presented in Table~\ref{Table:flux_vacua_IIB_O9} (and its continuation). Instead, one dilaton\footnote{Any of the four dilatons in the $\,\mathcal{N}=1\,$ RSTU-models can be chosen to be the flat direction of the scalar potential, namely, the unstabilised modulus. This can be made explicit by performing the field redefinitions
\begin{equation}
\phi_0 = \{A_4 \,,\, \mu_4 \,,\, A \,,\, \mu\}
\hspace{8mm},\hspace{8mm}
\phi_1 = \frac{\mu_4}{A_4}
\hspace{8mm},\hspace{8mm}
\phi_2=\frac{\mu_4}{A}
\hspace{8mm},\hspace{8mm}
\phi_3=\frac{\mu}{A} \ ,
\end{equation}
where $\,\phi_{0}\,$ is the dilaton chosen to be the unstabilised modulus. In this parameterisation, the selected $\,\phi_{0}\,$ drops out of the scalar potential, making its flat nature manifest. After stabilisation of $\,\phi_{1,2,3}\,$, the unstabilised modulus $\,\phi_{0}\,$ comes with a kinetic term $\,\sim \frac{1}{\phi_{0}^{2}} (\partial \phi_{0})^{2}$.} always remains unstabilised by the fluxes. Still, one could think of a two-step stabilisation procedure in which, first, fluxes stabilise all but one dilaton and, then, such a dilaton gets stabilised by other means, \textit{e.g.}, by the inclusion of non-perturbative effects in the spirit of \cite{Kachru:2003aw}. Even in this case, it can be shown that, under mild assumptions, a weakly-coupled and properly scale-separated regime cannot be achieved for any of the flux vacua in Table~\ref{Table:flux_vacua_IIB_O9}: the size of some internal cycles does not separate from the scale of the AdS$_{3}$ external spacetime.

Let us illustrate the above statement with an example. Let us consider $\textcolor{Purple}{\textbf{vac~2}}$ in Table~\ref{Table:flux_vacua_IIB_O9} with $\,\xi=0$, which, as explained below eq.(\ref{WO9}), is captured by the minimal $\,\mathcal{N}=1\,$ model in (\ref{Z2-even_fluxes_O9})-(\ref{WO9}). This AdS$_{3}$ vacuum requires
\begin{equation}
\label{Z2-even_fluxes_O9_example}
\omega_3=\omega_5=0
\,\,\,\, , \,\,\,\,
\omega_7=\dfrac{\omega_4^2}{\omega_6}
\,\,\,\, , \,\,\,\,
\omega_7 = \frac{f_{34}f_7}{4 \, \omega_4 \, \omega_6}
\,\,\,\, , \,\,\,\,
\omega_{11} = 2 \, \omega_4 
\,\, \,\,\,\, \textrm{ and }\,\,\,\, \,\, 
f_{32} = f_{35} = 0 \ ,
\end{equation}
so that there are four independent flux parameters $\,\left\lbrace \omega_{4} \,,\, \omega_{6} \,,\, f_{34} \,,\, f_{7} \right\rbrace$. In the standard picture, three of the dilaton VEV's are given by
\begin{equation}
\label{VEV's_example_O9}
A_{4} = \frac{f_{34}f_7}{4 \, \omega_4 \, \omega_6} \, \mu 
\hspace{8mm} , \hspace{8mm}
\mu_{4} = \frac{2 \, \omega_4^2}{\omega_6 f_7}   \, \mu 
\hspace{8mm} , \hspace{8mm}
A=\frac{f_{34}}{2 \, \omega_4}\, \mu  \ ,
\end{equation}
which depends on the unstabilised modulus $\,\mu$. The value of the scalar potential at the vacuum is nevertheless well-defined, namely, independent of $\,\mu$, and reads
\begin{equation}
\label{V_example_O9}
g^{-2} V_0 = -\frac{2}{(gL)^{2}} = - \frac{2 \, \omega_{4}^{6}}{(f_{7} \, f_{34})^{2}}
%= -\frac{f_7^2}{32}\left( \frac{\mu_4}{A_4} \right)^2  
\ ,
\end{equation}
where $\,(gL)\,$ is the radius of AdS$_{3}$. The moduli VEV's in (\ref{VEV's_example_O9}) and the vacuum energy in (\ref{V_example_O9}) determine all the quantities needed to assess whether a weakly-coupled and scale-separated regime can be achieved. Using the definitions in (\ref{vol7&gs}), the moduli dictionary in (\ref{dictionary_dilatons_O9}) and the VEV's in (\ref{VEV's_example_O9}), the expressions for the string coupling and the volume of the internal space (in the string frame, see (\ref{10D_metric})) are given by
\begin{equation}
\label{gs&vol7_example}
g_s^{2} = \frac{4}{\mu^4} \left( \frac{f_7}{f_{34}^3} \right) \left(\omega_4 \, \omega_6 \right)
\hspace{5mm} , \hspace{5mm}
\textrm{vol}_7 = \frac{1}{\mu^{3}}\left( \frac{f_{7}}{f_{34}} \right)^{\frac{7}{4}} \left( \frac{\omega_6}{\omega_4}\right)^{\frac{3}{4}}  \ ,
\end{equation}
whereas the characteristic size of the external AdS$_{3}$ spacetime (in the string frame, see (\ref{10D_metric})) is
\begin{equation}
\label{AdS3_scale}
\tau^{-1} \, (g L) = \frac{4}{\mu} \left( \frac{f_{7}}{f_{34}} \right)^{\frac{1}{4}} \left(\omega_{4}^{-5} \, \omega_{6}\right)^{\frac{1}{4}}  \ ,
\end{equation}
so that
\begin{equation}
\label{vol7/L_ratio}
\frac{\textrm{vol}_{7}^{\frac{1}{7}}}{\tau^{-1} \, (g L)} = \frac{\mu^{\frac{4}{7}}}{4} \left(\frac{\omega_4^{8}}{\omega_{6}} \right)^{\frac{1}{7}}\ .
\end{equation}

For the above AdS$_{3}$ vacuum to be in a weakly-coupled regime and to achieve scale separation between the internal space and the characteristic size of the AdS$_{3}$ external spacetime, we will make an ansatz for the fluxes in terms of a scaling parameter $\,N\,$ which will be taken arbitrarily large.\footnote{Recall that there is no possible flux-induced tadpole for O$9$/D$9$ sources (see (\ref{J_O9/D9=0})), so we do not have to worry about exceeding the number of such sources in the compactification when scaling the fluxes with $N$.} More concretely, we will take\footnote{We are being somewhat liberal in allowing metric fluxes to scale with $N$. This is done for the sake of generality, as one can always choose them not to scale by setting $\,\alpha=\beta=0\,$ in (\ref{flux_scaling_O9}).}
\begin{equation}
\label{flux_scaling_O9}
\omega_{4} \sim N^{\alpha}
\hspace{5mm} , \hspace{5mm}
\omega_{6} \sim N^{\beta}
\hspace{5mm} , \hspace{5mm}
f_{34} \sim N^{\delta}
\hspace{5mm} , \hspace{5mm}
f_{7} \sim N^{\gamma} \ ,
\end{equation}
with $\, N,\,\alpha,\,\beta,\,\delta,\,\gamma \in \mathbb{N}\,$ (fluxes must be quantised). Once the modulus $\,\mu\,$ acquires a VEV through a mechanism yet to be determined and the moduli stabilisation process is completed, it simply becomes a number which we are not assuming to be particularly large or small (it is expected to depend on the specific mechanism stabilising $\mu$). From (\ref{gs&vol7_example}) and (\ref{vol7/L_ratio}) it then follows that
\begin{equation}
g_{s}^{2} \sim \frac{1}{\mu^4}  \, N^{ \gamma + \alpha + \beta - 3 \delta}
\hspace{8mm} \textrm{ and } \hspace{8mm}
\frac{\textrm{vol}_{7}^{\frac{1}{7}}}{\tau^{-1} \, (g L)} \sim \mu^{\frac{4}{7}} \, N^{\frac{8\alpha-\beta}{7}}  \ ,
\end{equation}
and so scale separation between the internal volume and the external AdS$_{3}$ spacetime requires $\,8\,\alpha < \beta$. This implies that not only the gauge fluxes but also some metric flux scales with $\,N$. Finally, setting $\,\gamma + \alpha + \beta < 3 \delta\,$ yields $\,g_{s} \ll 1\,$ and $\,\tau^{-1} \, (g L) \gg \textrm{vol}_{7}^{\frac{1}{7}}\,$ in the limit $\,N \gg 1$. This implies that a weak coupling regime exists in which the internal volume is scale-separated from the external AdS$_{3}$ spacetime.

Still, one might worry about whether all internal cycles -- not just the overall volume -- are scale separated. A more refined analysis of the would-be one-cycles yields
\begin{equation}
\begin{array}{lcll}
\rho \, \ell_{a} &=& \left( \dfrac{f_{7}}{f_{34}} \right)^{\frac{1}{4}} \left( \omega_{4}^{-5} \,  \omega_{6}\right)^{\frac{1}{4}} \, \omega_4 & , \\[4mm]
\rho \, \ell_{i} &=& \dfrac{1}{\mu}  \left( \dfrac{f_{7}}{f_{34}} \right)^{\frac{1}{4}} \left( \omega_{4}^{-5} \, \omega_{6}\right)^{\frac{1}{4}} \, \omega_4 & , \\[4mm]
\rho \, \ell_{7} &=& \left( \dfrac{f_{7}}{f_{34}} \right)^{\frac{1}{4}} \left( \omega_{4}^{-5} \, \omega_{6} \right)^{\frac{1}{4}} \omega_{4}^{2} \, \omega_{6}^{-1} & , \\[4mm]
\end{array}
\end{equation}
which translates into an anisotropic scaling of the form
\begin{equation}
\label{one-form_scalings_O9_general}
\frac{\rho \, \ell_{a}}{\tau^{-1} \, (g L) } =  \mu \, \frac{\omega_4}{4} \sim \mu \, N^{\alpha}
\hspace{3mm} , \hspace{3mm}
\frac{\rho \, \ell_{i}}{\tau^{-1} \, (g L) } =  \frac{\omega_{4}}{4} \sim N^{\alpha}
\hspace{3mm} , \hspace{3mm}
\frac{\rho \, \ell_{7}}{ \tau^{-1} \, (g L)} = \mu \, \frac{\omega_{4}^{2}}{4 \, \omega_{6}} \sim \mu \, N^{2\alpha-\beta}\ .
\end{equation}
In the most favorable case, which is $\,\alpha=0$, (\ref{one-form_scalings_O9_general}) implies the existence of cycles in the internal space with the same characteristic size as the AdS$_{3}$ external spacetime. More concretely,
\begin{equation}
\label{one-form_scalings_O9}
\frac{\rho \, \ell_{a}}{\tau^{-1} \, (g L) } \sim \mu
\hspace{6mm} , \hspace{6mm}
\frac{\rho \, \ell_{i}}{\tau^{-1} \, (g L) } \sim \mathcal{O}(1)
\hspace{6mm} , \hspace{6mm}
\frac{\rho \, \ell_{7}}{ \tau^{-1} \, (g L)} \sim \mu \, N^{-\beta} \ ,
\end{equation}
thus making scale separation not fully satisfactory. This not only applies to one-cycles, but also to three- and four-cycles, which are the relevant ones when the internal space has G$_{2}$-structure. Note that a possibly large or small value of $\,\mu\,$ does not change this result, as $\,\rho \, \ell_{i} \sim \tau^{-1} \, (g L)\,$ is the best we can get. Regarding precisely the subsequent stabilisation of $\,\mu$, a word of caution is in order. The value of the flux-induced scalar potential in (\ref{V_example_O9}) scales as $\,g^{-1} |V_0|^{1/2} \sim N^{3\alpha-\delta-\gamma}$. For it to provide a reliable starting point for the stabilisation of $\,\mu\,$ via some (possibly non-perturbative) mechanism, \textit{i.e.}, $\,V = V_{0} + V_{\textrm{non-pert}}(\mu)$, one would like to impose $\,3\alpha - \delta - \gamma = 0\,$ so that $\,V_0\,$ does not scale with $\,N$. This requires taking $\,\alpha \neq 0\,$ which, as pointed out before, renders the one-cycles $\,\rho \, \ell_{i}\,$ in (\ref{one-form_scalings_O9_general}) dangerously larger than the characteristic AdS$_{3}$ length scale.

The same analysis can be repeated for the remaining AdS$_{3}$ flux vacua in Table~\ref{Table:flux_vacua_IIB_O9} which are captured by the minimal $\,\mathcal{N}=1\,$ model in (\ref{Z2-even_fluxes_O9})-(\ref{WO9}). Moving to the standard picture in which moduli VEV's are expressed in terms of free flux parameters, two different situations occur:
\begin{itemize}

\item[$i)$] \textit{One modulus remains unstabilised:} This is the case for the groups \textbf{vac~6,7} and \textbf{vac~36,37}. Actually, the two groups become degenerated when only $\mathbb{Z}_{2}^{*}$-even fluxes are activated, namely, in the minimal $\,\mathcal{N}=1\,$ setup.

\item[$ii)$]  \textit{Two moduli remain unstabilised:} This is the case for the groups \textcolor{ForestGreen}{\textbf{vac~8}}-\textcolor{ForestGreen}{\textbf{vac~11}} as well as \textbf{vac~12}-\textbf{vac~15} and \textbf{vac~16}-\textbf{vac~19}.  This time the last two groups become degenerated in the minimal $\,\mathcal{N}=1\,$ setup.

\end{itemize}

\noindent In both cases, the ratios $\,\frac{\rho \, \ell_{a}}{\tau^{-1} \, (g L)}\,$ and $\,\frac{\rho \, \ell_{i}}{\tau^{-1} \, (g L)}\,$ scale as $\,N^{f(\alpha,\beta)}\,$ where $\,f(\alpha,\beta) \ge 0\,$ is a function of the parameters $\,\alpha$, $\beta\,$ specifying the scalings of the metric fluxes. This was so in (\ref{one-form_scalings_O9_general}). However, unlike in  (\ref{one-form_scalings_O9_general}), it turns out that $\,\frac{\rho \, \ell_{7}}{\tau^{-1} \, (g L)} \sim N^{\gamma}\,$ with $\,\gamma\,$ being the parameter scaling the gauge flux $\,f_{7}$. Therefore, the overall internal volume cannot be decoupled from the AdS$_{3}$ characteristic scale. Still, the string coupling can be made $\,g_{s} \ll 1\,$ by scaling the gauge flux $\,f_{34}$.

Because fully satisfactory scale separation is not possible for the AdS$_{3}$ vacua of Table~\ref{Table:flux_vacua_IIB_O9}, the tower of KK modes is a priori not decoupled from the modes in the 3D supergravity.\footnote{This by no means implies that the 3D supergravity theory is not a consistent truncation of type IIB supergravity. Indeed, it is a consistent truncation rather than a reliable effective theory.} In this context, the type IIB vacua with O$9$-planes may be of particular interest. The presence of an unstabilised modulus, together with the fact that the type IIB with O$9$ setting can be realised as a generalised Scherk--Schwarz reduction of the O$(8,8)$-DFT (see Section~\ref{sec:IIB_O9}), makes it an excellent setting in which to test the distance conjecture \cite{Ooguri:2006in}. More concretely, the KK spectrometry techniques developed in \cite{Malek:2019eaz} and applied to three-dimensional gauged supergravity in \cite{Eloy:2020uix} would allow us to track the masses of the KK modes when approaching the boundary of the moduli space.

\subsubsection*{Mkw$_{3}$ vacuum and moduli stabilisation}

When expressing the moduli VEV's in terms of flux parameters, the \textbf{vac~1} in Table~{\ref{Table:flux_vacua_IIB_O9}} only stabilises the modulus $\,\mu=\sqrt{-\frac{\omega_5}{\omega_7}}$. The three other dilatons in the RSTU-model remain unstabilised. This is in agreement with (a three-dimensional version of) the massless Minkowski conjecture \cite{Andriot:2022yyj}.\\

To close the analysis of type IIB with O$9$, we have verified that the flux vacua of Table~\ref{Table:flux_vacua_IIB_O9} (and its continuation) belong to different orbits with respect to the scaling symmetry in Appendix~\ref{app:Flux_rescalings}. In other words, when moving to the standard picture in which moduli VEV's are expressed in terms of arbitrary flux parameters, only one vacuum solution is found for a specific choice of the fluxes.

\subsection{Type IIB with O$5$-planes}

This type IIB duality frame was originally studied in \cite{Arboleya:2024vnp} and presented there as an appetiser to the landscape of RSTU-models.

\subsubsection*{AdS$_{3}$ vacua: moduli stabilisation and scale separation}

We have shown that type IIB with O$5$ turns out to be the unique duality frame that, firstly, stabilises all the dilatons of the RSTU-models in an AdS$_{3}$ vacuum and, secondly, does it in a weakly-coupled and scale-separated regime. More concretely, this happens only for \textbf{vac~10} and \textbf{vac~11} of Table~\ref{Table:flux_vacua_IIB_O5}, which are not embeddable into maximal supergravity and come with a contribution $\,V_{\textrm{O$5$/D$5$}} < 0\,$ in the scalar potential, thus in agreement with \cite{Tringas:2025uyg}. The algebra of the isometry generators of the internal space is
\begin{equation}
\label{algebra_IIB_O5}
\left[ X_{i}, X_{7} \right]  = \omega_{3} \, \delta_{ia} \, X_{a}
\hspace{6mm} , \hspace{6mm}
\left[ X_{a},X_{7} \right] = - \omega_{2} \,  \delta_{ai} \, X_{i}  \ , 
\end{equation}
and is a $2$-step solvable algebra that can be viewed as three copies of the three-dimensional solvmanifold E$_2$ sharing a common generator $X_{7}$. It corresponds to a seven-dimensional solvmanifold compatible with a co-closed G$_2$-structure \cite{VanHemelryck:2025qok}.

For the specific choice of fluxes made in \cite{Arboleya:2025ocb,Arboleya:2025lwu}, where only gauge fluxes were assumed to scale with a scaling parameter $N$ without affecting the tadpole condition in (\ref{Tadpole_condition_O5/D5}), \textbf{vac~10} and \textbf{vac~11} give
\begin{equation}
\label{example_CORFU}
g_{s}^2 \sim N^{-4}
\hspace{7mm} , \hspace{7mm}
\textrm{vol}_{7} \sim  N^{\frac{31}2}
\hspace{7mm} , \hspace{7mm}
\tau^{-1} (gL) \sim N^{\frac{13}2} \ ,
\end{equation}
so that $\,g_{s} \ll 1\,$ and $\,\tau^{-1} \, (g L) \gg \textrm{vol}_{7}^{\frac{1}{7}}\,$ in the limit $\,N \gg 1$. Namely, a weak coupling regime exists in which the internal volume gets scale-separated from the external AdS$_{3}$ spacetime. In addition, the would-be one-cycles satisfy
\begin{equation}
\label{one-form_scalings_O5}
\frac{\rho \, \ell_{a}}{\tau^{-1} \, (g L) } \sim \frac{\rho \, \ell_{i}}{\tau^{-1} \, (g L) } \sim N^{-5}
\hspace{10mm} \textrm{ and } \hspace{10mm}
\frac{\rho \, \ell_{7}}{ \tau^{-1} \, (g L)}  \sim \mathcal{O}(1) \ .
\end{equation}
In contrast to what happened in (\ref{one-form_scalings_O9}), now it is the seventh internal direction the one that becomes of the same size as the AdS$_{3}$ external spacetime. This has the consequence that all the three- and four-cycles that are present when the internal space has G$_{2}$-structure remain scale-separated. Nevertheless, one might worry about the anisotropic scaling in (\ref{one-form_scalings_O5}) and the potential impact of $\,\rho \, \ell_{7} \sim \tau^{-1} \, (g L)\,$ on the lowest eigenvalue in the KK tower. A more thorough analysis is clearly required to clarify this issue. Also, since \textbf{vac~10} and \textbf{vac~11} are non-supersymmetric, it would be interesting to investigate mechanisms that make such flux vacua decay in light of the AdS conjecture \cite{Ooguri:2016pdq}. Lastly, as originally observed in \cite{Conlon:2021cjk,Apers:2022tfm} for the AdS$_{4}$ DGKT flux vacua \cite{DeWolfe:2005uu}, both scale-separated \textbf{vac~10} and \textbf{vac~11} come along with integer-valued conformal dimensions of the would-be dual CFT$_{2}$ operators \cite{Arboleya:2024vnp,Arboleya:2025ocb}. Moreover, by inspection of Table~\ref{Table:flux_vacua_IIB_O5_gravitini_scalar_spectrum}, one realises that in both vacua (as well as in their susy cousin in \cite{VanHemelryck:2025qok}, \textit{cf.} eq.(57) in \cite{Arboleya:2025ocb}), some of the $\,\Delta$'s obey the extremal arrangements $\,\Delta_{i} = \Delta_{j} + \Delta_{k}$ \cite{Bobev:2025yxp}. In light of the criterion put forward in \cite{Bobev:2025yxp}, it would be interesting to assess whether the cubic coupling of the scalar fields dual to such operators vanishes in 3D half-maximal supergravity, thus not leading to an ill-defined three-point function in the dual CFT$_{2}$ (at large $N$).

Leaving aside the scale-separated \textbf{vac~10} and \textbf{vac~11} in Table~\ref{Table:flux_vacua_IIB_O5}, the remaining AdS$_{3}$ flux vacua in this duality frame do not stabilise all the dilatons of the corresponding RSTU-models. Again there is either one or two moduli that remain unstabilised. By inspection of Table~\ref{Table:flux_vacua_IIB_O5}, these AdS$_{3}$ vacua are embeddable into maximal supergravity (they are marked with $*$) and, therefore, $\,V_{\textrm{O$5$/D$5$}} = 0$. Again, the arguments put forward in \cite{Tringas:2025uyg} make scale separation implausible at any of these vacua. A similar analysis to that performed in the type IIB with O$9$ case can also be carried out here, and indeed, the conclusions regarding moduli stabilisation and scale separation remain the same. This is consistent with the observation that most of the AdS$_{3}$ vacua in Table~\ref{Table:flux_vacua_IIB_O5} can be generalised to AdS$_{3}$ vacua in Table~\ref{Table:flux_vacua_IIB_O9} (recall the color code used in the tables).\footnote{This is not the case for \textbf{vac~6,7} in Table~\ref{Table:flux_vacua_IIB_O5} which do not have a generalisation to type IIB with O$9$. Assuming that a mechanism stabilises the unique modulus at these two AdS$_{3}$ vacua, the string coupling can be made $\,g_{s} \ll 1\,$ and the overall internal volume can be scale-separated from the AdS$_{3}$ characteristic size. However, at best one finds $\,\frac{\rho \, \ell_{i}}{ \tau^{-1} \, (g L)} \sim \frac{\rho \, \ell_{a}}{ \tau^{-1} \, (g L)} \sim \mathcal{O}(1)$, which eliminates the possibility of fully satisfactory scale separation: some internal three- and four-cycles do not decouple from the AdS$_3$ length scale.} In conclusion, even assuming a mechanism to fix the unstabilised modulus exists, scale separation can only be achieved at \textbf{vac~10} and \textbf{vac~11} of Table~\ref{Table:flux_vacua_IIB_O5}.

\subsubsection*{Mkw$_{3}$ vacua and moduli stabilisation}

Apart from the trivial \textcolor{red}{\textbf{vac~1}}, there are two Mkw$_{3}$ flux vacua in Table~\ref{Table:flux_vacua_IIB_O5}. In the standard picture where moduli VEV's are written in terms of arbitrary flux parameters one finds
\begin{equation}
\textbf{vac~2 : } \hspace{3mm}   A_4 = -\frac{f_{33}}{\omega_2} A  \,\,\,\,\,,\,\,\,\,\, \mu_4 = \frac{\omega_1}{f_{32}}\mu
\hspace{8mm} , \hspace{8mm}
\textbf{vac~3 : } \hspace{3mm} \mu_4 = \frac{\omega_2}{\omega_3 \, A} \ ,
\end{equation}
so both Mkw$_{3}$ vacua come with unstabilised moduli, in agreement with \cite{Andriot:2022yyj}.

\subsection{Type IIB with O$3$-planes}

Although the RSTU-dilatons are not all stabilised in the Mkw$_{3}$ vacuum of type IIB with O3-planes displayed in Table~\ref{Table:flux_vacua_IIB_O3}, interesting patterns of partial supersymmetry breaking from $\,\mathcal{N}=8\,$ down to $\,\mathcal{N}=0,1,2,3,4\,$ and $\,6\,$ emerge depending on the choice of flux parameters. Whenever the unique metric flux $\,\omega_{5}\,$ is non zero, the isometry algebra of the internal space is given by
\begin{equation}
\label{algebra_IIB_O3_bis}
[X_{i},X_{a}] = - \omega_{5} \, \delta_{ia} \, X_{7} \ ,
\end{equation}
and corresponds to the seven-dimensional nilmanifold compatible with a co-closed G$_2$-structure labeled ``$17$'' in Table~$3$ of \cite{VanHemelryck:2025qok}. 

Moving to the standard picture in which moduli VEV's are expressed in terms of flux parameters, the Mkw$_{3}$ vacua in Table~\ref{Table:flux_vacua_IIB_O3} are given by
\begin{equation}
A = -\frac{\omega_5}{f_5} \, A_{4}
\hspace{8mm} , \hspace{8mm}
\mu = \sqrt{-\frac{f_{34} h_{32}}{f_{33} h_{31}}}
\hspace{8mm} , \hspace{8mm}
\mu_4= \sqrt{-\frac{f_{34} h_{32}^3}{f_{33}^3 h_{31}}} \ ,
\end{equation}
and gauge fluxes are restricted to obey
\begin{equation}
h_{33}= \frac{f_{32} \, f_{34} \, h_{32}^2}{f_{33}^2 \, h_{31}}
\hspace{8mm} , \hspace{8mm}
h_{34}= \frac{f_{31} \, f_{34}^2 \, h_{32}^3}{f_{33}^3 \, h_{31}^2} \ .
\end{equation}
Once again, one dilaton remains unstabilised, in agreement with \cite{Andriot:2022yyj}. 
\\

To conclude, we have shown that any type IIB AdS$_{3}$ flux vacuum of the three-dimensional RSTU-models that admits an embedding into maximal supergravity, \textit{i.e.} $V_{\textrm{O$p$/D$p$}} = 0$, fails to stabilise all the dilatons. Assuming that some mechanism (possibly non-perturbative) fixes the unstabilised moduli, we have shown that a weak coupling regime can be achieved in which the overall internal volume is scale-separated from the characteristic size of the external AdS$_{3}$ spacetime. Still, some internal cycles stay unavoidably of the same size as the AdS$_{3}$ length scale. This is to be contrasted with the type IIA AdS$_{4}$ flux vacua of the four-dimensional STU-models investigated in \cite{Dibitetto:2011gm}. In this case, all the AdS$_{4}$ vacua turn out to be embeddable into maximal supergravity and, at the same time, succeed in stabilising all the dilatons. However, as shown in \cite{Balaguer:2024cyb}, they fail to realise scale separation. It would be interesting to further investigate this role swap between type IIB (3D) and type IIA (4D) flux vacua, as well as a possible connection between them via dimensional reduction on G$_{2}$-structure (3D) \textit{vs} $\textrm{SU}(3)$-structure (4D) manifolds. We leave this study for future work.

\section*{Acknowledgements}

We are grateful to Giuseppe Sudano for collaboration in related work. \'AA, AG and MM are partially supported by the grants from the Spanish government MCIU-22-PID2021-123021NB-I00 and MCIU-25-PID2024-161500NB-I00. The work of \'AA is also supported by the Severo Ochoa fellowship NAC-AT-PUB-ASV-2025 BP24-134. The work of GC is supported by the German Research Foundation through a German-Israeli Project Cooperation (DIP) grant “Holography and the Swampland”.

%\newpage

\appendix

\section{Type IIB dimensional reduction and scalar potential}
\label{App:dimensionsal_reduction}

The bosonic field content of type IIB supergravity consists of the universal NS-NS sector $\{ G \,,\, \Phi \,,\, B_{(2)}\}$ and the (democratic formulation of) R-R sector of gauge potentials $\,\lbrace{C_{(2 p)}\rbrace}\,$ with $\,p = 0, \ldots,4$. The ten-dimensional pseudo-action in the string frame is
\begin{equation} 
\label{S_IIB}
\begin{array}{rcl}
S_{\textrm{IIB}} &=&  \frac{1}{2\kappa^{2}} \displaystyle\int \textrm{d}^{10}X \, \sqrt{-G} \, \left[ e^{-2\Phi} \left(R^{(10)} \, + \, 4  (\partial\Phi)^{2} \, - \, \frac{1}{2 \cdot 3!} |H_{(3)}|^2 \right ) \, - \, \frac{1}{2} \sum_{p=0}^4 \frac{|F_{(2p + 1)}|^2}{2 \cdot (2p + 1)!}\right] \\[2mm]
& + & S_{\textrm{O$p$/D$p$}} \ ,
\end{array}
\end{equation}
where the DBI contribution for the O$p$/D$p$ sources is given by
\begin{equation}
\label{S_DBI}
S_{\textrm{O$p$/D$p$}} = (2N_{\textrm{O}p}-N_{\textrm{D}p}) \, T_{\textrm{D}p} \displaystyle\int_{\textrm{WV}(\textrm{O}p/\textrm{D}p)} \textrm{d}^{p+1}\sigma \ e^{- \Phi} \sqrt{- \textrm{det} [G_{\mathsf{\bar{M}\bar{N}}}]}  \ .
\end{equation}
We denoted $\,X^{\mathsf{M}}=\left(x^{\mu},y^{i},y^{a},y^{7}\right)\,$ the total spacetime coordinates, and $\,X^{\bar{\mathsf{M}}}\,$ the $\,(p+1)\,$ coordinates on the worldvolume (WV) of the spacetime-filling O$p$-planes/D$p$-branes. In addition, $\,(2\kappa^{2})^{-1}=(2\pi)(2 \pi \ell_{s})^{-8}\,$ where $\,\ell_{s}=\sqrt{\alpha'}\,$ is the string length scale, and $\,T_{\textrm{D}p}=(2\pi)(2\pi \ell_{s})^{-(p+1)}\,$ is the D$p$-brane tension ($T_{\textrm{O}p} = - 2 \, T_{\textrm{D}p}$). The pseudo-action (\ref{S_IIB}) must be supplemented by the ten-dimensional Hodge duality relations
\begin{equation}
\label{Duality_relations}
F_{(9)} = \star F_{(1)} 
\hspace{6mm} , \hspace{6mm}
F_{(7)} = - \star F_{(3)}
\hspace{6mm} , \hspace{6mm}
F_{(5)} = \star F_{(5)} \ ,
\end{equation}
to have the correct number of degrees of freedom. The NS-NS and RR field strengths satisfy the Bianchi identities in (\ref{Bianchi_id_10D}).

The starting point to perform the dimensional reduction of (\ref{S_IIB}) is to make an ansatz for the spacetime metric of the form
\begin{equation}
\label{10D_metric}
ds_{10}^2 = \tau^{-2} \, ds_{3}^2 + \rho^2 \, d\tilde{s}_7^2 \ ,
\end{equation}
with $\,ds_{3}^2=g_{\mu\nu} \, dx^{\mu} \, dx^{\nu}\,$ and $\,d\tilde{s}_{7}^2=\tilde{g}_{mn} \, \eta^{m} \, \eta^{n}\,$ being the line elements of the $D=3$ external spacetime and the seven-dimensional internal space with unit volume, \textit{i.e.} $\,\textrm{det}[\tilde{g}_{mn}]=1$. The two scalars $\,\tau(x)\,$ and $\,\rho(x)\,$ in (\ref{10D_metric}) determine the string coupling $\,g_{s}=e^{\Phi}\,$ and the internal volume as
\begin{equation}
\label{vol7&gs}
g_{s}^{2} =  \frac{\textrm{vol}_7}{\tau} 
\hspace{6mm} , \hspace{6mm} 
\textrm{vol}_7 = \rho^7 \ .
\end{equation}
The $\,\textrm{SO}(3)\,$ symmetry of the RSTU-models requires the metric on the internal seven-dimensional twisted torus to be of the form
\begin{equation}
\label{7D_metric}
d\tilde{s}_{7}^2 = {\ell_{i}}^2 \left[ \left(\eta^2\right)^{2} + \left(\eta^4\right)^{2} + \left(\eta^6\right)^{2} \right]+ {\ell_{a}}^2 \left[ \left(\eta^1\right)^{2} + \left(\eta^3\right)^{2} + \left(\eta^5\right)^{2} \right] +  \left(  \ell_a^{-3} \ell_i^{-3} \right)^2 \left(\eta^7\right)^2 \ ,
\end{equation}
with two scalars $\,\ell_{i}(x)\,$ and  $\,\ell_{a}(x)$.

\subsubsection*{Reduction of the Einstein--Hilbert term and the type IIB dilaton $\,\Phi$}

The ansatz (\ref{10D_metric}) for the ten-dimensional metric implies
\begin{equation}
\sqrt{-G} = \tau^{-3} \rho^{7} \ \sqrt{-g} \ .
\end{equation}
Then the dimensional reduction of the Ricci scalar and the type IIB dilaton kinetic term in (\ref{S_IIB}) gives
\begin{equation}
\frac{1}{2\kappa^{2}} \displaystyle\int \textrm{d}^{10}X \, \sqrt{-G} \, e^{-2\Phi} \, \left(R^{(10)} \, + \, 4  (\partial\Phi)^{2} \right) = \frac{1}{2\kappa^{2}} \displaystyle\int \textrm{d}^{3}x \, \sqrt{-g} \,  \big( R + L_{\textrm{kin}} - V_{\omega} \big) \ ,
\end{equation}
with
\begin{equation}
\label{L_kin}
L_{\textrm{kin}} = - \frac{1}{\tau^2} \partial_\mu \tau \partial^\mu \tau - \frac{7}{\rho^2} \partial_\mu \rho \partial^\mu \rho - \frac{12}{{\ell_i}^2} \partial_\mu {\ell_i} \partial^\mu {\ell_i} - \frac{18}{\ell_a \ell_i} \partial_\mu \ell_a \partial^\mu \ell_i - \frac{12}{{\ell_a}^2} \partial_\mu {\ell_a} \partial^\mu {\ell_a} \ ,
\end{equation}
and
\begin{equation}
\label{V_metric}
V_{\omega} = \tau^{-2} \, \rho^{-2} \left(\dfrac{1}{4} \, \tilde{g}_{qm} \, \tilde{g}^{nr} \, \tilde{g}^{ps} \omega_{np}{}^q \, \omega_{rs}{}^m + \dfrac{1}{2} \, \tilde{g}^{np} \, \omega_{mn}{}^q \, \omega_{qp}{}^m  \right) \ ,
\end{equation}
provided the identifications in (\ref{vol7&gs}) hold.

\subsubsection*{Reduction of the gauge fluxes}

The NS-NS three-form flux $\,H_{(3)}\,$ is expanded using the basis of one-forms as
\begin{equation}
H_{(3)}=\frac{1}{3!} H_{mnp} \, \eta^{m} \wedge \eta^{n} \wedge \eta^{p}  \ ,  
\end{equation}
with constant flux components $\,H_{mnp}$. Then the dimensional reduction of the $\,|H_{(3)}|^2\,$ term in (\ref{S_IIB}) gives
\begin{equation}
\frac{1}{2\kappa^{2}}  \int \textrm{d}^{10}X \, \sqrt{-G^{(10)}} \, \left(-\frac{1}{2 \cdot 3!}\, e^{-2\Phi}\, |H_{(3)}|^2 \right) = \frac{1}{2\kappa^{2}}  \int \textrm{d}^3 x \sqrt{-g} \, \big(-V_{H} \big) \ ,
\end{equation}
with
\begin{equation}
\label{V_H}
V_{H} =  \dfrac{1}{2 \cdot 3!} \, \tau^{-2} \, \rho^{-6} \, H_{mnp} \, H^{mnp}  \ .
\end{equation} 
In (\ref{V_H}) the internal space indices are lowered/raised using the internal metric in (\ref{7D_metric}) and its inverse. Similarly, the dimensional reduction of the R-R terms $\,|F_{(2p + 1)}|^2\,$ in (\ref{S_IIB}) gives
\begin{equation}
 \frac{1}{2\kappa^{2}} \displaystyle\int \textrm{d}^{10}X \, \sqrt{-G} \, \left[ - \frac{1}{2} \sum_{p=0}^3 \frac{|F_{(2p + 1)}|^2}{2 \cdot (2p + 1)!}\right] = \frac{1}{2\kappa^{2}}  \int \textrm{d}^3 x \sqrt{-g} \, \big(-V_{F} \big) \ ,
\end{equation}
with
\begin{equation}
\label{V_F}
V_{F} =  \displaystyle\sum_{p=0}^{3} \dfrac{1}{2 \cdot (2p+1)!} \, \tau^{-3} \, \rho^{7-2(2p+1)}  \, F_{m_1\ldots m_{2p+1}} \, F^{m_1\ldots m_{2p+1}} \ .
\end{equation}

\subsubsection*{Reduction of the DBI term}

The dimensional reduction of the DBI term in (\ref{S_DBI}) gives
\begin{equation}
S_{\textrm{O$p$/D$p$}} = \frac{1}{2\kappa^{2}}  \int \textrm{d}^3 x \sqrt{-g} \, \big(-V_{\textrm{O$p$/D$p$}} \big) \ ,
\end{equation}
with
\begin{equation}
\label{V_DBI}
\begin{array}{rcl}
V_{\textrm{O$9$/D$9$}}  &=& -  T_{\textrm{D}9} \, \big(2 N_{\textrm{O}9} - N_{\textrm{D}9} \big) \, \tau^{-\frac{5}{2}} \rho^{\frac{7}{2}}  \ , \\[2mm]
V_{\textrm{O$5$/D$5$}}  &=& -  T_{\textrm{D}5} \, \big(2 N_{\textrm{O}5} - N_{\textrm{D}5} \big) \, \tau^{-\frac{5}{2}} \rho^{-\frac{1}{2}} \, {\ell_i}^3 \ , \\[2mm]
V_{\textrm{O$3$/D$3$}}  &=& -  T_{\textrm{D}3} \, \big(2 N_{\textrm{O}3} - N_{\textrm{D}3} \big) \, \tau^{-\frac{5}{2}} \rho^{-\frac{5}{2}} \, {\ell_i}^{-3} \, {\ell_a}^{-3} \ ,
\end{array}
\end{equation}
for the spacetime-filling O$p$/D$p$ sources placed in the internal space as described in Table~\ref{Table:Op_planes}.

\subsubsection*{Matching the 3D half-maximal gauged supergravity}

The scalar potential obtained upon dimensional reduction takes the form
\begin{equation}
\label{V_reduction}
V = V_{\omega} + V_{H} + V_{F} + V_{\textrm{O$p$/D$p$}} \ ,
\end{equation}
with the contributions given by (\ref{V_metric}), (\ref{V_H}), (\ref{V_F}) and (\ref{V_DBI}). The kinetic terms (\ref{L_kin}) and the scalar potential (\ref{V_reduction}) precisely match the expressions for the RSTU-models obtained using the gauged supergravity formalism if the following dictionary between moduli fields applies:
\begin{itemize}

\item Type IIB with O$9$-planes

\begin{equation} 
\label{dictionary_dilatons_O9}
\textrm{Im}R = \frac{{\tau} \, {\rho}}{{\ell_i}^3 \, {\ell_a}^3}
\hspace{4mm} , \hspace{4mm} 
\textrm{Im}S = \frac{1}{{\tau}^{\frac12} \, {\rho}^{\frac52} \, {\ell_i}^3 \, {\ell_a}^3 }
\hspace{4mm} , \hspace{4mm} 
\textrm{Im}T = \frac{{\tau}^{\frac12} \, {\ell_i} \, {\ell_a}}{{\rho}^{\frac32}}
\hspace{4mm} , \hspace{4mm} 
\textrm{Im}U = \frac{{\ell_a}}{{\ell_i}} \ .
\end{equation}

\item Type IIB with O$5$-planes

\begin{equation} 
\label{dictionary_dilatons_O5}
\textrm{Im}R = \frac{\tau \, {\ell_i}^3 \, {\ell_a}^3}{\rho} 
\hspace{4mm} , \hspace{4mm} 
\textrm{Im}S = \frac{{\ell_a}^{3}}{\tau^{\frac{1}{2}} \, \rho^{\frac{1}{2}}}
\hspace{4mm} , \hspace{4mm} 
\textrm{Im}T = \frac{\tau^{\frac{1}{2}} \, \rho^{\frac{1}{2}}}{{\ell_i}^{2} \, {\ell_a}}
\hspace{4mm} , \hspace{4mm} 
\textrm{Im}U = \frac{1}{{\rho}^{2} \, {\ell_i} \, {\ell_a}} \ .
\end{equation}

\item Type IIB with O$3$-planes

\begin{equation} 
\label{dictionary_dilatons_O3}
\textrm{Im}R = \frac{\tau \, \rho}{{\ell_i}^3\, {\ell_a}^3}
\hspace{4mm} , \hspace{4mm} 
\textrm{Im}S = \frac{\rho^{\frac72}}{\tau^{\frac12}}
\hspace{4mm} , \hspace{4mm} 
\textrm{Im}T = {\tau}^{\frac12} \, {\rho}^{\frac12} \, {\ell_i}^2 \, {\ell_a}^2
\hspace{4mm} , \hspace{4mm} 
\textrm{Im}U = \frac{\ell_i}{\ell_a} \ .
\end{equation}

\end{itemize}

\noindent The matching of the scalar potential crucially requires the Bianchi identities in (\ref{Bianchi_id_10D}). In particular, the one associated with the single type of sources present in the compactification is given by
\begin{equation}
\label{BI_Fp}
\begin{array}{rcl}
\textrm{O$9$/D$9$} & : & 0 =  2\kappa^{2} \,  T_{\textrm{D}9} \, \big(2N_{\textrm{O}9} - N_{\textrm{D}9} \big) \ ,  \\[4mm]
\textrm{O$5$/D$5$} & : &  \left. dF_{(3)} \, \right|_{7abc} =  3\, \omega_{1} \, f_{32} - 3\,   \omega_{2} \, f_{33} =  2\kappa^{2} \,  T_{\textrm{D}5} \, \big(2N_{\textrm{O}5} - N_{\textrm{D}5} \big) \ , \\[4mm]
\textrm{O$3$/D$3$} & : &  \left. dF_{(5)} - H_{(3)} \wedge F_{(3)}  \, \right|_{aibjck} = - 3 \, \omega_5 \, f_5 - h_{31} \, f_{34} + 3 \, h_{32} \, f_{33} - 3 \, h_{33} \, f_{32} + h_{34} \, f_{31} \\[2mm]  
& & \hspace{4.28cm} = 2\kappa^{2} \,  T_{\textrm{D}3} \, \big(2N_{\textrm{O}3} - N_{\textrm{D}3} \big) \ ,
\end{array}
\end{equation}
for the three duality frames of relevance. Only when substituting (\ref{BI_Fp}) into the contributions (\ref{V_DBI}), the scalar potential arising from the dimensional reduction and the one of gauged supergravity exactly match.

\section{Flux rescaling symmetry}
\label{app:Flux_rescalings}

The scalar potential of the RSTU-models in a type II with O$p$-planes duality frame is invariant under the moduli rescaling in (\ref{scaling_moduli}) followed by a rescaling of the metric and gauge fluxes that depends on the specific value of $\,p$. In this appendix we present the explicit form of such flux rescalings for the various duality frames.

\subsubsection*{Type IIB with O$9$-planes}

The transformation of the metric fluxes is of the form
\begin{equation}
\label{IIB_O9_SL(2)_transformation_metric}
\begin{array}{c}
\omega_1' = \left( \frac{\lambda_R \lambda_T \lambda_U}{\lambda_S} \right)^{\frac12} \omega_1 
\hspace{3mm} , \hspace{3mm} 
\omega_2' = \lambda_R \, \omega_2 
\hspace{3mm} , \hspace{3mm}
\omega_3' = \left( \frac{\lambda_R \lambda_T \lambda_U^3}{\lambda_S} \right)^{\frac12}  \omega_3
\hspace{3mm} , \hspace{3mm}
\omega_4' =  \left( \frac{\lambda_R \lambda_T}{\lambda_S \lambda_U} \right)^{\frac12}  \omega_4 \ ,
\\[2mm]
\omega_5' =  \lambda_R \lambda_U \, \omega_5 
\hspace{3mm} , \hspace{3mm}
\omega_6' = \frac{\lambda_T}{\lambda_S} \, \omega_6
\hspace{3mm} , \hspace{3mm}
\omega_7' = \frac{\lambda_R}{\lambda_U} \, \omega_7 
\hspace{3mm} , \hspace{3mm}
\omega_8' =  \left( \frac{\lambda_R \lambda_T}{\lambda_S \lambda_U^3} \right)^{\frac12}  \omega_8 \ ,
\\[2mm]
\omega_9' = \left( \frac{\lambda_R \lambda_T \lambda_U}{\lambda_S} \right)^{\frac12} \omega_9
\hspace{3mm} , \hspace{3mm}
\omega_{10}' =  \lambda_R  \, {\omega_{10}}
\hspace{3mm} , \hspace{3mm}
\omega_{11}' = \left( \frac{\lambda_R \lambda_T}{\lambda_S \lambda_U} \right)^{\frac12} \omega_{11} \ ,
\end{array}
\end{equation}
whereas gauge fluxes transform as
\begin{equation}
\label{IIB_O9_SL(2)_transformation_gauge}
\begin{array}{c}
f_{31}' = \left( \frac{\lambda_R \lambda_T^{3} \lambda_U^3}{\lambda_S} \right)^{\frac12}  f_{31} 
\hspace{5mm} , \hspace{5mm} 
f_{32}' =  \left( \frac{\lambda_R \lambda_T^{3} \lambda_U}{\lambda_S} \right)^{\frac12}  f_{32}
\hspace{5mm} , \hspace{5mm}
f_{33}' =   \left( \frac{\lambda_R \lambda_T^{3}}{\lambda_S \lambda_U} \right)^{\frac12} f_{33} \ ,
\\[2mm]
f_{34}' =   \left( \frac{\lambda_R \lambda_T^{3}}{\lambda_S \lambda_U^3} \right)^{\frac12} f_{34}
\hspace{5mm} , \hspace{5mm}
f_{35}' = \lambda_R \lambda_T  f_{35}  \hspace{5mm} , \hspace{5mm} f_{7}' = \frac{\lambda_R}{\lambda_S}  f_{7}  \ .
\end{array}
\end{equation}

\subsubsection*{Type IIA with O$8$-planes}

The transformation of the metric fluxes is of the form
\begin{equation}
\label{IIA_O8_SL(2)_transformation_metric}
\begin{array}{c}
\omega_1' = \left( \frac{\lambda_R \lambda_T \lambda_U}{\lambda_S} \right)^{\frac12} \, \omega_1 
\hspace{5mm} , \hspace{5mm} 
\omega_2' = \left( \frac{\lambda_R \lambda_T \lambda_U^3}{\lambda_S} \right)^{\frac12} \, \omega_2 \hspace{5mm} , \hspace{5mm}
\omega_3' = \left( \frac{\lambda_R \lambda_T}{\lambda_S \lambda_U} \right)^{\frac12} \, \omega_3 \ ,\\[2mm]
\omega_4' =  \left( \frac{\lambda_R \lambda_T \lambda_U}{\lambda_S} \right)^{\frac12} \, \omega_4 \hspace{5mm} , \hspace{5mm}
\omega_5' = \left( \frac{\lambda_R \lambda_T}{\lambda_S \lambda_U^3} \right)^{\frac12} \, \omega_5 
\hspace{5mm} , \hspace{5mm} 
\omega_6' =  \left( \frac{\lambda_R \lambda_T}{\lambda_S \lambda_U} \right)^{\frac12} \, \omega_6
\ ,
\end{array}
\end{equation}
whereas gauge fluxes transform as
\begin{equation}
\label{IIA_O8_SL(2)_transformation_gauge}
\begin{array}{c}
h_3' = \frac{\lambda_T}{\lambda_S} \, h_3 
\hspace{5mm} , \hspace{5mm} 
f_2' = \lambda_R \lambda_T f_2
\hspace{5mm} , \hspace{5mm} 
f_{41}' = \left(\frac{\lambda_R \lambda_T^{3} \lambda_U^3}{\lambda_S} \right)^{\frac12} f_{41}
\hspace{5mm} , \hspace{5mm} 
f_{42}' =  \left(\frac{\lambda_R \lambda_T^{3} \lambda_U}{\lambda_S} \right)^{\frac12}  f_{42} \ ,\\[2mm]
f_{43}' = \left(\frac{\lambda_R \lambda_T^{3}}{\lambda_S \lambda_U} \right)^{\frac12}  f_{43}
\hspace{5mm} , \hspace{5mm} 
f_{44}' =  \left(\frac{\lambda_R \lambda_T^{3}}{\lambda_S \lambda_U^3} \right)^{\frac12} f_{44}  
\hspace{5mm} , \hspace{5mm} 
f_6' = \frac{\lambda_R}{\lambda_S} f_6 \ .
\end{array}
\end{equation}

\subsubsection*{Type IIA with O$6$-planes}

The transformation of the metric fluxes is of the form
\begin{equation}
\label{IIA_O6_SL(2)_transformation_metric}
\begin{array}{c}
\omega_1' = \left( \frac{\lambda_R \lambda_T \lambda_U}{\lambda_S} \right)^{\frac12} \, \omega_1 
\hspace{5mm} ,  \hspace{5mm}
\omega_3' = \left( \frac{\lambda_R \lambda_T \lambda_U}{\lambda_S} \right)^{\frac12} \, \omega_3 \hspace{5mm} , \hspace{5mm}
\omega_5' = \left( \lambda_R \lambda_S \lambda_T^3 \lambda_U \right)^{\frac12} \, \omega_5
\ ,
\end{array}
\end{equation}
whereas gauge fluxes transform as
\begin{equation}
\label{IIA_O6_SL(2)_transformation_gauge}
\begin{array}{c}
h_{31}' = \left( \frac{\lambda_R \lambda_T \lambda_U^{3}}{\lambda_S} \right)^{\frac12 }\, h_{31} 
\hspace{4mm} , \hspace{4mm} 
h_{32}' = \lambda_R \lambda_U h_{32}
\hspace{4mm} , \hspace{4mm} 
h_{33}' = \left( \lambda_R \lambda_S \lambda_T^3 \lambda_U^{3} \right)^{\frac12}  h_{33}   \ ,\\[2mm]
f_{0}' = \left(\frac{\lambda_R \lambda_T^3}{\lambda_S \lambda_U^3} \right)^{\frac12} f_{0} \hspace{4mm} , \hspace{4mm} 
f_{2}' =  \left(\frac{\lambda_R \lambda_T^3}{\lambda_S \lambda_U} \right)^{\frac12}  f_{2}
\hspace{4mm} , \hspace{4mm} 
f_{41}' = \frac{\lambda_R}{\lambda_S} \, f_{41}
\hspace{4mm} , \hspace{4mm} 
f_{42}' = \left( \frac{\lambda_R \lambda_T^{3} \lambda_U}{\lambda_S} \right)^{\frac12}  f_{42} \ ,\\[2mm]
f_{43}' = \lambda_R \lambda_T f_{43}
\hspace{4mm} , \hspace{4mm} 
f_6' = \left( \frac{\lambda_R \lambda_T^{3}\lambda_U^3 }{\lambda_S} \right)^{\frac12} f_6  \ .
\end{array}
\end{equation}

\subsubsection*{Type IIB with O$5$-planes}

The transformation of the metric fluxes is of the form
\begin{equation}
\label{IIB_O5_SL(2)_transformation_metric}
\begin{array}{c}
\omega_1' = \left( \frac{\lambda_R \lambda_S \lambda_T^3}{\lambda_U}\right)^{\frac12} \omega_1 
\hspace{5mm} , \hspace{5mm} 
\omega_2' = \lambda_T^2 \,  \omega_2
\hspace{5mm} , \hspace{5mm} 
\omega_3' = \frac{\lambda_T}{\lambda_S} \, \omega_3 \ , \\[3mm]
\omega_4' = \frac{\lambda_R}{\lambda_U} \, \omega_4
\hspace{5mm} , \hspace{5mm} 
\omega_5' = \left( \frac{\lambda_R \lambda_T}{\lambda_S \lambda_U}\right)^{\frac12} \omega_5
\hspace{5mm} , \hspace{5mm} 
\omega_6' = \left( \frac{\lambda_R \lambda_T}{\lambda_S \lambda_U} \right)^{\frac12} \omega_6 \ ,
\end{array}
\end{equation}
whereas gauge fluxes transform as
\begin{equation}
\label{IIB_O5_SL(2)_transformation_gauge}
\begin{array}{c}
h_{31}' = \left( \frac{\lambda_R \lambda_S \lambda_T^3}{\lambda_U^3}\right)^{\frac12} h_{31} 
\hspace{5mm} , \hspace{5mm} 
h_{32}' = \left( \frac{\lambda_R \lambda_T}{\lambda_S \lambda_U^3}\right)^{\frac12} h_{32} \ , \\[2mm]
f_{31}' = \frac{\lambda_R}{\lambda_S} \, f_{31}
\hspace{5mm} , \hspace{5mm}
f_{32}' = \left( \frac{\lambda_R \lambda_U \lambda_T^3}{\lambda_S}\right)^{\frac12} f_{32}
\hspace{5mm} , \hspace{5mm}
f_{33}' = \lambda_R \, \lambda_T \, f_{33}  \ , \\[2mm]
f_5' = \left( \frac{\lambda_R \lambda_T^{3}}{\lambda_S \lambda_U} \right)^{\frac12}  f_5 
\hspace{5mm} , \hspace{5mm}
f_7' = \left( \frac{\lambda_R \lambda_T^{3}}{\lambda_S \lambda_U^3} \right)^{\frac12} f_7  \ .
\end{array}
\end{equation}

\subsubsection*{Type IIB with O$3$-planes}

The transformation of the metric fluxes is of the form
\begin{equation}
\label{IIB_O3_SL(2)_transformation_metric}
\begin{array}{c}
\omega_1' = \lambda_R \,  \omega_1 
\hspace{3mm} , \hspace{3mm} 
\omega_2' =  \frac{\lambda_R}{\lambda_U} \, \omega_2
\hspace{3mm} , \hspace{3mm}
\omega_3' = \lambda_R  \lambda_U \, \omega_3
\hspace{3mm} , \hspace{3mm}
\omega_4' = \lambda_R \, \omega_4
\hspace{3mm} , \hspace{3mm}
\omega_5' =  \lambda_T^{2} \, \omega_5 \ ,
\end{array}
\end{equation}
whereas gauge fluxes transform as
\begin{equation}
\label{IIB_O3_SL(2)_transformation_gauge}
\begin{array}{c}
h_{31}' = \left( \frac{\lambda_R \lambda_S \lambda_T^3}{\lambda_U^3}\right)^{\frac12} h_{31} 
\hspace{5mm} , \hspace{5mm}
h_{32}' = \left( \frac{\lambda_R \lambda_S \lambda_T^3}{\lambda_U}\right)^{\frac12}  h_{32} \ ,
\\[2mm]
h_{33}' = \left( \lambda_R \lambda_S \lambda_T^3 \lambda_U \right)^{\frac12}   h_{33}
\hspace{5mm} , \hspace{5mm}
h_{34}' = \left(\lambda_R \lambda_S \lambda_T^3\lambda_U^3\right)^{\frac12} h_{34} \ ,
\\[2mm]
f_{31}' = \left(\frac{\lambda_R \lambda_T^3}{\lambda_S \lambda_U^3} \right)^{\frac12} \, f_{31}
\hspace{5mm} , \hspace{5mm}
f_{32}' = \left( \frac{\lambda_R \lambda_T^3}{\lambda_S \lambda_U}\right)^{\frac12}  f_{32} \ , \\[2mm]
f_{33}' = \left( \frac{\lambda_R \lambda_T^3 \lambda_U}{\lambda_S}\right)^{\frac12} f_{33} 
\hspace{5mm} , \hspace{5mm}
f_{34}' =  \left( \frac{\lambda_R \lambda_T^3  \lambda_U^3}{\lambda_S}\right)^{\frac12} f_{34}
\hspace{5mm} ,\\[2mm]
f_5' = \lambda_R  \lambda_T f_5 \ .
\end{array}
\end{equation}

\subsubsection*{Type IIA with O$2$-planes}

The transformation of the gauge fluxes is of the form
\begin{equation}
\label{IIA_O2_SL(2)_transformation_gauge}
\begin{array}{c}
h_{31}' = \left( \frac{\lambda_R \lambda_S \lambda_T^3}{\lambda_U^3}\right)^{\frac12} h_{31} 
\hspace{3mm} , \hspace{3mm} 
h_{32}' = \left( \frac{\lambda_R \lambda_S \lambda_T^3}{\lambda_U}\right)^{\frac12} h_{32}
\hspace{3mm} , \hspace{3mm}
h_{33}' = \left( \lambda_R \lambda_S \lambda_T^3 \lambda_U \right)^{\frac12} h_{33} \ , \\[3mm]
h_{34}' = \left( \lambda_R \lambda_S \lambda_T^3 \lambda_U^3 \right)^{\frac12}  h_{34}
\hspace{3mm} , \hspace{3mm}
h_{35}' = \lambda_T^2 \, h_{35} \hspace{5mm} , \hspace{5mm}
f_0' =  \frac{\lambda_R}{\lambda_S} f_0 \ , \\[2mm]
f_{41}' = \left(\frac{\lambda_R \lambda_T^3}{\lambda_S \lambda_U^3} \right)^{\frac12} \, f_{41}
\hspace{3mm} , \hspace{3mm}
f_{42}' = \left( \frac{\lambda_R \lambda_T^3}{\lambda_S \lambda_U}\right)^{\frac12}  f_{42}
\hspace{3mm} , \hspace{3mm}
f_{43}' = \left( \frac{\lambda_R \lambda_T^3\lambda_U}{\lambda_S}\right)^{\frac12} f_{43} \ , \\[2mm]
f_{44}' =  \left( \frac{\lambda_R \lambda_T^3\lambda_U^3}{\lambda_S}\right)^{\frac12}  f_{44}
\hspace{3mm} , \hspace{3mm}
f_{45}' = \lambda_R \lambda_T f_{45} \ .
\end{array}
\end{equation}
Note that metric fluxes are not permitted in the type IIA with O2 duality frame.

\newpage

\section{Mass spectra within half-maximal supergravity}
\label{App:Mass_spectra}

In this appendix we collect the gravitino and scalar mass spectra of all the inequivalent families of type IIB with O$9$, O$5$ and O$3$ flux vacua presented in the main text. In all the tables, the subscript in $\,n_{(s)}\,$ denotes the multiplicity $\,s\,$ of a given mass $\,n\,$ in the spectrum. Lastly, for the sake of presentation, we are omitting the boxes and the marks $\,*\,$ and $\,\dagger\,$ assigned to the flux vacua in the various tables appearing in the main text.

\subsection{Type IIB with O$9$-planes}
\label{App:IIB_O9_spectra}

Let us present the gravitino and scalar mass spectra for the type IIB with O$9$-planes flux vacua listed in Table~\ref{Table:flux_vacua_IIB_O9} (and its continuation).

\subsubsection*{Gravitino mass spectrum}

The gravitino mass spectra are collected in Tables~\ref{Table:flux_vacua_IIB_O9_gravitini_spectrum_Mkw} and \ref{Table:flux_vacua_IIB_O9_gravitini_spectrum_AdS}. For some of the flux vacua, the analytic expressions in terms of the flux parameters are rather lengthy. Let us present these special cases separately.
\\

\noindent - The gravitino mass spectrum for \textbf{vac~42,43} is given in terms of the functions $\mathtt{P}(\kappa_1,\kappa_2,\kappa_3)$ and $\mathtt{Q}(\kappa_1,\kappa_2,\kappa_3)$, whose expressions are quite involved. Explicitly,
\begin{equation}
\begin{aligned}
    \mathtt{P}(\kappa_1,\kappa_2,\kappa_3) =& \, 3 \kappa_3^5 \kappa_2 \left(5\kappa_1^4-10\kappa_1^2\kappa_2^2+\kappa_2^4 \right) + \kappa_3^4 P_5 - 2\kappa_3^3 \kappa_2 \left( \kappa_1^2+\kappa_2^2\right) P_4 + 2\kappa_3^2 \kappa_2^2 P_3 + \\[1mm]&- \kappa_3 \kappa_2^3 \left( \kappa_1^2+\kappa_2^2\right)^2 P_2 + \kappa_2^4 \left( \kappa_1^2+\kappa_2^2\right)^2 P_1 \ ,
\end{aligned}
\end{equation}
where
\begin{equation}
    \begin{array}{ccl}
         P_1 &=&  5 \left( \kappa_1^2 + \kappa_2^2 \right) - 3 \kappa_1 \sqrt{\mathcal{C}_1}\ , \\[2mm]
          P_2 &=& 12 \kappa_1^2 - 3\kappa_2^2 -20\kappa_1 \sqrt{\mathcal{C}_1}\ , \\[2mm]
          P_3 &=&  5 \left[ 3\kappa_1^6 + 5\kappa_1^4\kappa_2^2 + \kappa_2^4 \left( \kappa_1^2-\kappa_2^2\right)\right] - 9\kappa_1 \left( \kappa_1^4-\kappa_2^4\right) \sqrt{\mathcal{C}_1} \ , \\[2mm]
          P_4 &=& 6 \kappa_1^4 - 21\kappa_1^2\kappa_2^2 + 3 \kappa_2^4 -10\kappa_1 \left( \kappa_1^2-\kappa_2^2\right) \sqrt{\mathcal{C}_1} \ , \\[2mm]
          P_5 &=& 5 \left[ \kappa_1^6 -5\kappa_1^2\kappa_2^2\left(\kappa_1^2+\kappa_2^2\right)+\kappa_2^6 \right] - 3\kappa_1 \left( \kappa_1^4-10\kappa_1^2 \kappa_2^2 +5\kappa_2^4\right) \sqrt{\mathcal{C}_1} \ ,
    \end{array}
\end{equation}
and
\begin{equation}
    \mathtt{Q}(\kappa_1,\kappa_2,\kappa_3) = 6\kappa_3^3 \kappa_2 \left(3\kappa_1^2-\kappa_2^2 \right) + \kappa_3^2 P_8 - 2\kappa_3\kappa_2 \left(\kappa_1^2+\kappa_2^2\right) P_7 + \kappa_2^2 \left(\kappa_1^2+\kappa_2^2\right) P_6 \ ,
\end{equation}
where
\begin{equation}
    \begin{array}{ccl}
        P_6 &=&  11 \left( \kappa_1^2+\kappa_2^2 \right) - 6\kappa_1 \sqrt{\mathcal{C}_1} \ ,\\[2mm]
        P_7 &=& 6\kappa_1^2 - 3\kappa_2^2 -11\kappa_1 \sqrt{\mathcal{C}_1} \ , \\[2mm] 
        P_8 &=&  11 \left( \kappa_1^4-\kappa_2^4\right) - 6\kappa_1 \left( \kappa_1^2-3\kappa_2^2\right) \sqrt{\mathcal{C}_1} \ .
    \end{array}
\end{equation}

\noindent - The gravitino masses for \textcolor{ForestGreen}{\textbf{vac~46-49}} are given in terms of nine flux-dependent functions $\,N_{i}\,$ with $\,i = 1,\ldots,9$. They are defined as 
\begin{equation}
\begin{array}{rcl}
N_1 &=& \kappa_1^2 \left(\kappa_1-\kappa_3\right)^2 + \kappa_1 \kappa_2^2 \left( 3\kappa_1-2\kappa_3\right) + \kappa_2^4 \ , \\[2mm]
N_i(\kappa_1,\kappa_2,\kappa_3;c) &=& \kappa_1^4 \left(\kappa_1-\kappa_3\right)^2 + \kappa_1^2 \kappa_2^2 \left( 4\kappa_1^2-4\kappa_1\kappa_3+c \kappa_3^2\right) + 2\kappa_1\kappa_2^4 \left( 2\kappa_1-c\kappa_3\right) + c\kappa_2^6 \ ,
\end{array}
\end{equation}
in terms of the pairs $\,(i,c)=\left\lbrace(2,9),(3,\frac{37}{5}),(4,\frac{1}{4}),(5,\frac{5}{17}),(6,\frac{9}{4}),(7,\frac{37}{17}),(8,4),(9,\frac{17}{5}),(10,3)\right\rbrace$.

\subsubsection*{Scalar mass spectrum}

The scalar mass spectra are collected in Tables~\ref{Table:flux_vacua_IIB_O9_scalar_spectrum_Mkw} and \ref{Table:flux_vacua_IIB_O9_scalar_spectrum_AdS}. As before, we will present the special cases separately.
\\

\noindent - In order to present the scalar masses for \textbf{vac~42} we introduce three functions $\mathsf{R}_1(\kappa_1,\kappa_2,\kappa_3)$, $\mathsf{R}_2(\kappa_1,\kappa_2,\kappa_3)$ and $\mathsf{R}_3(\kappa_1,\kappa_2,\kappa_3)$. These are given by

\begin{equation}
\begin{array}{ccl}
    \mathsf{R}_1 &=& \left( \kappa_1^2 + \kappa_3^2 \right) \left( \kappa_2^2 - \kappa_3^2 \right)^2 \ , \\[2mm]
    \mathsf{R}_2 &=& \left( \kappa_1^2 + \kappa_2^2 \right) \left(\kappa_2\sqrt{ \mathcal{C}_1} + \kappa_1 \kappa_3 \right)^2  \ , \\[2mm]
    \mathsf{R}_3 &=& \kappa_2^6 - \kappa_2^5 \kappa_3 + \kappa_2^4 R_4 + \kappa_2^3 \kappa_3 R_3 + \kappa_1 \kappa_2^2 R_2 + \kappa_1^2 \kappa_2 \kappa_3 R_1 + R_0  \ ,
\end{array}
\end{equation}
where
\begin{equation}
\begin{array}{c}
R_0 =  \kappa_1^3 \kappa_3^2 \left( \kappa_1 + \sqrt{\mathcal{C}_1} \right)
\,\,\,\,\,,\,\,\,\,\,
R_1 =  2 \kappa_1^2 - 3 \kappa_3^2 + 2 \kappa_1 \sqrt{\mathcal{C}_1}
\,\,\,\,\,,\,\,\,\,\,
R_2 = \kappa_1^3 + \kappa_1^2 \sqrt{\mathcal{C}_1} - 3\kappa_3^2 \sqrt{\mathcal{C}_1} \ , \\[2mm]
R_3 =  \kappa_1^2  + \kappa_3^2 + 2\kappa_1 \sqrt{\mathcal{C}_1}
\,\,\,\,\,,\,\,\,\,\,
R_4 = 2 \kappa_1^2 - \kappa_3^2 + \kappa_1 \sqrt{\mathcal{C}_1} \ .
\end{array}
\end{equation}

\noindent - In order to present the scalar masses for \textbf{vac~43}, we introduce two functions $\mathtt{R}_1(\kappa_1,\kappa_2,\kappa_3)$ and $\mathtt{R}_2(\kappa_1,\kappa_2,\kappa_3)$, of the form
\begin{equation}
\begin{array}{ccl}
    %\mathtt{R}_2 &=& - 64 \, \mathtt{R}_1 \ , \\[2mm]
    \mathtt{R}_1 &=& \left( \kappa_1^2 + \kappa_2^2 \right)^2 \left[ \left( \kappa_2^2-\kappa_3^2\right)^5 \kappa_2^{10} +  \left( \kappa_2^2-\kappa_3^2\right)^4 R_4  +  \left( \kappa_2^2-\kappa_3^2\right)^3 R_3 + \left( \kappa_2^2-\kappa_3^2\right)^2 R_2 + R_1 \right] \ , \\[2mm]
    \mathtt{R}_2 &=& \, \kappa_2^2 \left( \kappa_2^2 - \kappa_3^2 \right)^2 \left[ \left( \kappa_2^2-\kappa_3^2\right)^5 \kappa_2^{10} +  \left( \kappa_2^2-\kappa_3^2\right)^4 R_9  +  \left( \kappa_2^2-\kappa_3^2\right)^3 R_8 + \left( \kappa_2^2-\kappa_3^2\right)^2 R_7 + R_6 \right] \\[1mm]&& + \,  R_5 \ ,
\end{array}    
\end{equation}
where
\begin{equation}
    \begin{array}{ccl}
        R_1 &=& \kappa_1^7 \left( 2 \kappa_2 \kappa_3 S_1 \sqrt{\mathcal{C}_1} + \kappa_1 S_2 \right) \ ,\\[2mm]
        R_2 &=&  2\kappa_1^5 \kappa_2^4 \left[ 6 \kappa_2 \kappa_3 \left( 5\kappa_2^4 + 30 \kappa_2^2\kappa_3^2 + 21 \kappa_3^4\right) \sqrt{\mathcal{C}_1} + 5 \kappa_1 \left( \kappa_2^6 + 27 \kappa_2^4\kappa_3^2 + 63\kappa_2^2 \kappa_3^4 +21 \kappa_3^6 \right) \right] \ ,\\[2mm]
        R_3 &=&  10 \kappa_1^3 \kappa_2^6 \left[ 4 \kappa_2\kappa_3 \left( \kappa_1^2 + 3\kappa_3^2\right) \sqrt{\mathcal{C}_1} + \kappa_1 \left( \kappa_2^4 + 18 \kappa_2^2 \kappa_3^2 + 21\kappa_3^4 \right) \right]\ ,\\[2mm]
        R_4 &=& 5 \kappa_1 \kappa_2^8 \left[ 2 \kappa_2 \kappa_3 \sqrt{\mathcal{C}_1} + \kappa_1 \left( \kappa_2^2 + 9\kappa_3^2 \right)\right] \ , \\[2mm]
        R_5 &=& \kappa_1^{11} \left(4\kappa_2 \kappa_3 S_3 \sqrt{\mathcal{C}_1} + \kappa_1 S_4 \right)\ ,\\[2mm]
        R_6 &=&  \kappa_1^7 \left( 4 \kappa_2 \kappa_3  S_5 \sqrt{\mathcal{C}_1} + 3\kappa_1 S_6 \right)\ ,\\[2mm]
        R_7 &=&  4 \kappa_1^5 \kappa_2^4 \left[ 2\kappa_2\kappa_3 \left( 15\kappa_2^4+110\kappa_2^2\kappa_3^2 + 99\kappa_3^4 \right) \sqrt{\mathcal{C}_1} + 5\kappa_1 \left(\kappa_2^6+33\kappa_2^4\kappa_3^2+99\kappa_2^2\kappa_3^4 + \frac{231}{5}\kappa_3^6 \right)\right],\\[2mm]
        R_8 &=&  5\kappa_1^3 \kappa_2^6 \left[ 4\kappa_2\kappa_3 \left( 3\kappa_2^2+11\kappa_3^2\right) \sqrt{\mathcal{C}_1} + 3\kappa_1 \left(\kappa_2^4 + 22\kappa_2^2\kappa_3^2 + 33\kappa_3^4 \right) \right] \ , \\[2mm]
        R_9 &=& 6\kappa_1 \kappa_2^8 \left[ 2\kappa_2 \kappa_3 \sqrt{\mathcal{C}_1} + \kappa_1 \left( \kappa_2^2+11\kappa_3^2 \right) \right] \ , \\[2mm]
    \end{array}
\end{equation}
and
\begin{equation}
\begin{array}{ccl}
    S_1 &=& 5\kappa_2^8 \left( \kappa_1^2+ 4\kappa_2^2 \right) + 20 \kappa_2^6 \kappa_3^2 \left( 3\kappa_1^2+ 8\kappa_2^2 \right) + 18 \kappa_2^4 \kappa_3^4 \left( 7\kappa_1^2 + 4\kappa_2^2 \right)   \\[2mm]
    &&  +  12 \kappa_2^2 \kappa_3^6 \left( 5\kappa_1^2 - 16 \kappa_2^2 \right)+ 5\kappa_3^8 \left( \kappa_1^2-12\kappa_2^2\right) \ ,\\[3mm]
    S_2 &=& \, \kappa_2^{10} \left( \kappa_1^2+ 5\kappa_2^2 \right) + 5 \kappa_2^8 \kappa_3^2 \left( 9\kappa_1^2+ 35\kappa_2^2 \right) + 30 \kappa_2^6 \kappa_3^4 \left( 7\kappa_1^2 + 15\kappa_2^2 \right)   \\[2mm]
        &&  +  210 \kappa_2^4 \kappa_3^6 \left( \kappa_1^2 -  \kappa_2^2 \right)+ 45 \kappa_2^2\kappa_3^8 \left( \kappa_1^2-10\kappa_2^2\right) + \kappa_3^{10} \left( \kappa_1^2 - 45\kappa_2^2 \right)\ ,\\[3mm]
    S_3 &=& 3\left( \kappa_2^2 + \kappa_3^2 \right) \left( \kappa_2^4 + 2\kappa_2^2 \kappa_3^2 + \kappa_3^4\right) \left( \kappa_2^4 + 14\kappa_2^2 \kappa_3^2 + \kappa_3^4\right) \ , \\[3mm]
    S_4 &=& \left( \kappa_2^4 + 6\kappa_2^2 \kappa_3^2 + \kappa_3^4\right) \left( \kappa_2^8 + 60\kappa_2^6 \kappa_3^2 + 134 \kappa_2^4\kappa_3^4 + 60 \kappa_2^2 \kappa_3^6 + \kappa_3^8 \right) \ , \\[3mm]
    S_5 &=& \, 15 \kappa_2^8 \left( \kappa_1^2 + 2\kappa_2^2 \right) + 20 \kappa_2^6 \kappa_3^2 \left( 11\kappa_1^2 + 15 \kappa_2^2 \right) + 66 \kappa_2^4 \kappa_3^4 \left( 9\kappa_1^2 + 4\kappa_2^2\right)  \\[2mm]
    && + 396 \kappa_2^2 \kappa_3^6 \left( \kappa_1^2 - \kappa_2^2\right)+ 11\kappa_3^8 \left( 5\kappa_1^2-18\kappa_2^2 \right) \ ,\\[3mm]
    S_6 &=& \, \kappa_2^{10} \left( 2\kappa_1^2 + 5\kappa_2^2 \right) + 5 \kappa_2^8 \kappa_3^2 \left( 22\kappa_1^2 + 43 \kappa_2^2 \right) + 110 \kappa_2^6 \kappa_3^4 \left( 6\kappa_1^2 + 7\kappa_2^2 \right)  \\[2mm]
    &&+ 66 \kappa_2^4 \kappa_3^6 \left( 14 \kappa_1^2 - \kappa_2^2\right)+ 33 \kappa_2^2 \kappa_3^8 \left(10\kappa_1^2-23\kappa_2^2 \right) + 11\kappa_3^{10} \left(2\kappa_1^2-15\kappa_2^2 \right) \ .
\end{array}
\end{equation}
We remind the reader that the various flux-dependent quantities $\,\mathcal{C}'s\,$ were introduced in the caption of the continuation of Table~\ref{Table:flux_vacua_IIB_O9}.

\vspace{20mm}

\begin{table}[h!]
\begin{center}
%\scalebox{0.78}{
\renewcommand{\arraystretch}{1.8}
\begin{tabular}{!{\vrule width 1.5pt}c!{\vrule width 1pt}c!{\vrule width 1pt}c!{\vrule width 1pt}cccccc!{\vrule width 1pt}cc!{\vrule width 1pt}ccc!{\vrule width 1pt}c!{\vrule width 1pt}c!{\vrule width 1.5pt}}
\noalign{\hrule height 1.5pt}
     ID & Gravitino spectrum  \\
     \noalign{\hrule height 1pt}
     $\textbf{vac~1}$ &  $g^{-2}\,m^2_{3/2} = \left(\frac{\kappa^2}{16}\right)_{(6)}, \, \left(\frac{9\kappa^2}{16}\right)_{(2)} $ \\ 
\noalign{\hrule height 1.5pt}
\end{tabular}
%}
\caption{Gravitino masses at the type~IIB with O$9$-planes Mkw$_3$ vacuum of Table~\ref{Table:flux_vacua_IIB_O9}.} 
\label{Table:flux_vacua_IIB_O9_gravitini_spectrum_Mkw}
\end{center}
\end{table}

\vspace{20mm}

\begin{table}[h!]
\begin{center}
%\scalebox{0.78}{
\renewcommand{\arraystretch}{1.8}
\begin{tabular}{!{\vrule width 1.5pt}c!{\vrule width 1pt}c!{\vrule width 1pt}c!{\vrule width 1pt}cccccc!{\vrule width 1pt}cc!{\vrule width 1pt}ccc!{\vrule width 1pt}c!{\vrule width 1pt}c!{\vrule width 1.5pt}}
\noalign{\hrule height 1.5pt}
     ID & Scalar spectrum  \\
     \noalign{\hrule height 1pt}
     $\textbf{vac~1}$ &  $g^{-2}\,m^2 = \left(\frac{\kappa^2}{4}\right)_{(18)}, \, 0_{(46)} $ \\ 
\noalign{\hrule height 1.5pt}
\end{tabular}
%}
\caption{Scalar masses at the type~IIB with O$9$-planes Mkw$_3$ vacuum of Table~\ref{Table:flux_vacua_IIB_O9}.} 
\label{Table:flux_vacua_IIB_O9_scalar_spectrum_Mkw}
\end{center}
\end{table}

\newpage

\begin{table}[h!]
\begin{center}
\scalebox{0.68}{
\renewcommand{\arraystretch}{2.4}
\begin{tabular}{!{\vrule width 1.5pt}c!{\vrule width 1pt}c!{\vrule width 1.5pt}c!{\vrule width 1pt}c!{\vrule width 1.5pt}}
\noalign{\hrule height 1.5pt}
\multicolumn{4}{!{\vrule width 1.5pt}c!{\vrule width 1.5pt}}{Gravitino spectrum} \\
\noalign{\hrule height 1pt}
     ID & $ m^2_{3/2} \, L^2$ & ID & $ m^2_{3/2} \, L^2$  \\
     \noalign{\hrule height 1pt}
     $\textcolor{Purple}{\textbf{vac~2}}$ & $  25_{(7)} $, $9_{(1)}$& $ \textbf{vac~26} $ &  \multirow{2}{*}{$m^2_{3/2} \, L^2  = \left( 1 \pm \sqrt{20+\frac{32\kappa\xi}{\kappa^2+\xi^2}} \right)^2_{(3)}, \, 9_{(1)}, \, 1_{(1)}$}\\
    \cline{1-3}
     $\textcolor{Purple}{\textbf{vac~3}}$ & $  25_{(1)}$, $9_{(7)}$ & \textbf{vac~27} & \\
    \hline
     $\textcolor{Purple}{\textbf{vac~4}} $ & $  49_{(3)}, \, 9_{(4)}, \, 1_{(1)} $ & $\textbf{vac~28}$ & \multirow{2}{*}{$  25_{(4)}, \, 9_{(4)} $} \\
     \cline{1-3} 
     $\textcolor{Purple}{\textbf{vac~5}}$ & $  25_{(3)}, \, 9_{(1)}, \, 1_{(4)}$ &  $ \textbf{vac~29} $ & \\
\noalign{\hrule height 1pt}
     $\textbf{vac~6} $ & $   25_{(4)}$, $9_{(4)}$ & $\textbf{vac~30}$ & $ \left( 1 \pm \sqrt{10+\frac{12\kappa\xi}{\kappa^2+\xi^2}} \right)^2_{(4)}$\\
     \hline
     $\textbf{vac~7}$ & $   9_{(4)}$, $1_{(4)}$  & $ \textbf{vac~31} $ & $   \left( 1 \pm \sqrt{26+\frac{20\kappa\xi}{\kappa^2+\xi^2}} \right)^2_{(3)}  , \, \left( 1 \pm \sqrt{10-\frac{12\kappa\xi}{\kappa^2+\xi^2}} \right)^2_{(1)} $ 
     \\
     %\noalign{\hrule height 1pt}
     \clineB{1-2}{2}\cline{3-4}
      \textcolor{ForestGreen}{$\textbf{vac~8}$} & $    \left(1\pm2\sqrt{\frac{4\kappa^2+9\xi^2}{\kappa^2+\xi^2}} \right)^2_{(3)}, \, \left(1\pm2\sqrt{\frac{4\kappa^2+ \xi^2}{\kappa^2+\xi^2}} \right)^2_{(1)} $ & $\textbf{vac~32} $ & $   \left( 1 \pm \sqrt{10-\frac{12\kappa\xi}{\kappa^2+\xi^2}} \right)^2_{(4)}$ \\ 
     \hline
     \textcolor{ForestGreen}{$\textbf{vac~9}$} & $   \left(1\pm2\sqrt{\frac{\kappa^2+ 4\xi^2}{\kappa^2+\xi^2}} \right)^2_{(4)} $ & $ \textbf{vac~33} $ & $  \left( 1 \pm \sqrt{26-\frac{20\kappa\xi}{\kappa^2+\xi^2}} \right)^2_{(3)}  , \, \left( 1 \pm \sqrt{10+\frac{12\kappa\xi}{\kappa^2+\xi^2}} \right)^2_{(1)} $\\
     \cline{1-2}\clineB{3-4}{2} 
     \textcolor{ForestGreen}{$\textbf{vac~10}$} & $  25_{(4)}, \, 9_{(4)} $ & $ \textbf{vac~34} $ & \multirow{2}{*}{$ \left( 1 \pm \sqrt{10+\frac{12\kappa\xi}{\kappa^2+\xi^2}} \right)^2_{(4)} $} \\ 
     \cline{1-3} 
     \textcolor{ForestGreen}{$\textbf{vac~11}$} & $   \left(1\pm2\sqrt{\frac{\kappa^2+9\xi^2}{\kappa^2+\xi^2}} \right)^2_{(3)} , \, 9_{(1)}, \, 1_{(1)} $ & \textbf{vac~35} & \\
     \noalign{\hrule height 1pt} 
     $ \textbf{vac~12} $ & $\left(1\pm2\sqrt{\frac{\kappa^2+ 4\xi^2}{\kappa^2+\xi^2}} \right)^2_{(4)}$ & $\textbf{vac~36}$ & $ 25_{(4)}, \, 9_{(4)}$   
     \\ 
     \hline
     $ \textbf{vac~13} $ & $\left(1\pm2\sqrt{\frac{4\kappa^2+ \xi^2}{\kappa^2+\xi^2}} \right)^2_{(4)}$ & $ \textbf{vac~37} $ & $   9_{(4)}, \, 1_{(4)}$ \\
     \cline{1-2}\clineB{3-4}{2}
     $ \textbf{vac~14} $ &  $ 25_{(4)}$, $9_{(4)}$ & $ \textbf{vac~38} $ & $ \left( 1 \pm \sqrt{10+\frac{6\kappa\xi}{\kappa^2+\xi^2}} \right)^2_{(4)} $ \\ 
     \hline
     $ \textbf{vac~15} $ & $  9_{(4)}$, $1_{(4)} $ & $ \textbf{vac~39} $ & $ \left( 1 \pm \sqrt{10-\frac{6\kappa\xi}{\kappa^2+\xi^2}} \right)^2_{(4)} $ \\ 
     \noalign{\hrule height 1pt} 
     $\textbf{vac~16}$ & $ \left(1\pm2\sqrt{\frac{9\kappa^2+4\xi^2}{\kappa^2+\xi^2}} \right)^2_{(3)}, \, \left(1\pm2\sqrt{\frac{\kappa^2+4\xi^2}{\kappa^2+\xi^2}} \right)^2_{(1)} $ & $ \textbf{vac~40} $ & $   \left( 1\pm2\sqrt{\frac{\kappa^2+4\xi^2}{\kappa^2+\xi^2}} \right)^2_{(4)}$\\
    \hline
     $\textbf{vac~17}$ & $  \left(1\pm2\sqrt{\frac{4\kappa^2+\xi^2}{\kappa^2+\xi^2}} \right)^2_{(4)} $ & $\textbf{vac~41}$ & $ \left( 1\pm2\sqrt{\frac{4\kappa^2+9\xi^2}{\kappa^2+\xi^2}} \right)^2_{(3)} , \, \left( 1\pm2\sqrt{\frac{4\kappa^2+\xi^2}{\kappa^2+\xi^2}} \right)^2_{(1)} $ \\
     \cline{1-2}\clineB{3-4}{2}
     $\textbf{vac~18}$ & $  25_{(4)}, \, \,9_{(4)} $ & $ \textbf{vac~42} $ &  $\left[ \frac{\mathtt{Q}(\kappa_1,\kappa_2,\kappa_3) \pm 2\sqrt{2} \sqrt{\left( \kappa_1^2+\kappa_2^2\right) \mathtt{P}(\kappa_1,\kappa_2,\kappa_3)}}{\left( \kappa_1^2+\kappa_2^2\right) \mathcal{C}_4} \right]_{(4)} $ \\
    \hline
     $\textbf{vac~19}$ & $ \left(1\pm2\sqrt{\frac{9\kappa^2+\xi^2}{\kappa^2+\xi^2}} \right)^2_{(3)} , 9_{(1)}, \, 1_{(1)} $ & $\textbf{vac~43} $ &   $\left[ \frac{\mathtt{Q}(-\kappa_1,\kappa_2,\kappa_3) \pm 2\sqrt{2} \sqrt{\left( \kappa_1^2+\kappa_2^2\right) \mathtt{P}(-\kappa_1,\kappa_2,\kappa_3)}}{\left( \kappa_1^2+\kappa_2^2\right) \mathcal{C}_5} \right]_{(4)} $ \\
    \clineB{1-2}{2}\cline{3-4}
     $\textbf{vac~20}$ & $  \left(1\pm2\sqrt{\frac{\kappa^2+4\xi^2}{\kappa^2+\xi^2}} \right)^2_{(4)}$& \textcolor{RoyalBlue}{$\textbf{vac~44}$} & $   9_{(4)}, \, 1_{(4)}$ \\
     \hline
    $\textbf{vac~21}$ & $ \left(1\pm2\sqrt{\frac{4\kappa^2+9\xi^2}{\kappa^2+\xi^2}} \right)^2_{(3)} , \, \left(1\pm2\sqrt{\frac{4\kappa^2+\xi^2}{\kappa^2+\xi^2}} \right)^2_{(1)}$ & \textcolor{RoyalBlue}{$\textbf{vac~45}$} & $   25_{(4)}, \, 9_{(4)}$\\
    \noalign{\hrule height 1pt}
     $\textbf{vac~22}$ & $  25_{(4)}$, $9_{(4)}$ & \textcolor{ForestGreen}{$\textbf{vac~46}$} & $ \left( \frac{5N_3}{\left(\kappa_1^2+\kappa_2^2 \right) N_1} \pm 4 \sqrt{\frac{N_2}{\left(\kappa_1^2+\kappa_2^2 \right) N_1}} \right)_{(3)}, \, 9_{(1)}, \, 1_{(1)} $ \\
    \hline
     $\textbf{vac~23}$ & $  \left(1\pm2\sqrt{\frac{\kappa^2+9\xi^2}{\kappa^2+\xi^2}} \right)^2_{(3)} , \, 9_{(1)} , \, 1_{(1)}$ & \textcolor{ForestGreen}{$\textbf{vac~47}$} & $   25_{(4)}, \, 9_{(4)} $\\
     \clineB{1-2}{2}\cline{3-4}
     $ \textbf{vac~24} $ & $  \left( 1\pm \sqrt{10-\frac{6\xi}{\sqrt{\kappa^2+\xi^2}}}\right)^2_{(4)} $ & \textcolor{ForestGreen}{$\textbf{vac~48}$} & $   \left[ \frac{17 N_7}{\left( \kappa_1^2 + \kappa_2^2 \right) N_1} \pm 4 \sqrt{\frac{4 N_6}{\left( \kappa_1^2 + \kappa_2^2 \right) N_1}} \right]_{(3)} , \, \left[ \frac{17 N_5}{\left( \kappa_1^2 + \kappa_2^2 \right) N_1} \pm 4 \sqrt{\frac{4 N_4}{\left( \kappa_1^2 + \kappa_2^2 \right) N_1}}\right]_{(1)}$\\
     \hline
     $\textbf{vac~25} $ & $ \left( 1\pm \sqrt{10+\frac{6\xi}{\sqrt{\kappa^2+\xi^2}}}\right)^2_{(4)} $ & \textcolor{ForestGreen}{$\textbf{vac~49}$} & $  \left[ \frac{5 N_{9}}{\left( \kappa_1^2+\kappa_2^2\right) N_1} \pm 4 \sqrt{\frac{N_8}{\left( \kappa_1^2+\kappa_2^2\right)N_1}}\right]_{(4)}$
     \\
\noalign{\hrule height 1.5pt}
\end{tabular}}
\caption{Gravitino masses at the type~IIB with O$9$-planes AdS$_3$ vacua of Table~\ref{Table:flux_vacua_IIB_O9}.} 
\label{Table:flux_vacua_IIB_O9_gravitini_spectrum_AdS}
\end{center}
\end{table}

\begin{table}[h!]
\begin{center}
\scalebox{0.68}{
\renewcommand{\arraystretch}{1.8}
\begin{tabular}{!{\vrule width 1.5pt}c!{\vrule width 1pt}c!{\vrule width 1.5pt}c!{\vrule width 1pt}c!{\vrule width 1.5pt}}
\noalign{\hrule height 1.5pt}
\multicolumn{4}{!{\vrule width 1.5pt}c!{\vrule width 1.5pt}}{Scalar spectrum} \\
\noalign{\hrule height 1pt}
     ID & $ m^2 \, L^2$ & ID & $ m^2 \, L^2$  \\
     \noalign{\hrule height 1pt}
     $\textcolor{Purple}{\textbf{vac~2}}$ &\multirow{4}{*}{$\begin{array}{ccl} m^2 \, L^2 &=& 15_{(8)},\, 8_{(19)},\, 3_{(8)},\, 0_{(29)} \\ \Delta &=& 5_{(8)},\, 4_{(19)},\, 3_{(8)},\, 2_{(29)} \end{array}$} & $ \textbf{vac~26} $ &  \multirow{2}{*}{$ 16 \left( \frac{\kappa^2+\kappa\xi+\xi^2}{\kappa^2+\xi^2} \right)_{(10)}, \, 4\left[\frac{\left(\kappa+\xi\right)^2}{\kappa^2+\xi^2}\right]_{(18)}, \, 4\left[\frac{\left(\kappa-\xi\right)^2}{\kappa^2+\xi^2}\right]_{(2)}, \, 8_{(7)}, \, 0_{(27)} $}\\
    \cline{1-1} \cline{3-3}
     $\textcolor{Purple}{\textbf{vac~3}}$ &  & \textbf{vac~27} & \\
    \cline{1-1} \cline{3-4}
     $\textcolor{Purple}{\textbf{vac~4}} $ &  & $\textbf{vac~28}$ & \multirow{2}{*}{$16 \left( \frac{\kappa^2-\kappa\xi+\xi^2}{\kappa^2+\xi^2} \right)_{(10)}, \, 4\left[\frac{\left(\kappa+\xi\right)^2}{\kappa^2+\xi^2}\right]_{(2)}, \, 4\left[\frac{\left(\kappa-\xi\right)^2}{\kappa^2+\xi^2}\right]_{(18)}, \, 8_{(7)}, \, 0_{(27)}$} \\
     \cline{1-1} \cline{3-3}
     $\textcolor{Purple}{\textbf{vac~5}}$ & &  $ \textbf{vac~29} $ & \\
\noalign{\hrule height 1pt}
     $\textbf{vac~6} $ & \multirow{2}{*}{$4 \left(1 \pm \sqrt{\frac{\kappa^2}{\kappa^2+4\xi^2}}\right)_{(18)}, \, 8_{(1)}, \, 0_{(27)}$} & $ \textbf{vac~30} $ & \multirow{2}{*}{$\begin{array}{rcl}16 \left( \frac{\kappa^2+\kappa\xi+\xi^2}{\kappa^2+\xi^2} \right)_{(5)},  &  4\left[\frac{\left(\kappa+\xi\right)^2}{\kappa^2+\xi^2}\right]_{(12)}, & 8_{(13)}, \\[-2mm] 12\left[\frac{\left(\kappa+\xi\right)^2}{\kappa^2+\xi^2}\right]_{(5)}, & 4\left[\frac{\left(\kappa-\xi\right)^2}{\kappa^2+\xi^2}\right]_{(1)},  & 0_{(28)}\end{array}$} \\
     \cline{1-1}\cline{3-3}
     $\textbf{vac~7}$ &   & $ \textbf{vac~31} $ &  
     \\
     \clineB{1-2}{2}\cline{3-4}
      \textcolor{ForestGreen}{$\textbf{vac~8}$} & \multirow{2}{*}{$\begin{array}{rcl}8\left(\frac{\kappa^2+3\xi^2}{\kappa^2+\xi^2}\right)_{(5)}, &  24\left(\frac{\xi^2}{\kappa^2+\xi^2}\right)_{(5)}, & 8_{(13)}, \\[-2mm] 8\left(\frac{\kappa^2}{\kappa^2+\xi^2}\right)_{(1)}, &  8\left(\frac{\xi^2}{\kappa^2+\xi^2}\right)_{(12)},&  0_{(28)}\end{array}$} & $\textbf{vac~32} $ & \multirow{2}{*}{$\begin{array}{rcl}16 \left( \frac{\kappa^2-\kappa\xi+\xi^2}{\kappa^2+\xi^2} \right)_{(5)}, & 4\left[\frac{\left(\kappa+\xi\right)^2}{\kappa^2+\xi^2}\right]_{(1)}, & 8_{(13)}, \\[-2mm] 12\left[\frac{\left(\kappa-\xi\right)^2}{\kappa^2+\xi^2}\right]_{(5)}, & 4\left[\frac{\left(\kappa-\xi\right)^2}{\kappa^2+\xi^2}\right]_{(12)},   & 0_{(28)}\end{array}$} \\ 
     \cline{1-1}\cline{3-3}
     \textcolor{ForestGreen}{$\textbf{vac~9}$} &  & $ \textbf{vac~33} $ & \\
     \cline{1-2}\clineB{3-4}{2} 
     \textcolor{ForestGreen}{$\textbf{vac~10}$} & \multirow{2}{*}{$8\left(\frac{\kappa^2+3\xi^2}{\kappa^2+\xi^2}\right)_{(10)},\, 8\left(\frac{\xi^2}{\kappa^2+\xi^2}\right)_{(18)}, \, 8\left(\frac{\kappa^2}{\kappa^2+\xi^2}\right)_{(2)}, \,  8_{(7)},\, 0_{(27)}$}  & $ \textbf{vac~34} $ & \multirow{2}{*}{$4\left[\frac{\left(\kappa+\xi\right)^2}{\kappa^2+\xi^2}\right]_{(9)}, \, 4\left[\frac{\left(\kappa-\xi\right)^2}{\kappa^2+\xi^2}\right]_{(9)}, \, 8_{(10)}, \, 0_{(36)}$} \\ 
     \cline{1-1}\cline{3-3} 
     \textcolor{ForestGreen}{$\textbf{vac~11}$} &  & \textbf{vac~35} & \\
     \noalign{\hrule height 1pt} 
     $ \textbf{vac~12}$ & \multirow{2}{*}{$8\left(\frac{\xi^2}{\kappa^2+\xi^2}\right)_{(9)},  8\left(\frac{\kappa^2}{\kappa^2+\xi^2}\right)_{(9)}, \, \, 8_{(10)}, \, 0_{(36)}$} & $\textbf{vac~36}$ & \multirow{2}{*}{$4\left[\frac{\left(\kappa+\xi\right)^2}{\kappa^2+\xi^2}\right]_{(18)}, \, 4\left[\frac{\left(\kappa-\xi\right)^2}{\kappa^2+\xi^2}\right]_{(18)}, \, 8_{(1)}, \, 0_{(27)}$}  
     \\ 
     \cline{1-1}\cline{3-3}
     $ \textbf{vac~13} $ &  & $ \textbf{vac~37} $ & \\
     \cline{1-2}\clineB{3-4}{2}
     $ \textbf{vac~14} $ & \multirow{2}{*}{$8\left(\frac{\xi^2}{\kappa^2+\xi^2}\right)_{(18)}, \, 8\left(\frac{\kappa^2}{\kappa^2+\xi^2}\right)_{(18)},  \, 8_{(1)}, \, 0_{(27)}$}  & $ \textbf{vac~38} $ & \multirow{2}{*}{$ 4\left(1+\frac{\kappa}{\sqrt{\kappa^2+\xi^2}}\right)_{(9)}, \,4\left(1-\frac{\kappa}{\sqrt{\kappa^2+\xi^2}}\right)_{(9)}, \, 8_{(10)}, \, 0_{(36)}$} \\ 
     \cline{1-1}\cline{3-3}
     $ \textbf{vac~15} $ &  & $ \textbf{vac~39} $ &  \\ 
     \noalign{\hrule height 1pt} 
     $\textbf{vac~16}$ & \multirow{2}{*}{$\begin{array}{rcl} 24\left(\frac{\kappa^2}{\kappa^2+\xi^2}\right)_{(5)}, & 8\left(\frac{3\kappa^2+\xi^2}{\kappa^2+\xi^2}\right)_{(5)}, & 8_{(13)},\\[-2mm] 8\left(\frac{\kappa^2}{\kappa^2+\xi^2}\right)_{(12)}, & 8\left(\frac{\xi^2}{\kappa^2+\xi^2}\right)_{(1)}, & 0_{(28)} \end{array}$} & $ \textbf{vac~40} $ & \multirow{2}{*}{$\begin{array}{rcl} 24\left(\frac{\xi^2}{\kappa^2+\xi^2}\right)_{(5)}, &  8\left(\frac{\kappa^2+3\xi^2}{\kappa^2+\xi^2}\right)_{(5)}, & 8_{(13)}, \\[-2mm] 8\left(\frac{\xi^2}{\kappa^2+\xi^2}\right)_{(12)}, & 8\left(\frac{\kappa^2}{\kappa^2+\xi^2}\right)_{(1)}, & 0_{(28)}\end{array}$}\\
    \cline{1-1}\cline{3-3}
     $\textbf{vac~17}$ &  & $\textbf{vac~41}$ & \\
     \cline{1-2}\clineB{3-4}{2}
     $\textbf{vac~18}$ & \multirow{2}{*}{$8\left(\frac{3\kappa^2+\xi^2}{\kappa^2+\xi^2}\right)_{(10)},\, 8\left(\frac{\kappa^2}{\kappa^2+\xi^2}\right)_{(18)}, \, 8\left(\frac{\xi^2}{\kappa^2+\xi^2}\right)_{(2)}, \,  8_{(7)},\, 0_{(27)}$} & $ \textbf{vac~42} $ & $4 \left[  \frac{\mathsf{R}_3(\kappa_1,\kappa_2,\kappa_3)}{\mathsf{R}_1} \right]_{(9)} , \, 4 \left[ \frac{\mathsf{R}_3(-\kappa_1,\kappa_2,-\kappa_3)}{\mathsf{R}_1} \right]_{(9)} , \, 8 \left[ \frac{\mathsf{R_2}}{\mathsf{R_1}} \right]_{(10)}, \, 0_{(36)}$ \\
    \cline{1-1}\cline{3-4}
     $\textbf{vac~19}$ &  & $\textbf{vac~43} $ & $\left[ 4\left( 1\pm \sqrt{1-\frac{\mathtt{R}_2}{\mathtt{R}_1}} \right) \right]_{(9)}, \, 8_{(10)}, \, 0_{(36)}$  \\
    \clineB{1-2}{2}\cline{3-4}
      \multirow{2}{*}{$\textbf{vac~20}$} &\multirow{4}{*}{$\begin{array}{rcl} 24\left(\frac{\xi^2}{\kappa^2+\xi^2}\right)_{(5)}, & 8\left(\frac{\kappa^2+3\xi^2}{\kappa^2+\xi^2}\right)_{(5)}, & 8_{(13)}, \\[5mm] 8\left(\frac{\xi^2}{\kappa^2+\xi^2}\right)_{(12)}, & 8\left(\frac{\kappa^2}{\kappa^2+\xi^2}\right)_{(1)}, & 0_{(28)}\end{array}$} &  \multirow{2}{*}{\textcolor{RoyalBlue}{$\textbf{vac~44}$}} &\multirow{4}{*}{$\begin{array}{rl}4 \left[ 1 + \frac{\kappa_1^2 \left( \kappa_1-\kappa_2\right) + \kappa_3^2 \left( 3\kappa_1+\kappa_2 \right)}{\left(\kappa_1^2+\kappa_3^2\right) \sqrt{\left(\kappa_1-\kappa_2 \right)^2+4\kappa_3^2}} \right]_{(18)} , & 8_{(1)}, \\[5mm] 4 \left[ 1 - \frac{\kappa_1^2 \left( \kappa_1-\kappa_2\right) + \kappa_3^2 \left( 3\kappa_1+\kappa_2 \right)}{\left(\kappa_1^2+\kappa_3^2\right) \sqrt{\left(\kappa_1-\kappa_2 \right)^2+4\kappa_3^2}} \right]_{(18)}, &  0_{(27)} \end{array}$} \\
      &&&\\
     \cline{1-1}\cline{3-3}
     \multirow{2}{*}{$\textbf{vac~21}$} &  &  \multirow{2}{*}{\textcolor{RoyalBlue}{$\textbf{vac~45}$}} & \\
     &&&\\
    \noalign{\hrule height 1pt}
      \multirow{2}{*}{$\textbf{vac~22}$} &\multirow{4}{*}{$\begin{array}{crl}
       8\left(\frac{\kappa^2}{\kappa^2+\xi^2}\right)_{(2)}, & \multicolumn{2}{c}{8\left(\frac{\kappa^2+3\xi^2}{\kappa^2+\xi^2}\right)_{(10)},} \\[5mm] 8\left(\frac{\xi^2}{\kappa^2+\xi^2}\right)_{(18)}, & 8_{(7)}, & 0_{(27)}\end{array}$} &  \multirow{2}{*}{\textcolor{ForestGreen}{$\textbf{vac~46}$}} & \multirow{4}{*}{$\begin{array}{cc} 8\left[ \frac{\kappa_2^2 \left( \kappa_2^2-\kappa_1\kappa_3\right)^2}{\left(\kappa_1^2 + \kappa_2^2 \right) N_1}\right]_{(18)}, &   8\left[ \frac{\kappa_1^2  \left( \kappa_1^2-\kappa_1\kappa_3 + 2\kappa_2^2\right)^2}{\left(\kappa_1^2 + \kappa_2^2 \right) N_1}\right]_{(2)} , \\[5mm] 8\left[ \frac{N_{10}}{\left(\kappa_1^2 + \kappa_2^2 \right) N_1} \right]_{(10)}, & 8_{(7)} , \hspace{10mm} 0_{(27)}\end{array}$} \\
      &&&\\
    \cline{1-1}\cline{3-3}
      \multirow{2}{*}{$\textbf{vac~23}$} & &  \multirow{2}{*}{\textcolor{ForestGreen}{$\textbf{vac~47}$}} & \\
      &&&\\
     \noalign{\hrule height 1pt}
      \multirow{2}{*}{$ \textbf{vac~24} $} & \multirow{4}{*}{$4\left( 1 \pm \sqrt{\frac{\xi^2}{\kappa^2 + \xi^2}} \right)_{(9)} , \, 8_{(10)} , \, 0_{(36)}$} &  \multirow{2}{*}{\textcolor{ForestGreen}{$\textbf{vac~48}$}} & \multirow{4}{*}{$\begin{array}{rcl} 8 \left[ \frac{\kappa_1^2 \left( \kappa_1^2 -\kappa_1\kappa_3 + 2\kappa_2^2\right)^2}{\left( \kappa_1^2+\kappa_2^2 \right) N_1}\right]_{(1)}, & 8 \left[ \frac{\kappa_2^2 \left( \kappa_2^2 -\kappa_1\kappa_3\right)^2}{\left( \kappa_1^2+\kappa_2^2 \right) N_1}\right]_{(12)}, & 8_{(13)} ,\\[5mm] 24 \left[ \frac{\kappa_2^2 \left( \kappa_2^2 -\kappa_1\kappa_3\right)^2}{\left( \kappa_1^2+\kappa_2^2 \right) N_1}\right]_{(5)}, & 8\left[ \frac{N_{10}}{\left( \kappa_1^2+\kappa_2^2 \right) N_1} \right]_{(5)}, &  0_{(28)}\end{array}$} \\
      &&&\\
     \cline{1-1}\cline{3-3}
      \multirow{2}{*}{$\textbf{vac~25}$} &  &  \multirow{2}{*}{\textcolor{ForestGreen}{$\textbf{vac~49}$}} & \\
      &&&\\
\noalign{\hrule height 1.5pt}
\end{tabular}}
\caption{Scalar masses at the type~IIB with O$9$-planes AdS$_3$ vacua of Table~\ref{Table:flux_vacua_IIB_O9}. For \textcolor{Purple}{\textbf{vac~2}-\textbf{vac~5}}, we also report the conformal dimensions $\Delta$'s of the would-be CFT$_2$ dual operators, which are all integer-valued.} 
\label{Table:flux_vacua_IIB_O9_scalar_spectrum_AdS}
\end{center}
\end{table}

\newpage

\subsection{Type IIB with O$5$-planes}
\label{App:IIB_O5_spectra}

The gravitino and scalar mass spectra for the type IIB with O$5$-planes flux vacua listed in Table~\ref{Table:flux_vacua_IIB_O5} are presented in Table~\ref{Table:flux_vacua_IIB_O5_gravitini_scalar_spectrum}.
\vspace{10mm}

\begin{table}[h!]
\begin{center}
\scalebox{0.8}{
\renewcommand{\arraystretch}{2.3}
\begin{tabular}{!{\vrule width 1.5pt}c!{\vrule width 1.5pt}c!{\vrule width 1pt}c!{\vrule width 1.5pt}}
\noalign{\hrule height 1.5pt}
     ID & Gravitini spectrum & Scalar spectrum  \\
\noalign{\hrule height 1pt}
    $ \textcolor{red}{\textbf{vac~1}} $ & $g^{-2}m_{3/2}^2=0_{(8)}$ & $g^{-2}m^2=0_{(64)}$\\[2mm] 
     \hline     
     $ \textbf{vac~2} $ & $g^{-2} m^2_{3/2} = \left[\frac{(\kappa \pm \xi)^{2}}{16}\right]_{(3)} , \left[\frac{9(\kappa \pm \xi)^{2}}{16}\right]_{(1)}$ & $\begin{array}{cccccc}
     g^{-2} \, m^2 &=&  \left[ \frac{\left( \kappa + 2\xi \right)^2}{16} \right]_{(3)}, & \left(\frac{\kappa^2}{16} \right)_{(9)}, & \left(\frac{\kappa^2}4\right)_{(9)} , & 0_{(30)}, \\[-1mm]
     && \left[ \frac{\left( \kappa - 2\xi \right)^2}{16} \right]_{(3)}, &  \left(\frac{9\kappa^2}{16} \right)_{(1)}, & \left(\frac{\xi^2}4\right)_{(9)}, & \end{array}$  \\[5mm] 
     \hline 
     $ \textbf{vac~3} $ & $g^{-2} m^2_{3/2} = \left(\frac{9 \kappa^{2}}{16}\right)_{(2)} , \left(\frac{\kappa^{2}}{16}\right)_{(6)}$ & $g^{-2} \, m^2 =  \left( \frac{\kappa^2}{4} \right)_{(15)} , 0_{(49)} $ \\[2mm] 
\noalign{\hrule height 1pt}
     \textcolor{RoyalBlue}{$\textbf{vac~4}$} & $m^2_{3/2} L^{2} = 1_{(4)} , 9_{(4)}$ & \multirow{2}{*}{$\begin{array}{ccl}
        m^2 L^2 &=&  8_{(19)}, \ 0_{(45)}  \\[-3mm]
         \Delta &=&  4_{(19)}, \ 2_{(45)} 
     \end{array}$}
     \\[1mm] 
    \cline{1-2} 
     \textcolor{RoyalBlue}{$\textbf{vac~5}$} &  $m^2_{3/2} L^{2} = 9_{(4)} , 25_{(4)}$ &
     \\[1mm]
\noalign{\hrule height 1pt}
     $ \textbf{vac~6} $ & $m^2_{3/2} L^{2} = 1_{(3)} , 9_{(4)} , 25_{(1)}$ & \multirow{2}{*}{$\begin{array}{ccl}
     m^2 L^2 &=&  8_{(10)}, \ 4_{(18)} , \ 0_{(36)}
     %\\[-3mm] \Delta &=&  4_{(10)}, \ (1+\sqrt{5})_{(18)} , \ 2_{(36)}
     \end{array}$} \\[1mm] 
     \cline{1-2} 
     $ \textbf{vac~7} $ & $m^2_{3/2} L^{2} = 1_{(1)} , 9_{(4)} , 25_{(3)}$  & \\[1mm]
\noalign{\hrule height 1pt}
     \textcolor{ForestGreen}{$\textbf{vac~8}$} & $m^2_{3/2} L^{2} = 1_{(1)} , 9_{(1)} , 25_{(3)}  , 49_{(3)}$ & \multirow{2}{*}{$\begin{array}{ccl}
     m^2 L^2 &=& 24_{(10)}, \ 8_{(25)}, \ 0_{(29)} \\[-3mm]
      \Delta &=& 6_{(10)}, \ 4_{(25)}, \ 2_{(29)} \end{array}$} \\[1mm] 
     \cline{1-2} 
     \textcolor{ForestGreen}{$\textbf{vac~9}$} &  $m^2_{3/2} L^{2} = 9_{(4)} , 25_{(4)}$ & \\[1mm]
\noalign{\hrule height 1pt}
     $ \textbf{vac~10} $ & $m^2_{3/2} L^{2} = 9_{(3)} , 49_{(3)} , 81_{(1)}  , 169_{(1)}$ & $\begin{array}{ccl}
     m^2 L^2 &=&  80_{(3)}, \ 48_{(9)}, \ 24_{(4)}, \ 8_{(7)}, \ 0_{(41)} \\[-3mm]
     \Delta &=&  10_{(3)}, \ 8_{(9)}, \ 6_{(4)}, \ 4_{(7)}, \ 2_{(41)}
     \end{array}$  \\[1mm]
     \hline 
     $ \textbf{vac~11} $ & $m^2_{3/2} L^{2} = 25_{(6)} , 49_{(1)} , 225_{(1)}$  & $\begin{array}{ccl} 
     m^2 L^2 &=& 48_{(15)}, \ 8_{(13)}, \ 0_{(36)} \\[-3mm]
     \Delta &=& 8_{(15)}, \ 4_{(13)}, \ 2_{(36)}
     \end{array}$ \\[1mm]
\noalign{\hrule height 1pt} 
     \textcolor{Purple}{$\textbf{vac~12}$} &  $m^2_{3/2} L^{2} = 1_{(4)} , 9_{(1)} , 25_{(3)}$ & \multirow{4}{*}{$\begin{array}{ccl}
     m^2 L^2 &=&  15_{(8)}, \ 8_{(19)}, \ 3_{(8)}, \ 0_{(29)} \\[-3mm]
     \Delta &=&  5_{(8)}, \ 4_{(19)}, \ 3_{(8)}, \ 2_{(29)}
     \end{array}$} \\[1mm]
     \cline{1-2}
     \textcolor{Purple}{$\textbf{vac~13}$} & $m^2_{3/2} L^{2} = 1_{(1)} , 9_{(4)} , 49_{(3)}$ &  \\[1mm]
     \cline{1-2}
     \textcolor{Purple}{$\textbf{vac~14}$} &  $m^2_{3/2} L^{2} = 9_{(7)} , 25_{(1)}$ & \\[1mm]
     \cline{1-2}
     \textcolor{Purple}{$\textbf{vac~15}$} &  $m^2_{3/2} L^{2} = 9_{(1)} , 25_{(7)}$ & \\[1mm]
\noalign{\hrule height 1.5pt}
\end{tabular}}
\caption{Gravitino and scalar masses at the type~IIB with O$5$-planes vacua of Table~\ref{Table:flux_vacua_IIB_O5}. We also report the conformal dimensions $\Delta$'s of the would-be CFT$_2$ dual operators for those AdS$_{3}$ vacua for which all of them are integer-valued.}
\label{Table:flux_vacua_IIB_O5_gravitini_scalar_spectrum}
\end{center}
\end{table}

\newpage

\subsection{Type IIB with O$3$-planes}
\label{App:IIB_O3_spectra}

The gravitino and scalar mass spectra for the type IIB with O$3$-planes Mkw$_{3}$ flux vacuum in Table~\ref{Table:flux_vacua_IIB_O3} are presented in Table~\ref{Table:flux_vacua_IIB_O3_gravitini_scalars_spectrum}. In order to present the spectra, we have introduced the flux-dependent quantities 
\begin{equation}
\label{M_functions_O3}
\begin{array}{lclc}
M_{1} =  \left(\kappa_{1}+\kappa_{3}\right)^2 + \left(\kappa_{4}+\kappa_{2}\right)^2 
& \,\,\,,\,\,\, &
M_{3} =   4 (\kappa_2^2+\kappa_3^2) + (\kappa_4-\kappa_2)^2 + (\kappa_1-\kappa_3)^2 & , \\[2mm]
M_{2} = \left(\kappa_{1}-3\kappa_{3}\right)^2 + \left(\kappa_{4}-3\kappa_{2}\right)^2   & \,\,\,,\,\,\, &
M_{4} =  4 (\kappa_1^2+\kappa_4^2) + (\kappa_1+3\kappa_3)^2 + (\kappa_4+3\kappa_2)^2 & .
\end{array}
\end{equation}

\begin{table}[h]
\begin{center}
\scalebox{0.83}{
\renewcommand{\arraystretch}{2}
\begin{tabular}{!{\vrule width 1.5pt}c!{\vrule width 1pt}c!{\vrule width 1pt}c!{\vrule width 1.5pt}}
\noalign{\hrule height 1.5pt}
     ID & Gravitino spectrum  & Scalar spectrum \\
\noalign{\hrule height 1pt}
     $ \textbf{vac~1} $ &  $\begin{array}{rcl}16 \, g^{-2} m^2_{3/2} & = & \left(\sqrt{M_1} \pm \kappa_5\right)^{2}_{(3)} \, , \\
     &&\left(\sqrt{M_2} \pm 3\kappa_5\right)^{2}_{(1)}\end{array}$ & $\begin{array}{rclll}32 \, g^{-2} \, m^2 &=& \left( M_4 \pm 3\sqrt{M_1 M_2}\right)_{(1)} \, , & \multicolumn{2}{l}{2\left( \sqrt{M_1} \pm 2\kappa_5 \right)^{2}_{(3)} \,,} \\
     && \left( M_3 \pm \sqrt{M_1 M_2}\right)_{(9)} \,, & 8\left(\kappa_5^2\right)_{(9)} \,, &  0_{(29)}\end{array}$\\[2mm]  
\noalign{\hrule height 1.5pt}
\end{tabular}}
\caption{Gravitino and scalar masses at the type~IIB with O$3$-planes vacuum of Table~\ref{Table:flux_vacua_IIB_O3}.}
\label{Table:flux_vacua_IIB_O3_gravitini_scalars_spectrum}
\end{center}
\end{table}

\FloatBarrier

\bibliographystyle{JHEP}
\bibliography{references}

\end{document}